\documentclass[aps,prd,twocolumn,
               notitlepage,mathrsfs,
               amssymb,amsmath,amsfonts,
               superscriptaddress,
               longbibliography,
               nofootinbib,floatfix]{revtex4-2}

\usepackage{aas_macros}
\usepackage{natbib}
\usepackage{color}
\usepackage{graphicx}
\usepackage{dcolumn}
\usepackage{bm}
\usepackage{pifont}
\usepackage{comment}

\usepackage{url}
\usepackage[linktocpage,breaklinks]{hyperref}
\usepackage[capitalize]{cleveref}
\usepackage[usenames,dvipsnames]{xcolor}
 \usepackage{ulem}
\hypersetup{colorlinks=true,
            citecolor=NavyBlue,
            linkcolor=magenta,
            urlcolor=magenta}

\begin{document}

\title{Neutrino pair annihilation driven jets from black-hole torus systems}

\author{Kyohei Kawaguchi}
\affiliation{Max Planck Institute for Gravitational Physics (Albert Einstein Institute), Am M\"{u}hlenberg 1, Potsdam-Golm, 14476, Germany}
\affiliation{Center of Gravitational Physics and Quantum Information,
 Yukawa Institute for Theoretical Physics, 
Kyoto University, Kyoto, 606-8502, Japan}

\author{Sho Fujibayashi}
\affiliation{Frontier Research Institute for Interdisciplinary Sciences, Tohoku University, Sendai 980-8578, Japan}
\affiliation{Astronomical Institute, Graduate School of Science, Tohoku University, Sendai 980-8578, Japan}
\affiliation{Max Planck Institute for Gravitational Physics (Albert Einstein Institute), Am M\"{u}hlenberg 1, Potsdam-Golm, 14476, Germany}

\author{Masaru Shibata}
\affiliation{Max Planck Institute for Gravitational Physics (Albert Einstein Institute), Am M\"{u}hlenberg 1, Potsdam-Golm, 14476, Germany}
\affiliation{Center of Gravitational Physics and Quantum Information,
 Yukawa Institute for Theoretical Physics, 
Kyoto University, Kyoto, 606-8502, Japan} 

\newcommand{\angstrom}{\text{\normalfont\AA}}
\newcommand{\rednote}[1]{{\color{red} (#1)}}
\newcommand{\addms}[1]{{\color{blue} {#1}}}
\newcommand{\addkk}[1]{{\color{magenta} #1}}
\newcommand{\sho}[1]{{\color{green} #1}}

\begin{abstract}
We perform axisymmetric general relativistic radiation-viscous hydrodynamics simulations of black hole (BH)-torus systems with full Boltzmann Monte-Carlo neutrino transport to investigate the role of neutrino-antineutrino pair annihilation in launching relativistic outflows. Our models span a wide range of BH spins, torus masses, and viscosity parameters. We find that the pair annihilation leads to the formation of relativistic fireballs in most cases, except for those with low black-hole spin and high viscosity. The isotropic-equivalent energies of these outflows reach $\lesssim 10^{51}\,{\rm erg}$ with durations $\lesssim 0.2\,{\rm s}$. While this is insufficient to explain the brightest short gamma-ray bursts (sGRBs), our results suggest that the pair annihilation may account for some low-luminosity sGRBs and GRB precursors. We also provide updated scaling relations for the pair annihilation energy deposition rate as a function of accretion rate, and discuss the sensitivity of outflow properties to numerical resolution and floor density.
\end{abstract}


\maketitle

\section{Introduction}\label{sec:intro}

Gamma-ray bursts (GRBs) represent the most energetic and extreme astrophysical events observed in the Universe. They produce relativistic outflows and are considered to be among the systems in which the strongest gravitational and magnetic fields as well as the highest density and temperature environments in the Universe are realized. It is an interesting system where all the fundamental interactions ---general relativistic gravity, equation of state of dense nuclear matter, neutrino radiation, and magnetohydrodynamic (MHD) effects--- play an important role. Therefore, GRBs provide a unique opportunity to investigate extreme environmental physics that cannot be reproduced on Earth. However, the definitive understanding of the central engine of GRBs has not yet been achieved. The recent observations~\citep{1986ApJ...308L..43P,1986ApJ...308L..47G,2010ApJ...708....9F,2022ApJ...940...56F,2022ApJ...940...57N,2022MNRAS.515.4890O,1998Natur.395..672I,2003ApJ...591L..17S,1999ApJ...524..262M} have yielded the hypotheses that GRBs are associated with neutron star (NS)  merger~\citep{1986ApJ...308L..43P,1986ApJ...308L..47G,2010ApJ...708....9F,2022ApJ...940...56F,2022ApJ...940...57N,2022MNRAS.515.4890O} and the rapidly-rotating massive star collapse, so-called collapsar~\citep{1998Natur.395..672I,2003ApJ...591L..17S,1999ApJ...524..262M,Shibata:2025gix}. In particular, relatively short duration GRBs ($< 2\,{\rm s}$) have been observationally suggested to be driven by NS mergers~\citep{Nakar:2007yr,Berger:2013jza}. However, long-duration GRBs with kilonova-like transients, which are thought to be specifically associated with NS mergers, were recently observed~\citep{GRB211211AG,GRB211211AR,GRB211211AT,GRB230307AL}, and complicate the interpretation of GRB progenitors.

One of the primary mechanisms proposed for relativistic outflow formation is via strong magnetic fields. In particular, the Blandford-Znajek (BZ) mechanism~\citep{Blandford:1982di} is one of the promising mechanisms for launching relativistic outflows, in which the rotational energy of a spinning black hole (BH) is extracted as electromagnetic energy through the magnetic field penetrating the BH and converted into plasma kinetic energy~\cite{Komissarov:2005wj,McKinney:2006tf,Barkov:2007us,Komissarov_2009,Tchekhovskoy:2011zx,Bromberg:2015wra,Gottlieb:2021srg,Hayashi:2024jwt}. The magnetic field amplified via dynamo process in a strongly magnetized rapidly-rotating NS formed in NS mergers can also be the source for launching a jet~\citep{Metzger:2011,Mosta:2020hlh,Most:2023sft,Kiuchi:2023obe}. 

Another proposed mechanism is the energy deposition via neutrino-antineutrino pair annihilation~\citep{1993ApJ...405..278M,  1996A&A...305..839J,1999ApJ...518..356P}. In this scenario, the fractional energy of high-luminosity neutrinos and antineutrinos emitted from the accretion disk or a hot NS is deposited in the plasma in the vicinity of the central object due to neutrino pair annihilation. This energy deposition can lead to the formation of an over-pressured region—or a neutrino-driven fireball—near the polar axis. As the surrounding baryon density decreases over time, this fireball can expand along the funnel, accelerating matter to relativistic speeds.

To date, there is no definitive consensus on the major mechanism of driving the GRB jets. Moreover, reflecting the progenitors and formation process, the BH/NS accretion disk systems can have diversity in their properties~\cite{Shibata:2019wef,Kyutoku:2021icp}, including whether the central object is a BH or a NS. The quantitative dependence of the jet formation process on the central objects has not been well understood. 

Theoretical studies of these mechanisms require accurately solving the equations that incorporate the effects of fluid dynamics, magnetic fields, and neutrino radiation within the framework of general relativity. Since many physical processes come into play in the central engines of GRBs, numerical simulations in realistic setups are indispensable for quantitatively elucidating their launching mechanism. In the last decade, numerical simulations of such systems have made great progress. For example, nowadays several numerical relativity (NR) simulations have been performed to investigate accretion disk formation, jet formation via magnetic fields, and matter ejection consistently with the effects of magnetic fields and neutrino radiation and comprehensively starting from NS mergers~\citep{Shibata:2017xdx,Fujibayashi:2017puw,Fujibayashi:2020qda,Hayashi:2021oxy,Hayashi:2022cdq,Kiuchi:2022nin,Just:2023wtj,Shibata:2021xmo,Shibata:2021bbj,Hayashi:2024jwt} and gravitational collapse of massive stars~\cite{Muller:2010ymw,Fujibayashi:2021wvv,Fujibayashi:2022xsm,Fujibayashi:2023oyt,Shibata:2023tho,Kuroda:2023mzi,Shibagaki:2023tmh,Kuroda:2024xbe,Issa:2024sts, Shibata:2025gix}. These studies have greatly advanced our understanding of the mechanisms of jet formation, particularly through magnetic fields, and have enabled us to predict consistent and realistic multi-messenger signals.

However, there is still room for improvement. One of the most important aspects is to improve the treatment of neutrino radiative transfer. Given the significant dependence of neutrino reaction rates on both neutrino energy and angular distribution~\citep{Bruenn:1985en,2004StellarCollapse}, it is essential to accurately solve the distribution functions of neutrinos for the quantification of radiation transfer effects. However, the Boltzmann equation governing the evolution of neutrino radiation fields has 6 degrees of freedom, 3 in space and 3 in momentum space, and it is currently computationally challenging to incorporate them completely due to the large number of degrees of freedom. For this reason, neutrino radiative transfer is often treated only approximately~\citep{Galeazzi:2013mia,Neilsen:2014hha,2013ApJ...772..127T,2014MNRAS.439..503S,2014MNRAS.441.3177M,2015MNRAS.447...49S,Sekiguchi:2015dma,2015PhRvD..91l4021F,Sekiguchi:2016bjd,Radice:2016dwd,Radice:2021jtw,2022arXiv220504487K,Sun:2022vri,Werneck:2022exo}, and the systematic errors due to this approximate treatment of radiative transfer remain uncertain (see, e.g.,~\cite{Foucart:2024npn} for the comparison work of the radiative transfer schemes in the inspiral phase and first few ms after the onset of a binary NS (BNS) merger). 

Moreover, jet launching due to neutrino-antineutrino pair annihilation (hereafter referred to as pair annihilation) has been little investigated. While there have been several studies that estimate pair annihilation effects post-processively from snapshots of density and temperature profiles obtained from the numerical simulations~\cite{Ruffert:1998qg,Setiawan:2004xy,Setiawan:2005ah,Richers:2015lma,Perego:2017fho,Sumiyoshi:2020bdh}, there are only a few numerical studies (still with an approximate radiative transfer method) in which the energy deposition rates and jet formation are computed dynamically and self-consistently incorporating pair annihilation~\citep{Just:2015dba,Fujibayashi:2017xsz,Fujibayashi:2017puw,Sun:2022vri}. In particular, there is no study that has dynamically investigated how and in which situations pair annihilation and their interplay with the magnetic fields lead to jet formation, and what are the effects on the multi-messenger signals.

Recently, we developed a new neutrino radiation viscous-hydrodynamics code employing the so-called Monte-Carlo (MC) method~\citep{2012ApJ...755..111A,2015ApJS..217....9R,2015ApJ...807...31R,2018MNRAS.475.4186F,2019PhRvD.100b3008M,2019ApJS..241...30M,Foucart:2020qjb,2021ApJ...920...82F,2022ApJ...933..226R,Izquierdo:2023fub}, which enables us to study the evolution of BH accretion disk systems fully solving the Boltzmann equation~\cite{Kawaguchi:2022tae,Kawaguchi:2024naa}. In this code, pair annihilation can be consistently incorporated. In fact, the formation of relativistic outflows from BH accretion disks due to pair annihilation has been demonstrated for the first time fully solving the Boltzmann equation~\cite{Kawaguchi:2024naa}. In this paper, we extend our study for a variety of setups of BH-torus systems, which can be formed as a result of NS mergers, focusing on their properties of neutrino emission, energy deposition due to pair annihilation, and relativistic outflow. In particular, pair annihilation including both electron and heavy lepton type neutrinos are considered by taking the effect of the non-thermal distribution functions into account. 

This paper is organized as follows: In Sec.~\ref{sec:method}, we describe the numerical methods employed in our code. In Sec.~\ref{sec:model}, we present our model setup of a BH-torus system studied in this paper. In Sec.~\ref{sec:diag}, we describe the definitions of several key quantities used for presenting the results. Sec.~\ref{sec:res} presents the results of the simulations for a BH-torus system. In Sec.~\ref{sec:dis}, we discuss the implication of our results obtained in this work. Finally, Sec.~\ref{sec:sum} is devoted to a summary. Throughout this paper, $c$ and $G$ denote the speed of light and gravitational constant, respectively, and the geometrical units of $c=G=1$ are employed unless otherwise stated.

\section{Numerical Methods}\label{sec:method}
We investigate the relativistic outflow formation in BH-torus systems driven by neutrino pair annihilation by performing axisymmetric general relativistic neutrino-radiation viscous-hydrodynamics simulations. We employ the numerical code developed in~\cite{Kawaguchi:2024naa}. In this code, the fixed background metric of a spinning BH described in Kerr-Schild coordinates is employed~\cite{1998ApJ...507L..67F}, and the equatorial plane symmetry as well as axial symmetry is imposed when evolving the matter and neutrino radiation fields. 
The equation of state includes contributions from relativistic particles composed of photons, electrons, and positrons, and non-relativistic particles composed of free protons, free neutrons, and $\alpha$-particles. The matter field is treated as a viscous fluid following the formulation of general-relativistic viscous hydrodynamics~\cite{Shibata:2017jyf} approximately taking the angular momentum transport and heating effect induced by magneto-hydrodynamics turbulence into account~\cite{Balbus:1998ja} (see Sec.~\ref{sec:dis-obs} for a brief speculative discussion on the potential impact of full magnetohydrodynamics effects). The viscous coefficient $\nu$ is modeled as $\nu=\alpha_{\rm vis} c_{\rm s} H_{\rm vis}$ following the so-called $\alpha$-viscosity description~\cite{1973A&A....24..337S}, with $\alpha_{\rm vis}$, $H_{\rm vis}$, and $c_{\rm s}$ being the dimensionless $\alpha$-viscous parameter, a scale height, and the sound speed, respectively.  We set $\alpha_{\rm vis}=0.05$ and $H_{\rm vis}=9\,{\rm km}\,\left(M_{\rm BH}/3\,M_\odot\right)\approx 2 M_{\rm BH}$ as the default setting where $M_{\rm BH}$ is the BH mass. 

Neutrino radiative transfer is solved by a MC method, which allows accurate modeling of non-thermal effects in neutrino spectra. The resulting radiative force feedback and electron number change are fully coupled with the hydrodynamics evolution. Three species of neutrinos are considered in this study: electron neutrino ($\nu_e$), electron antineutrino (${\bar \nu}_e$), and the other neutrino species ($\nu_x$), which represents all of the heavy lepton neutrinos, ($\nu_\mu$, ${\bar \nu}_\mu$, $\nu_\tau$, and ${\bar \nu}_\tau$). For neutrino weak interaction processes, we consider electron/positron captures by free protons/neutrons and electron-positron pair annihilation for neutrino emission. In particular, pair annihilation is taken into account with the effect of the non-thermal distributions. For the scattering process, iso-energy scattering by free protons and neutrons is considered. The expressions for the interaction rates are taken from~\cite{Bruenn:1985en,2004StellarCollapse,Horowitz:2001xf}. 

Note that the effect of the finite electron mass is consistently neglected for the equation of state and neutrino interaction rates. This simplification is also qualitatively reasonable in the BH-torus system studied in this paper since the typical matter temperature is above $1\,{\rm MeV}$.

\begin{table*}
\caption{BH-torus models studied in this paper: model name, initial torus mass ($M_{\rm torus}$), initial angular momentum profile parameter ($n$), BH mass ($M_{\rm BH}$), dimensionless BH spin ($\chi_{\rm BH}$), $\alpha$-viscous parameter ($\alpha_{\rm vis}$), initial radii of the inner and outer edges of the torus ($r_{\rm in}$ and $r_{\rm out}$, respectively), the radius up to which neutrino radiation is solved ($r_{\rm rad}$), mass averaged cylindrical torus radius ($\left<R\right>$), mass averaged torus scale height ratio ($\left<H/R\right>$), finest grid size ($\Delta x_{\rm min}$), grid number per axis $N_{\rm grid}$, and target MC packet number parameter ($N_{\rm trg}$). The initial entropy per baryon is set to $6\,k_{\rm B}$ for all models, where $k_{\rm B}$ is the Boltzmann constant. The model MT01s08DF01 shares the same setup as the fiducial model MT01s08, except that it employs a floor density that is smaller by a factor of 10.}
\centering
\begin{tabular}{c|c|c|c|c|c|c|c|c|c|c|c}\hline
Model	& $M_{\rm torus}\,[M_\odot]$  &   $n$   &   $M_{\rm BH}\,[M_\odot]$ & $\chi_{\rm BH}$ & $\alpha_{\rm vis}$ & $(r_{\rm in},r_{\rm out},r_{\rm rad})[M_{\rm BH}]$  & $\left<R\right>\,[M_\odot]$& $\left<H/R\right>$&$ \Delta x_{\rm min}\,{\rm [}M_{\rm BH}{\rm ]}$ & $ N_{\rm grid}$& $ N_{\rm trg}$ \\\hline\hline
MT01s08     &   $0.10$ &   $1/7$     &   $3$   & $0.8$& 0.05& $(2.7,40,100)$&36&0.164&0.2 & 320  &$120$\\
MT01s08LR     &   $0.10$ &   $1/7$     &   $3$   & $0.8$& 0.05& $(2.7,40,100)$&36&0.164& 0.23 & 280  &$120$\\
MT01s08HR     &   $0.10$ &   $1/7$     &   $3$   & $0.8$& 0.05& $(2.7,40,100)$&36&0.164&0.13 & 480 &$120$\\
MT01s08DF01     &   $0.10$ &   $1/7$     &   $3$   & $0.8$& 0.05& $(2.7,40,100)$&36&0.164&0.2 & 320  &$120$\\\hline
MT01s08v002     &   $0.10$ &   $1/7$     &   $3$   & $0.8$& 0.02& $(2.7,40,100)$&36&0.164& 0.2 & 320 &$120$\\
MT01s08v010     &   $0.10$ &   $1/7$     &   $3$   & $0.8$& 0.1& $(2.7,40,100)$&36&0.164& 0.2 & 320 &$120$\\
MT01s08v015     &   $0.10$ &   $1/7$     &   $3$   & $0.8$& 0.15& $(2.7,40,100)$&36&0.164& 0.2 & 320 &$120$\\\hline
MT01s08ni10 &   $0.10$ &   $1/10$    &   $3$   & $0.8$& 0.05& $(2.5,27,100)$&28&0.172& 0.2 & 320 &$120$\\
MT01s08ni5  &   $0.10$ &   $1/5$     &   $3$   & $0.8$& 0.05& $(2.8,102,110)$&59&0.160& 0.2 & 320 &$120$ \\
MT03s08     &   $0.30$ &   $1/7$     &   $3$   & $0.8$& 0.05& $(2.5,61,100)$&46&0.198& 0.2 & 320 &$120$\\
\hline
MT01s0      &   $0.10$ &   $1/7$     &   $3$ & $0.0$& 0.05& $(5.2,134,140)$&92&0.173& 0.2 & 320 & $120$\\
MT01s095    &   $0.11$ &   $1/7$     &   $3$ & $0.95$& 0.05& $(1.9,21,100)$&21&0.173& 0.13 & 480 &$36$\\
MT03s095    &   $0.30$ &   $1/7$     &   $3$ & $0.95$& 0.05& $(1.9,34,100)$&28&0.208& 0.13 & 480 &$36$\\
\hline
MT02s08MB6  &   $0.20$ &   $1/7$     &   $6$   & $0.8$& 0.05& $(2.6,24,10 0)$&53&0.129& 0.2  & 320 &$120$\\
MT02s08MB6v0025  &   $0.20$ &   $1/7$     &   $6$   & $0.8$& 0.025& $(2.6,24,10 0)$&53&0.129& 0.2  & 320 &$120$\\
\hline
\end{tabular}
\label{tb:model}
\end{table*}

\section{Model}\label{sec:model}
We construct the initial condition of BH-torus systems following the same procedures as described in~\cite{Kawaguchi:2024naa}. A variety of BH-torus configurations are considered to investigate the dependence of the pair annihilation rate and the properties of relativistic outflows on the system parameters.

Table~\ref{tb:model} lists the models of BH-torus systems studied in this paper. As a fiducial model, we employed the same configuration (MT01s08) as in~\cite{Kawaguchi:2024naa}. MT01s08v002, MT01s08v010, and MT01s08v015 are the models with the same configuration as in the fiducial model but with $\alpha_{\rm vis}=0.02$, 0.10, and 0.15, respectively. MT01s08ni10 and MT01s08ni5 have different angular momentum profiles. Specifically, the specific angular momentum, $l$, is defined in relation to the angular velocity, $\Omega$, by the scaling law $l \propto \Omega^{-n}$, with $n = 1/10$ and $1/5$ for the former and latter models. Here, $n=1/3$ corresponds to the Kepler orbital motion in the Newtonian limit, and $n=1/7$ is employed for the fiducial model. The torus becomes less compact and thinner for $n \rightarrow 1/3$. In fact, the average cylindrical radius and scale height of the torus defined by
\begin{align}
\left<R\right>=\frac{1}{M_{\rm torus}}\int \rho_* x d^3x
\end{align}
and
\begin{align}
\left<\frac{H}{R}\right>=\frac{1}{M_{\rm torus}}\int \rho_* \left|\frac{z}{x}\right|d^3x
\end{align}
become large and small, respectively, for $n \rightarrow 1/3$. Here, $\rho_*=\rho \alpha u^t \sqrt{\gamma}=\rho w \sqrt{\gamma}$ is the conserved baryon mass density of the matter with $w(=\alpha u^t)$, $\alpha$, $u^t$, and $\gamma$ being the Lorentz factor, lapse function, (upper) time component of the four velocity, and determinant of the spatial metric, respectively.

MT03s08 is a model with the same configuration as in the fiducial model, except that the torus mass is increased by a factor of 3. MT01s0 and MT01s095 are models with dimensionless BH spins of 0 and 0.95, respectively. MT03s095 is a model with the same configuration as in MT01s095 with the torus mass increased by a factor of 3. MT02s08MB6 and MT02s08MB6v0025 are models in which both the torus and BH masses are twice as large as the fiducial model. These models mimic a post-merger remnant of a BH-NS merger with the NS being tidally disrupted at the onset of the merger. $\alpha_{\rm vis} = 0.05$ and $0.025$ are employed for MT02s08MB6 and MT02s08MB6v0025, respectively.

Note that the torus mass deviates slightly from the target value due to the relaxation procedure used to determine the electron fraction, $Y_e$, and radiation field profiles during the initial data generation (see~\cite{Kawaguchi:2024naa} for details). Nevertheless, the deviation remains below 8\% in all cases, and the resulting systematic error is expected to be small compared to other sources of uncertainty (see Sec.~\ref{sec:dis}).

The cylindrical coordinate system is employed for solving the viscous-hydrodynamics, and $x$ and $z$ are assigned to the cylindrical radius and the vertical coordinate, respectively. The grid is set to be non-uniform in both $x$ and $z$ directions, with the grid-spacing increasing outward with a constant rate $\eta$, which is $\eta=1.0125$ by default. 
The simulations with low ($\eta=1.0143$) and high ($\eta=1.0083$) grid resolutions are performed for the fiducial model to check the numerical convergence of the results (MT01s08LR and MT01s08HR, respectively). The innermost (and hence the finest) grid spacing, $\Delta x_\mathrm{min}$, is determined so that for each coordinate the grid covers from 0 to $2500\,M_\odot \approx 3750\,{\rm km}$ with $320$ grid-cells by default, while 280 and 480 grid-cells are employed for low and high resolution runs, respectively.
\footnote{In this study, our primary goal is to investigate the semi-quantitative behavior of neutrino pair annihilation across a wide range of physical setups, while acknowledging the presence of various sources of errors, such as the choice of the equation of state and initial conditions. Therefore, we do not conduct a full convergence study, also due to computational limitations. Nevertheless, we note that such an analysis will be essential for future work aimed at fully quantitative predictions.}
We set the time interval of the simulation as $\Delta t=\Delta x_\mathrm{min}/2$. 

When employing a shock-capturing scheme for hydrodynamics, vacuum cannot be handled and an artificial atmosphere is necessary. We initialize the atmosphere with the rest-mass density, temperature, and $Y_e$ of $10\,{\rm g/cm^3}$, $0.036\,{\rm MeV}$, and 0.5, respectively, for $r\leq 300\,M_\odot$\footnote{This initial artificial atmosphere is applied throughout the computational domain for models MT01s08 and MT01s095, as their initial data were constructed following the setup in the previous study~\citep{Kawaguchi:2024naa}. Nevertheless, the difference in the initial atmosphere setup has a negligible effect on the results.}, and impose the following floor values for the rest of the computational domain. Here, $r=\sqrt{x^2+z^2}$ is the coordinate spherical radius. The floor values for the conserved rest-mass density ($\rho_*$) and temperature are set to $\rho_{\rm floor,0}\times {\rm min}\left[(r/300\,M_\odot)^{-3},1\right]$ and $300\,{\rm eV}$, respectively, as lower bounds during the evolution. We employ $\rho_{\rm floor,0}=1\,{\rm g/cm^3}$ for all the models. To evaluate the effect due to the presence of the floor density, we also run a simulation with the 10 times smaller value of $\rho_{\rm floor,0}$ (MT01s08FD01). The results of models with varied grid resolution and density floor settings are presented in detail in App.~\ref{app:err}.

For solving the radiation fields, we set $N_{\rm trg}=120$ and $r_{\rm abs}=0.1$ (see~\cite{Kawaguchi:2024naa} for details). With this setup, $\approx 10^7$--$10^8$ MC packets are solved in each time step. Exceptionally, we employ $N_{\rm trg}=36$ for MT01s095 and MT03s095 to reduce the computational cost, while we confirm in the previous study that the results are approximately unchanged (see Appendix C in~\cite{Kawaguchi:2024naa}). To reduce the computational cost, we limit the region of solving the radiation fields within $r\leq r_{\rm rad} = 100 M_{\rm BH}$ by default in this work. The larger value of $r_{\rm rad}$ is employed for the model with a large radial extension of the torus (see Table~\ref{tb:model}), so as to initially cover the entire torus. We justify this treatment by checking that the absorption and emission time scales are always more than an order of magnitude longer than the simulation time.

The simulations are performed on the Momiji cluster at Max Planck Computing and Data Facility, where each node is equipped with two Intel Xeon Platinum 8360Y CPUs (a total of 72 cores per node), using 16 nodes. Depending on the model and grid resolution, each simulation required between $\approx 100$ and $\approx 500$ hours of computation time.

\section{Diagnostics}\label{sec:diag}
We briefly summarize various quantities used in the analysis of our simulations. The mass accretion rate onto the BH, ${\dot M}_{\rm fall}$, is determined by integrating the mass flux on the event horizon as 
\begin{align}
    {\dot M}_{\rm fall}&=-\int_{{\rm EH}} dS_i\, \rho_* v^i,
\end{align}
where $dS_i$ is a surface element and $v^i$ is the spatial coordinate velocity defined by $u^i/u^t$ with $u^\mu$ being the fluid four velocity. The total mass accreted onto the BH, $M_{\rm fall}$, is given by integrating ${\dot M}_{\rm fall}$ over time.

Neutrino luminosity and number emission rate for the $\nu_i$ neutrino species ($L_{\nu_i}$ and ${\dot N}_{\nu_i}$, respectively) are obtained from the MC packets that pass through at $r=r_{\rm rad}$:
\begin{align}
    L_{\nu_i}&=-\frac{1}{\Delta t}\sum_{k,{\rm escape}} w_k^{\nu_i} p_{(k),t},\\
    {\dot N}_{\nu_i}&=\frac{1}{\Delta t}\sum_{k,{\rm escape}} w_k^{\nu_i},
\end{align}
where 
$w_k^{\nu_i}$ and $p_{(k),t}$ denote the weight and lower time components of the neutrino 4-momentum for $k${\it -th} MC packets, respectively.

The total pair annihilation energy deposition rate is calculated by
\begin{align}
    L_{\rm pair}=\sum_{{\nu_i}}\int d^3x\sqrt{\gamma}\,{\rm max}\left(Q_{\rm pair},0\right)
\end{align}
with $Q_{\rm pair}=-G_{{\nu_i},t}^{{\rm pair}}$, where $G_{{\nu_i},t}^{{\rm pair}}$ denotes the lower time component of the radiation 4-force density due to neutrino pair process. Note that  $G_{{\nu_i},t}^{{\rm pair}}$ denotes the net deposition rate in which both heating and cooling due to pair annihilation and pair-creation of neutrinos are taken into account. We also note that the total pair annihilation deposition rates for MT01s095 and MT03s095 are calculated only for $\rho\leq 10^{11}\,{\rm g/cm^3}$ to avoid fluctuations due to the MC shot noise in the high density and high temperature part of the torus which does not contribute to driving relativistic outflows.

Since there is a time delay until emitted neutrinos are observed and reflected in the luminosities due to the propagation to the extraction radius, for direct comparison with other instantaneously determined quantities, we shift the time origin of the neutrino luminosities by the propagation time scale, i.e., $t\rightarrow t-r_{\rm rad}$. On the other hand, for the quantities that can be instantaneously calculated, 
the time shift is not applied.

The terminal Lorentz factor of the matter is estimated by $hw$ where $h$ is the specific enthalpy. Note that the actual terminal Lorentz factor can be smaller than $hw$ due to radiative loss of the internal energy before it contributes to acceleration~\cite{Meszaros:1999gb}. 
We define the outflow luminosity of the relativistic components of which the terminal Lorentz factor exceeds $\Gamma_\infty$ by 
\begin{align}
    L_{\rm jet,>\Gamma_\infty}=\int_{r=r_{\rm rad}} dS_i\, (F_0^i-\rho_* h_{\rm min} v^i)\Theta\left(hw-\Gamma_\infty\right),\label{eq:Ljetinf}
\end{align}
and the total relativistic outflow energy, $E_{\rm jet,>\Gamma_\infty}$, is determined by integrating $L_{\rm jet,>\Gamma_\infty}$ in time. Here, $h_{\rm min}$ and $\Theta$ denote the minimum specific enthalpy and Heaviside step function, respectively, and 
\begin{equation}
F_0^i=\rho_* \left(hw-\frac{P}{\rho w}\right)v^i+P\sqrt{\gamma}\left(v^i+\beta^i\right)
\end{equation}
denotes the conserved energy flux of the fluid with $P$ and $\beta^i$ being the pressure and shift vector, respectively. Note that $\rho_* h_{\rm min} v^i$ in~\eqref{eq:Ljetinf} corresponds to the contribution of the rest-mass energy to the energy flux. 

Since we pay attention to the fluid elements with $w \gg 1$ in calculating the outflow luminosity, for the numerical accuracy, we approximate $F_0^i \approx S_0v^i$. Here, $S_0 = T_\mathrm{fl}{}_{\mu\nu} n^\mu n^\nu$, where $n^\mu$ is the time-like unit vector normal to the spatial hypersurface and $T_\mathrm{fl}{}_{\mu\nu}$ is the energy-momentum tensor of the viscous fluid (see \cite{Kawaguchi:2024naa} for its definition). 
This is because the energy flux derived directly from a fundamental variable $S_0$, which is defined in our hydrodynamics scheme, gives a more accurate estimation of the energy flux than that derived from the primitive variables, such as $\rho$, $h$, and $P$. 
We also approximate $h_{\rm min}\approx 1$ for simplicity since $hw\gg1$.

The outflow luminosity is determined by the energy flux evaluated at a surface normal to the $z$-axis with $z=z_{\rm ext}=3000\,{\rm km}$. This choice is made to measure the luminosity of the relativistic outflow, which is expected to be launched toward the polar direction. As is the case for neutrino luminosity, the time origin is shifted by the propagation time scale, i.e., $t\rightarrow t-z_{\rm ext}$ approximately assuming the propagation speed to be the speed of light. We note that the difference between the values of the total outflow energy evaluated at $z=z_{\rm ext}=3000\,{\rm km}$ and $z=z_{\rm ext}=1000\,{\rm km}$ is always less than $10\%$.

We pay attention to the neutrino luminosity, pair annihilation rate, and relativistic outflow luminosity only after $t=20\,{\rm ms}$, which approximately corresponds to the orbital period of the outer edge of the torus in the fiducial model. This is because the earlier data are significantly affected by the transient behavior during the relaxation of the initial profile (see below).

To discuss the collimation of the relativistic outflow, we define the (half) opening angle of the average relativistic outflow by 
\begin{align}
    \theta_{\rm jet,>\Gamma_\infty}=\frac{\displaystyle \int_{z_{\rm ext}/2}^{z_{\rm ext}} dz \max_{hw>\Gamma_\infty} \theta}{z_{\rm ext}/2},
\end{align}
where $\theta$ is the angle measured from the $z$-axis. We then define the isotropic luminosity and the time-averaged opening angle of the relativistic outflow, respectively, by
\begin{align}
    L_{\rm jet,>\Gamma_\infty}^{\rm iso}=\frac{L_{\rm jet,>\Gamma_\infty}}{1-{\rm cos}~[{\rm max}(\theta_{\rm jet,>\Gamma_\infty},\theta_{\rm min})] }
\end{align}
and
\begin{align}
    \left<\theta_{\rm jet,>\Gamma_\infty}\right>=\frac{\displaystyle \int \theta_{\rm jet,>\Gamma_\infty} L_{\rm jet,>\Gamma_\infty} dt}{\int  L_{\rm jet,>\Gamma_\infty} dt}.
\end{align}
Here, to avoid divergence in the beaming factor, we set a minimum angle of $\theta_{\rm min}=0.01\,{\rm rad}$, which is smaller than the typical value of $\theta_{\rm jet,>\Gamma_\infty}$ (see Sec.~\ref{sec:res}). Finally, the total isotropic outflow energy of the relativistic components and the $90\%$ energy duration, $E_{\rm jet,>\Gamma_\infty}^{\rm iso}$ and $t_{90}^{\rm iso >\Gamma_\infty}$, respectively, are determined by integrating $L_{\rm jet,>\Gamma_\infty}^{\rm iso}$ over time and the time at which $90\%$ of $E_{\rm jet,>\Gamma_\infty}^{\rm iso}$ is emitted.

\section{Results}\label{sec:res}

\subsection{Overview}

\begin{figure*}
 	 \includegraphics[width=\linewidth]{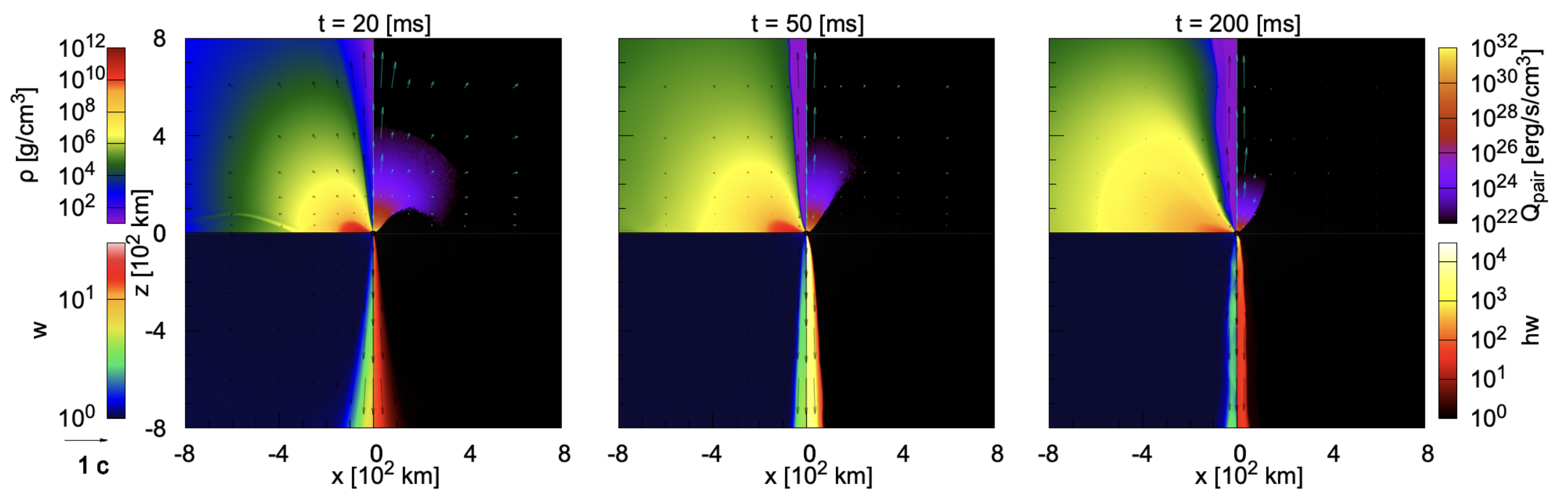}
 	 \caption{Snapshots of the fiducial model (MT01s08) at $t=20\,{\rm ms}$ (left panel), $50\,{\rm ms}$ (middle panel), and $200\,{\rm ms}$ (right panel). In each panel, the rest-mass density, pair annihilation deposition rate, Lorentz factor, and terminal Lorentz factor are presented in the top left, top right, bottom left, and bottom right corners, respectively.}
	 \label{fig:prof_fid}
\end{figure*}

We give a brief overview for the evolution of the system by showing the result of the fiducial model (MT01s08). The basic picture of the system evolution is the same for other models. Figure~\ref{fig:prof_fid} shows the snapshots of several quantities for the fiducial model at $t=20\,{\rm ms}$, $50\,{\rm ms}$, and $200\,{\rm ms}$. In the first few ms of the simulation, the inner edge of the torus rapidly falls into the BH while at the same time the matter blows up due to the strong viscous heating and neutrino irradiation. We note that this early transient process should be regarded as an artifact of the relaxation from the initial condition, as it is not clear whether such ejecta components are present in the results when consistently following the system from the merger phase.

We also note that the mass ejection due to the neutrino irradiation in this initial transient phase is enhanced by taking the multi-energy effect of neutrino radiative transfer into account. Neutrino radiation around the surface of the torus is dominated by that emitted and diffused out from the inner high density and temperature region of the torus. As a result, the mean neutrino energy close to the surface of the torus is by more than a factor of 2 larger than that estimated from the local matter temperature. Due to the strong neutrino energy dependence of the opacity, the radiative heating and resulting mass ejection are enhanced around the surface of the torus by taking this multi-energy effect of neutrino radiative transfer into account. In fact, we find that the mass ejection in this phase is strongly suppressed for a test simulation with the gray-approximation opacity employing the Planck-mean value with respect to the local matter temperature. 

After a few tens of ms, the matter accretes onto the BH and the torus gradually expands in a quasi-stationary manner due to the viscous angular momentum transport. The mass accretion rate is as large as $\sim 1\,M_\odot/{\rm s}$ at $10\,{\rm ms}$, drops broadly as $\propto t^{-1}$ due to the decrease in the torus mass and the increase in the viscous time scale associated with the torus expansion. In particular at late times ($\gtrsim 0.1\,{\rm s}$), the mass accretion rate approximately follows a scaling $\propto t^{-4/3}$, consistent with self-similar solutions~\citep{Metzger:2008av} (see also Fig.~\ref{fig:mdot}). The mass ejecta driven by the viscous heating is suppressed for $\alt 1\,{\rm s}$ due to the efficient neutrino cooling (see Fig.~6 in~\cite{Kawaguchi:2024naa}).

The energy deposition by neutrino pair annihilation is more significant in the earlier epoch, reflecting the rapid matter accretion onto the BH and the resulting large neutrino luminosity (see also Fig.~\ref{fig:lnu}). The region with the positive deposition rate is wide only in the earlier epoch, and it gradually shrinks as the neutrino luminosity decreases. Note that outside the positive-deposition region, the deposition rate is negative because the cooling due to neutrino pair creation dominates the heating by neutrino pair annihilation.

The matter with the relatively low rest-mass density ($\ll 10^6\,{\rm g/cm^3}$) present in the polar region becomes relativistic due to the energy deposition by the neutrino pair annihilation. This is confirmed by the fact that the relativistic components are absent in a test simulation without the $\nu_e$-${\bar \nu}_e$ pair annihilation (see Fig.9 in~\cite{Kawaguchi:2024naa}). Unlike the energy deposition of the pair annihilation, the relativistic outflow is most significantly developed at $\approx 50\,{\rm ms}$ and not in the first few tens of ms. This is due to the presence of the relatively high density matter ($\gtrsim 10^{6}\,{\rm g/cm^3}$) in the polar region in the early phase. Such matter is supplied from the inner edge of the torus due to the strong viscous heating during the early phase of the evolution, and it prevents the development of a pair-annihilation-powered fireball. It also prevents the matter from being accelerated to relativistic speeds by the baryon loading. However, as the torus expands (i.e., the mass infall to the polar region decreases) and also the matter in the polar region accretes onto the BH, the baryon loading becomes less significant. This helps the fireball to expand in a larger region and drives the matter to be relativistic. After $\approx 50\,{\rm ms}$, the relativistic outflow starts to decline following the decreases in the pair annihilation deposition luminosity. We find that the relativistic outflow is terminated at $\sim 0.1\,{\rm s}$ for all models studied in this paper (see below).

\subsection{Accretion rate and neutrino luminosity}

\begin{figure*}
 	 \includegraphics[width=0.49\linewidth]{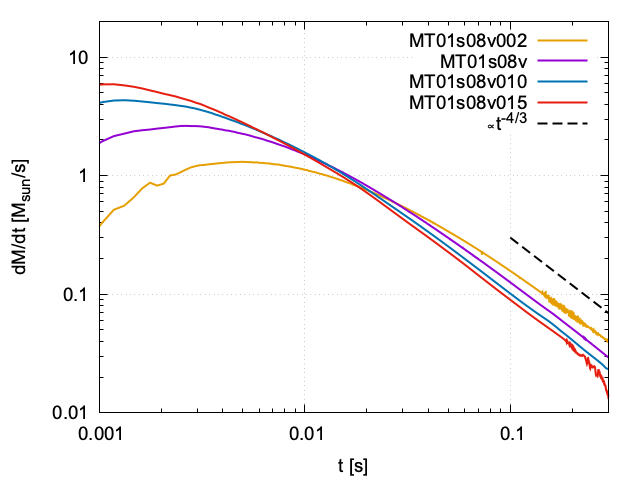}
 	 \includegraphics[width=0.49\linewidth]{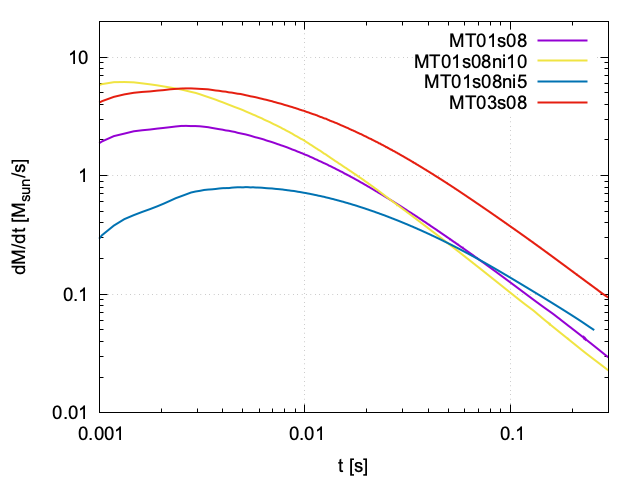}\\
 	 \includegraphics[width=0.49\linewidth]{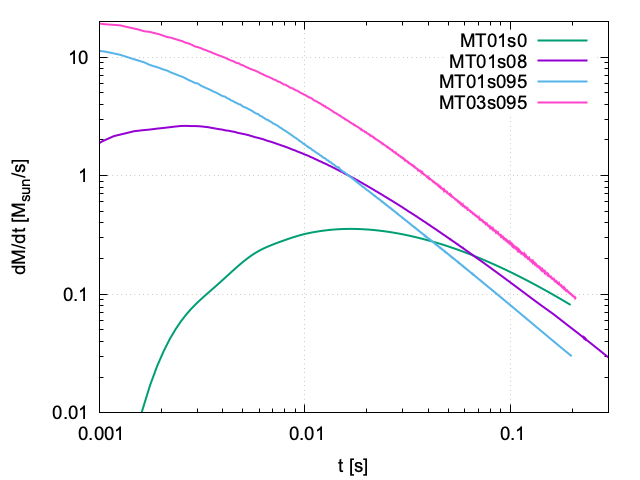}
 	 \includegraphics[width=0.49\linewidth]{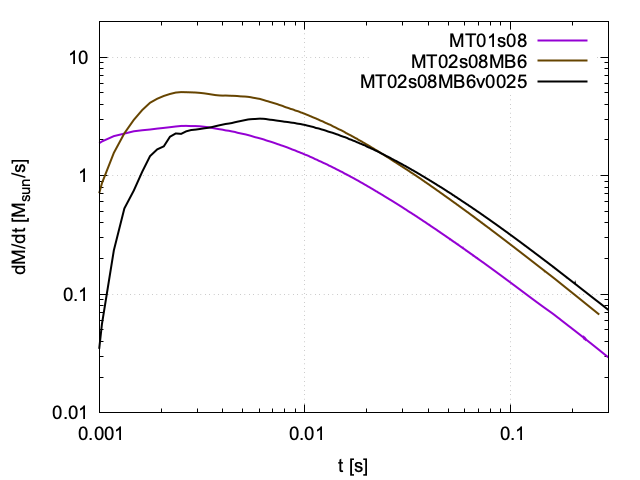}\\
 	 \caption{Mass accretion rate onto the BH as a function of time. The top left, top right, bottom left, and bottom right panels show the results with the same setups as the fiducial model but with different viscous parameters (MT01s08v002, MT01s08, MT01s08v010, and MT01s08v015), different angular momentum profiles and torus mass (MT01s08ni10, MT01s08ni5, and MT03s08), different BH spins (MT01s0, MT01s095, and MT03s095), and larger BH mass and torus mass (MT02s08MB6 and MT02s08MB6v0025), respectively. The result of the fiducial model (MT01s08) is always shown as a reference.} 
	 \label{fig:mdot}
\end{figure*}

Figure~\ref{fig:mdot} shows the mass accretion rate onto the BH as a function of time for various models. The results and their dependence on the torus properties agree broadly with those in~\cite{Fujibayashi:2020qda}. First, we pay attention to the models with different viscous coefficients (MT01s08v002, MT01s08, MT01s08v010, and MT01s08v015; the top left panel of Fig.~\ref{fig:mdot}). For larger viscous coefficients, the mass accretion rate onto the BH is larger in the early epoch but smaller in the later epoch (see the top left panel of Fig.~\ref{fig:mdot}). This reflects the viscous time scale, which is given by
\begin{align}
    \tau_{\rm vis}=\frac{R^2}{h\nu}\sim 0.3\,{\rm s}&\left(\frac{\alpha_{\rm vis}}{0.05}\right)^{-1}\left(\frac{h c_s}{0.08\,c^3}\right)^{-1}\nonumber\\
    &\times\left(\frac{H_{\rm vis}}{9\,{\rm km}}\right)^{-1}\left(\frac{R}{53\,{\rm km}}\right)^{2}.\label{eq:tauvis}
\end{align}
Note that the mass accretion rate reaches its peak broadly on the viscous time scale of the radius of the innermost stable circular orbit (ISCO), $R\sim r_{\rm ms}$, at which the density peak of the torus is approximately located.

In the top right panel of Fig.~\ref{fig:mdot}, we compare the results with different initial angular momentum profiles and mass of the torus (MT01s08ni10, MT01s08ni5, and MT03s08). The model with a larger value of $n$ shows a smaller mass accretion rate in the early epoch with a shallower decline. As is pointed out above, the torus becomes less compact as the angular momentum profile becomes closer to the Newtonian Keplerian profile ($n=1/3$). Equation~\eqref{eq:tauvis} suggests that the viscous time scale is longer for the torus with larger radius, and hence, the model with the larger value of $n$ has a smaller mass accretion rate with a shallower decline. The model with the torus mass larger by a factor of 3 (MT03s08) shows a larger mass accretion rate than the fiducial model by only a factor of $\approx2$, which is also the consequence of having larger torus radius that leads to the elongation of the viscous time scale (the same holds for MT03s095 relative to MT01s095).

The bottom left panel of Fig.~\ref{fig:mdot} shows that the model with a larger BH spin has a larger mass accretion rate in the early phase and the rate declines rapidly. This behavior of the mass accretion rate with respect to the BH spin can also be understood by the dependence of the viscous time scale on the torus radius: In our models, the averaged cylindrical radius of the torus becomes small as the BH spin parameter becomes large. As a result, the viscous time scale becomes shorter for the model with larger BH spins resulting in a large mass accretion rate in the early phase with a rapid decline. The difference in the mass accretion rate for the models with larger BH and torus masses (MT02s08MB6 and MT02s08MB6v0025) compared to the fiducial model is also consistent with differences in the torus mass and the viscous time scale (the bottom right panel of Fig.~\ref{fig:mdot}): the larger accretion rate due to larger torus mass and longer viscous time scale due to large torus radius and small viscous coefficient.

\begin{figure*}
 	 \includegraphics[width=0.49\linewidth]{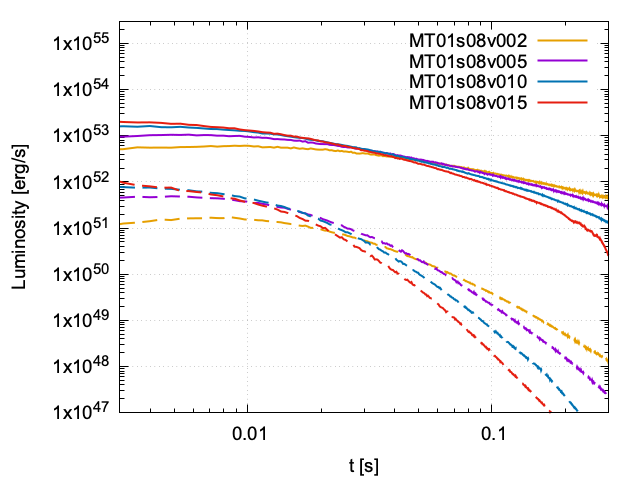}
 	 \includegraphics[width=0.49\linewidth]{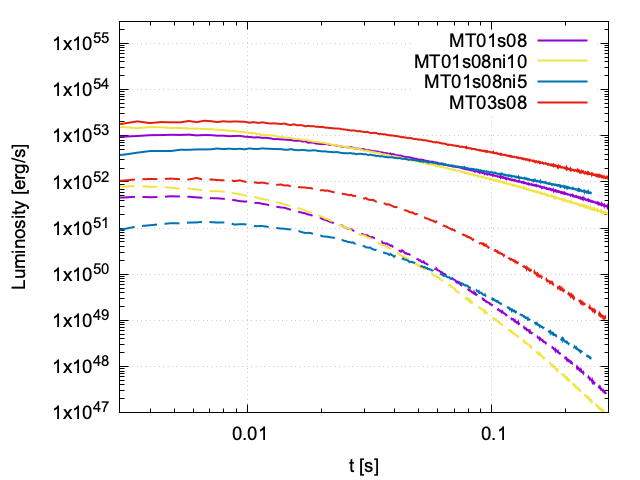}\\
 	 \includegraphics[width=0.48\linewidth]{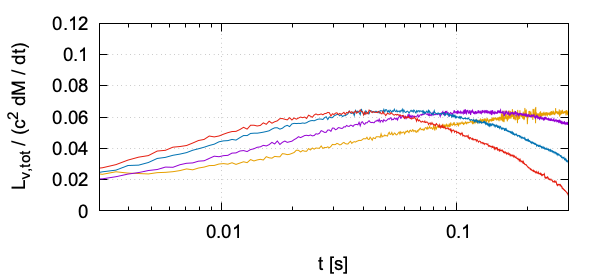}
 	 \includegraphics[width=0.48\linewidth]{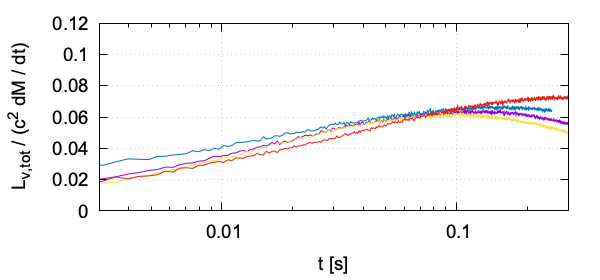}\\
   
 	 \includegraphics[width=0.49\linewidth]{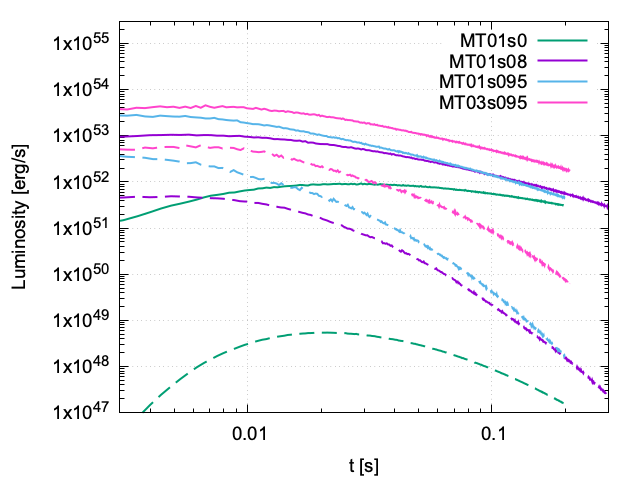}
 	 \includegraphics[width=0.49\linewidth]{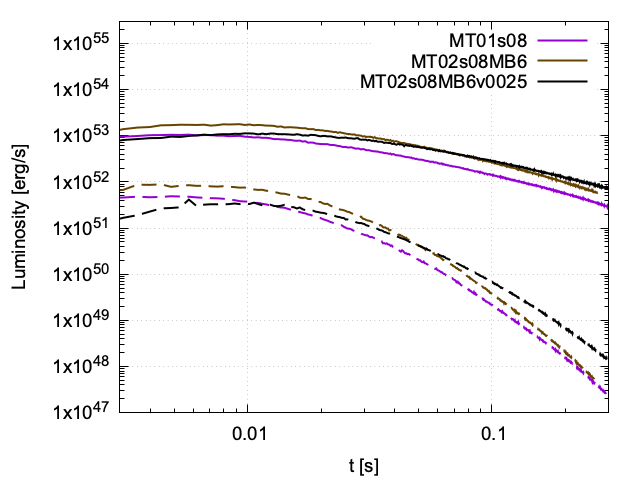}\\   
 	 \includegraphics[width=0.48\linewidth]{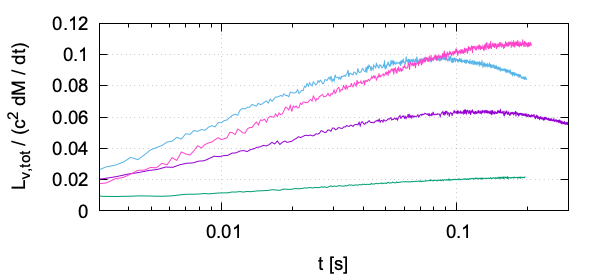}
 	 \includegraphics[width=0.48\linewidth]{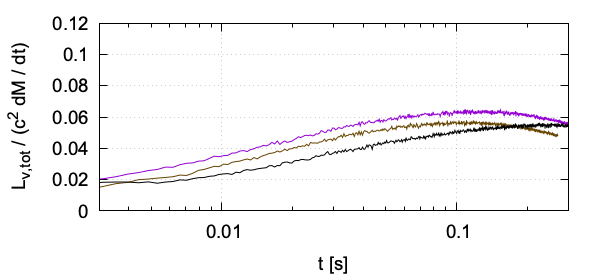}
 	 \caption{The same as Fig.~\ref{fig:mdot} but for the total neutrino luminosity (solid curves) and total pair annihilation deposition rate (dashed curves). The efficiency of the total neutrino luminosity with respect to the mass accretion rate is also shown in the lower part of each panel. Note that, for MT01s095 and MT03s095, the total pair-annihilation energy deposition rates are calculated only for regions with $\rho \leq 10^{11}\,{\rm g/cm^3}$ in order to avoid fluctuations caused by MC shot noise in the high-density and high-temperature regions of the torus.}
	 \label{fig:lnu}
\end{figure*}

Figure~\ref{fig:lnu} shows the total neutrino luminosity and total pair annihilation deposition rate as functions of time for various models. The results are again broadly in agreement with those found in~\cite{Fujibayashi:2020qda}. The total neutrino luminosity is larger in the earlier epoch and gradually decreases as the mass accretion rate drops. The relative magnitude of the luminosity among the models broadly reflects the relative magnitude of the mass accretion rate (see Fig.~\ref{fig:mdot}). 

The efficiency of the total neutrino luminosity with respect to the mass accretion rate for our models varies in time within the range of $0.01$--$0.1$. The efficiency reaches $\approx0.06$ at its peak for the fiducial model, and the peak value does not depend strongly on the viscous parameter (MT01s08v002, MT01s08, MT01s08v010, and MT01s08v015; the top left panel of Fig.~\ref{fig:lnu}). The models with the different angular momentum profiles also show approximately the same efficiency at the peak (MT01s08ni10 and MT01s08ni5; the top right panel of Fig.~\ref{fig:lnu}). The model with the same BH mass and spin but with the larger torus mass (MT03s08) shows slightly higher efficiency. This is because the density and temperature are higher for the model with the larger torus mass, which leads to a larger effective volume of the emission region and a more efficient conversion of internal energy into neutrino radiation. This is also the reason that the models with the larger BH and torus masses (MT02s08MB6 and MT02s08MB6v0025; the bottom right panel of Fig.~\ref{fig:lnu}) show slightly lower peak efficiency than the fiducial model, in which the typical matter density and temperature are smaller. However, this effect is minor compared with the effect by the BH spin.

The comparison among the models with different BH spins (MT01s0, MT01s095, and MT03s095; the bottom left panel of Fig.~\ref{fig:lnu}) show that the peak efficiency depends appreciably on the BH spin as also found in previous studies~\citep{Chen:2006rra,1999ApJ...518..356P,Fujibayashi:2020qda}. This reflects that $r_{\rm ms}$ is smaller for the larger BH spin, and thus, the gravitational binding energy available for the viscous dissipation, which is roughly $\propto r_{\rm ms}^{-1}$, is larger. 
We find that the peak efficiency of our results is approximately $\propto r_{\rm ms}^{-3/2}$. The steeper dependence may be the result of the higher typical density and temperature of the torus of the models with larger BH spins.

The time at which the peak efficiency is reached depends approximately linearly on the viscous coefficient among the models with different viscous coefficients. In contrast, for the other models, the peak efficiency time does not necessarily scale linearly with the viscous coefficient, though a correlation is still observed. This suggests that both the detailed structure of the torus and the properties of the central BH play important roles in determining the efficiency.

\subsection{Pair annihilation deposition rate}

The total pair annihilation rate is always by more than an order of magnitude smaller than the total neutrino luminosity (compare the solid and dashed curves in Fig.~\ref{fig:lnu}). The difference between them is more pronounced for lower total neutrino luminosity. The relative magnitude of the total pair annihilation rate among the models is aligned with the neutrino luminosity, $L_{\rm \nu,tot}$, but the relative differences are more pronounced for the pair annihilation rate. These properties reflect the fact that the total pair annihilation rate is proportional to the neutrino energy flux with the exponent larger than 2 (see below).

We do not observe the slower decline in the total pair annihilation rate than the total neutrino luminosity, in contrast to the finding in~\cite{Setiawan:2005ah}. We suspect that this is due to the higher initial $Y_e$ profile of the torus than the profile employed in~\cite{Setiawan:2005ah}. It is indeed discussed in~\cite{Setiawan:2005ah} that the initial neutron-richness of the torus is responsible for sustaining the total pair annihilation deposition rate.

\begin{figure*}
 	 \includegraphics[width=0.49\linewidth]{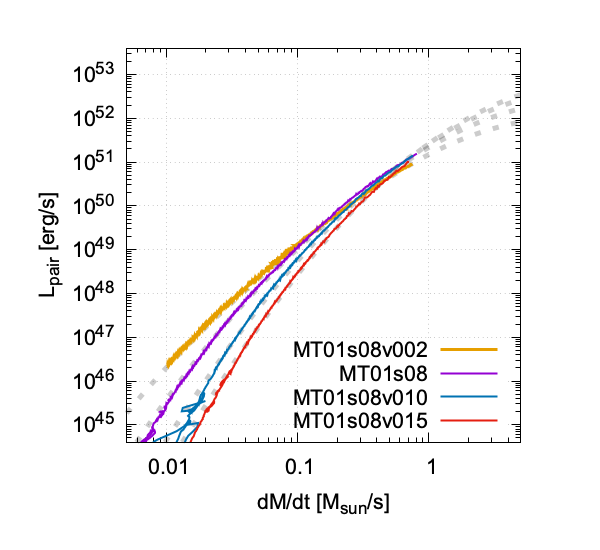}
 	 \includegraphics[width=0.49\linewidth]{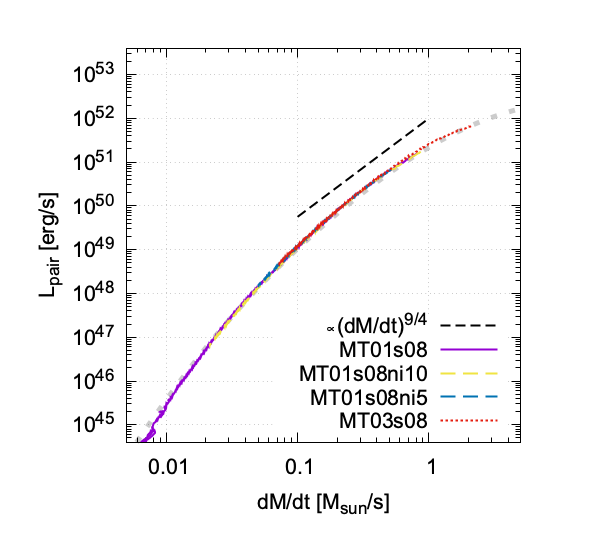}\\
     \vspace{-4mm}
 	 \includegraphics[width=0.49\linewidth]{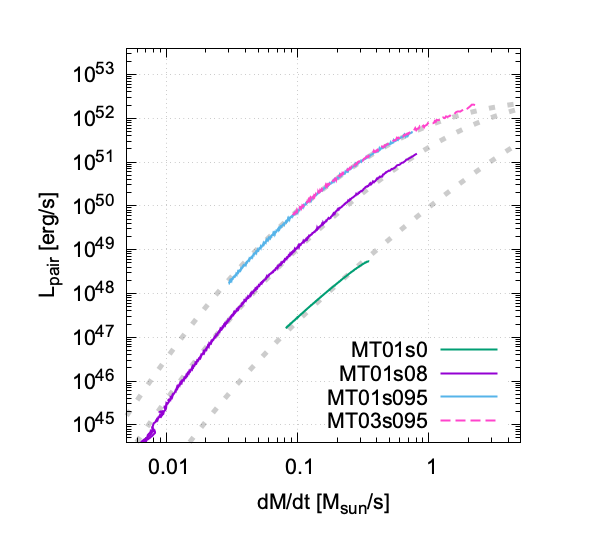}
 	 \includegraphics[width=0.49\linewidth]{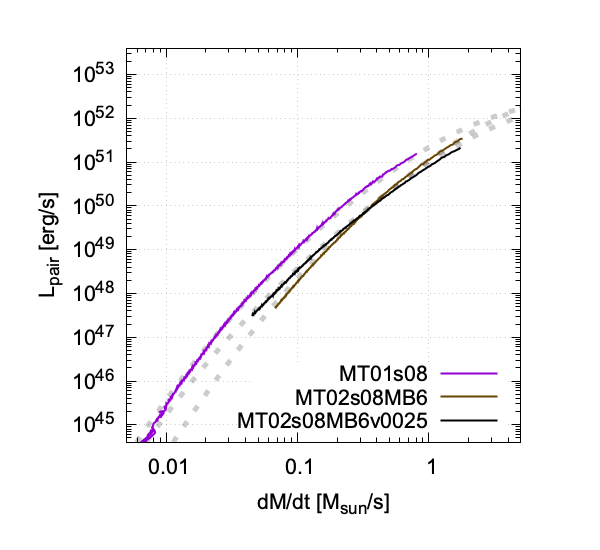}
     \vspace{-3mm}
 	 \caption{Total pair annihilation deposition rate as a function of the accretion rate to the BH for $t\ge 20\,{\rm ms}$. The top left panel shows the results with various viscous parameters (MT01s08v002, MT01s08, MT01s08v010, and MT01s08v015). The top right panel shows for the same BH spin but with different torus angular momentum profiles (MT01s08ni10 and MT01s08ni5) and mass (MT03s08). The bottom left panel shows the results with different BH spins (MT01s0, MT01s095, and MT03s095). The bottom right panel shows the results with the BH mass of $6\,M_\odot$ and the torus mass of $0.2\,M_\odot$ (MT02s08MB6 and MT02s08MB6v0025). The result of the fiducial model is always plotted as a reference. The dotted curves denote the fitting function denoted by Eq.~\eqref{eq:lpair-dmdt-fit}.}
	 \label{fig:lpair-dmdt}
\end{figure*}

Figure~\ref{fig:lpair-dmdt} shows the total pair annihilation deposition rate, $L_\mathrm{pair}$, as a function of the accretion rate onto the BH for $t\gtrsim 20\,{\rm ms}$. We find that our results are broadly consistent with the findings in~\cite{Zalamea:2010ax}. Generally, $L_\mathrm{pair}$ monotonically drops as the mass accretion rate drops. The slope of the curves, $dL_\mathrm{pair}/d\dot M$, becomes steeper for smaller mass accretion rate particularly for $<0.1\,M_\odot/{\rm s}$ for which the accretion disk becomes optically thin to neutrinos. This reflects that the neutrino emissivity is sensitive to the temperature of the matter (approximately $\propto T^6$). The slope of $dL_\mathrm{pair}/d\dot M$ in the high mass accretion rates ($>1\,M_\odot/{\rm s}$) becomes shallower due to the neutrino trapping effect~\cite{Zalamea:2010ax}.

The top left panel of Fig.~\ref{fig:lpair-dmdt} compares $L_\mathrm{pair}$ with different viscous coefficients. For the models with smaller viscous coefficients, $L_\mathrm{pair}$ at high mass accretion rates ($\sim 1\,M_\odot/{\rm s}$) becomes lower. By contrast, it exhibits a shallower decline as the accretion rate decreases below $\sim 0.1\,M_\odot/{\rm s}$. This reflects more efficient neutrino emission at later times for low-viscosity models (see Fig.~\ref{fig:lnu}), consistent with the findings of~\cite{Zalamea:2010ax}.

The top right and bottom left panels of Fig.~\ref{fig:lpair-dmdt} also show that the curves of $L_{\rm pair}$ as a function of the accretion rate ${\dot M}$ agree surprisingly well among the models with the same BH mass and spin but with different initial torus masses and angular momentum profiles. 
Our interpretation for this is that the total pair-annihilation rate is determined primarily by the accretion structure at the innermost part of the torus near the BH, which is primarily governed by the BH properties. This is because the matter profile and neutrino radiation there are likely in a quasi-stationary state for $t\gtrsim 20\,{\rm ms}$ (see Fig.~\ref{fig:mdot}).

The bottom left panel of Fig.~\ref{fig:lpair-dmdt} shows that $L_{\rm pair}$ depends strongly on ${\dot M}$ for the models with different BH spins. The curves for $\chi_{\rm BH}=0$ (MT01s0) and $0.95$ (MT01s095 and MT03s095) broadly agree with the results of~\cite{Zalamea:2010ax} (see Fig.~4 in~\cite{Zalamea:2010ax}). The result suggests that the pair annihilation rate is approximately proportional to $r_{\rm ms}^{-5}$, and this power-law index is also consistent with~\cite{Zalamea:2010ax} ($\propto r_{\rm ms}^{-4.8}$). 

The bottom right panel of Fig.~\ref{fig:lpair-dmdt} shows the models for $M_\mathrm{BH}=6\,M_\odot$ and initial torus mass of $0.2\,M_\odot$ with $\chi_\mathrm{BH}=0.8$ (MT02s08MB6 and MT02s08MB6v0025). $L_\mathrm{pair}$ depends on the mass accretion rate in a similar manner to the fiducial model, but with a value smaller by a factor of $\approx 3$. This suggests that $L_\mathrm{pair}$ decreases with $M_\mathrm{BH}$.

To summarize, $L_\mathrm{pair}$ depends on the mass accretion rate, ISCO radius, and BH mass. The dependence for $\sim 0.1$--$1\,M_\odot/{\rm s}$ can be understood by the following estimation, which is similar to that done in~\cite{Zalamea:2010ax}: Since the innermost part of the torus is marginally optically thick for the mass accretion rate with $\sim 0.1$--$1\,M_\odot/{\rm s}$, the neutrino luminosity from there, $L_{\rm \nu,ms}$, is approximately given by $L_{\rm \nu,ms}\approx 4\pi r_{\rm ms}^2 (F_{\rm \nu, ms}+F_{\rm {\bar \nu}, ms})\propto r_{\rm ms}^2 T_{\rm eff}^4$ with $F_{\rm \nu, ms}\sim F_{\rm {\bar \nu}, ms}$ being the neutrino and antineutrino energy flux, respectively, and $T_{\rm eff}$ being the effective temperature of the neutrino emission. Alternatively, $L_{\rm \nu,ms}$ can be written as $L_{\rm \nu,ms}\approx f M_{\rm BH}{\dot M}/r_{\rm ms}$ with ${\dot M}$ and $f$ being the mass accretion rate and efficiency releasing gravitational binding energy into neutrino emission. Assuming that $f\propto {\hat r}_{\rm ms}^{-1/2}=(r_{\rm ms}/M_{\rm BH})^{-1/2}$ based on the finding in Fig.~\ref{fig:lnu}, we obtain $T_{\rm eff}\propto M_{\rm BH}^{3/8}r_{\rm ms}^{-7/8}{\dot M}^{1/4}$. The total pair annihilation can be estimated by $L_{\rm pair}\approx V_{\rm ms}\left<\epsilon\right>{\dot n}_{\rm pair}$~\citep{Zalamea:2010ax} with $V_{\rm ms}\propto r_{\rm ms}^3$, $\left<\epsilon\right>\approx T_{\rm eff}$, and ${\dot n}_{\rm pair}\propto F_{\rm \nu,ms}F_{\rm {\bar \nu},ms}$ being the effective volume of the pair annihilation region, typical energy of neutrinos (antineutrinos) which involve in pair annihilation, and pair annihilation rate density, respectively. This leads to $L_{\rm pair}\propto r_{\rm ms}^{-39/8}M_{\rm BH}^{27/8}{\dot M}^{9/4}\propto {\hat r}_{\rm ms}^{-39/8}M_{\rm BH}^{-3/2}{\dot M}^{9/4}$. 

As found in~\cite{Zalamea:2010ax}, an approximation relation of $L_{\rm pair}\propto {\dot M}^{9/4}$ is satisfied up to $\dot M\approx 1\,M_\odot/{\rm s}$ in our results. The steep dependence of $L_\mathrm{pair}$ on the ISCO radius ($L_{\rm pair}\propto r_{\rm ms}^{-39/8}\approx r_{\rm ms}^{-4.9}$) is also consistent with our results~\cite{Zalamea:2010ax}. A factor of $\approx 3$ smaller value of $L_\mathrm{pair}$ found for MT02s08MB6 and MT02s08MB6v0025 compared to the fiducial model can be explained by dependence of $L_{\rm pair}$ on the BH mass ($M_{\rm BH}^{-3/2}$). 

We find that $L_{\rm pair}$ as a function of the accretion rate ${\dot M}$ can be well approximated by a function
\begin{align}
    L_{\rm pair}^{\rm fit}({\dot M})=\frac{L_{0}}{\left[1+\left(\frac{{\dot M}_{\rm 1}}{{\dot M}}\right)\right]^{5/2}\left[1+\left(\frac{{\dot M}_{\rm 2}}{{\dot M}}\right)\right]^{5/2}},\label{eq:lpair-dmdt-fit}
\end{align}
where $L_{0}$, ${\dot M}_{\rm 1}$, and ${\dot M}_{\rm 2}$ are the fitting parameters given as follows as functions of $\alpha_{\rm vis}$, $\chi_{\rm BH}$, and $M_{\rm BH}$:
\begin{align}
L_{0}&=4.2\times 10^{52}\,{\rm erg/s}\,\alpha_{\rm vis,0.05}^{}r_{\rm ms,0.8}^{5/8}M_{\rm BH,3},\\
{\dot M}_{\rm 1}&=0.020\,M_\odot/{\rm s}\,\alpha_{\rm vis,0.05}^{5/3}M_{\rm BH,3}^{4/3},\\
{\dot M}_{\rm 2}&=2.2\,M_\odot/{\rm s}\,\alpha_{\rm vis,0.05}^{1/3}r_{\rm ms,0.8}^{12/5}M_{\rm BH,3}.
\end{align}
Here, $\alpha_{\rm vis,0.05}=\alpha_{\rm vis}/0.05$, $r_{\rm ms,0.8}=r_{\rm ms}(\chi_{\rm BH})/r_{\rm ms}(0.8)$ and $M_{\rm BH,3}=M_{\rm BH}/(3\,M_\odot)$. ${\dot M}_{\rm 1}$ and ${\dot M}_{\rm 2}$ can be interpreted as the mass accretion rates at which the neutrino emission ignites due to the increase in the temperature (${\dot M}_{\rm ign}$) and neutrino radiation starts to be trapped to the accreted matter (${\dot M}_{\rm trap}$), respectively~\citep{Zalamea:2010ax}. Accordingly, the exponents of $\alpha_{\rm vis}$ for ${\dot M}_{\rm 1}$ and ${\dot M}_{\rm 2}$ are chosen to match those of ${\dot M}_{\rm ign}$ and ${\dot M}_{\rm trap}$ in~\cite{Zalamea:2010ax}, within fitting errors. The exponent of $M_{\rm BH}$ for ${\dot M}_{\rm 1}$ is also chosen to match that in~\citep{Agarwal:2025gbw}. The remaining exponents are assigned as simple rational numbers that approximately reproduce the numerical fits. 

We find that the results of the simulations for $L_{\rm pair}\geq 10^{46}\,{\rm erg/s}$ are reproduced within $30\%$ error by Eq.~\eqref{eq:lpair-dmdt-fit}, except for the model with $\chi_{\rm BH}=0$ for which the maximum fitting error is $35\%$. ${\dot M}_{\rm 1}$ and ${\dot M}_{\rm 2}$ broadly agree with those mass accretion rates given in~\cite{Zalamea:2010ax}. $L_0$ appears to represent the maximum pair annihilation deposition rate in the limit of a high mass accretion rate; however, the regime with ${\dot M} \gtrsim 1\,M_\odot/{\rm s}$ is not well calibrated in our model and should be interpreted with care.

\begin{figure*}[t]
 	 \includegraphics[width=0.49\linewidth]{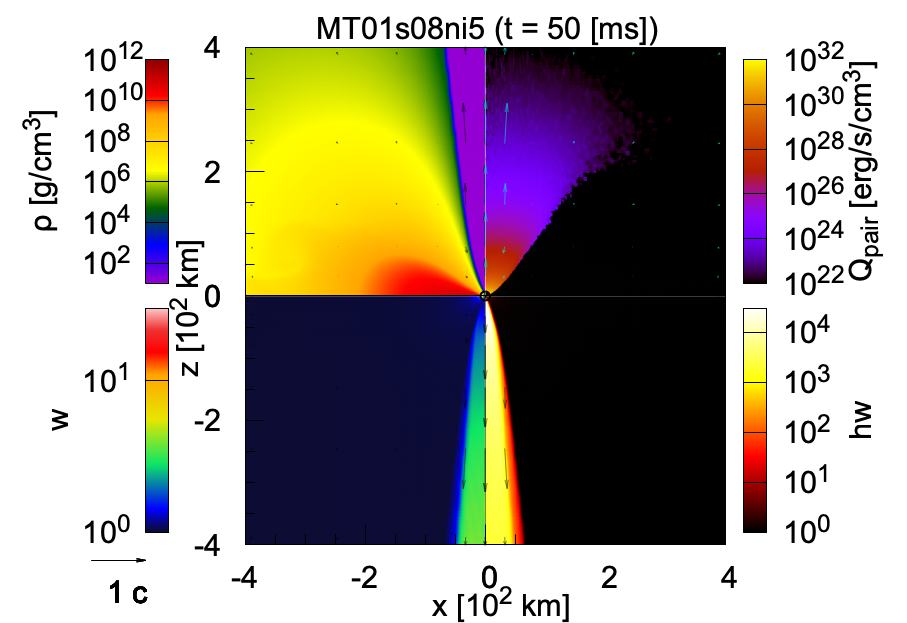}
 	 \includegraphics[width=0.49\linewidth]{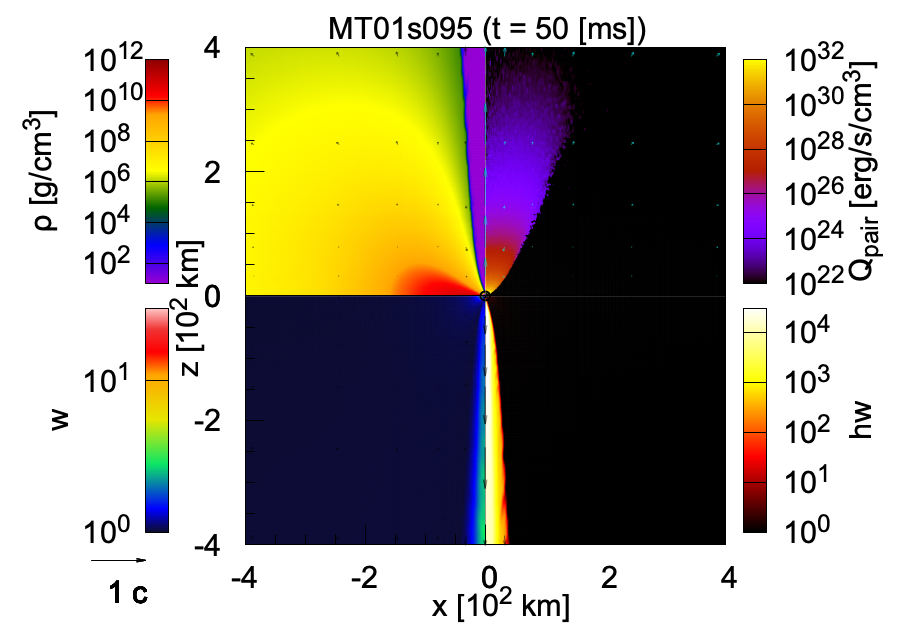}\\
 	 \includegraphics[width=0.49\linewidth]{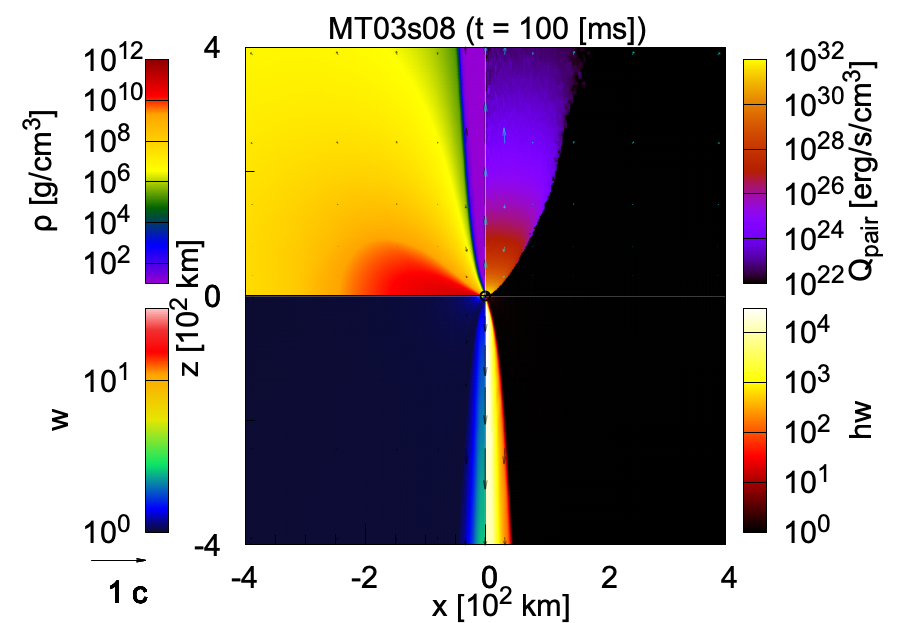}
 	 \includegraphics[width=0.49\linewidth]{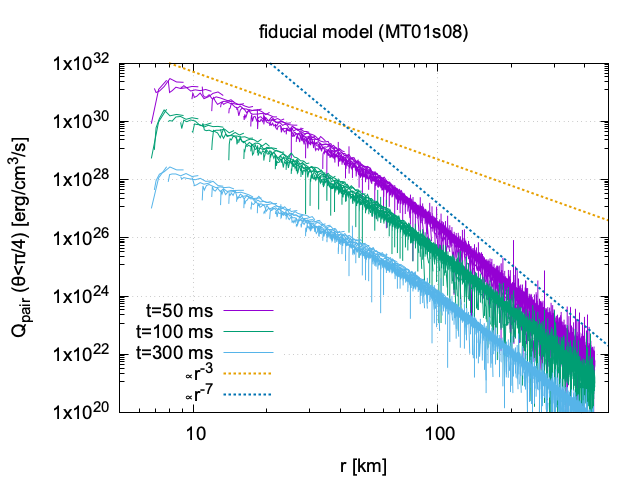}
 	 \caption{Snapshots of $\rho$, $Q_\mathrm{pair}$ (pair annihilation deposition rate), $w$, and $hw$ for the models with $n=1/5$ and $\chi=0.8$ (MT01s08ni5; top left) and $\chi_{\rm BH}=0.95$ (MT01s095; top right) at $t=50\,{\rm ms}$ and the model with $M_{\rm torus}=0.3\,M_\odot$ and $\chi=0.8$ at $t=100\,{\rm ms}$ (MT03s08; bottom left). The bottom right panel shows the radial profile of the pair annihilation deposition rate within $45^\circ$ measured from the pole for the fiducial model.}
	 \label{fig:qpair-prof}
\end{figure*}

We next pay attention to the spatial profile of the pair annihilation deposition rate. Figure~\ref{fig:qpair-prof} shows the snapshots of the models with $n=1/5$ (MT01s08ni5; top left), $\chi_{\rm BH}=0.95$ (MT01s095; top right) at $t=50\,{\rm ms}$, and the model with $M_{\rm torus}=0.3\,M_\odot$ at $t=100\,{\rm ms}$ (MT03s08; bottom left) (see Fig.~\ref{fig:prof_fid} for the fiducial model; MT01s08). The epoch for each model is chosen when the contribution of pair annihilation  energy deposition to the relativistic outflow is approximately the largest. Figure~\ref{fig:qpair-prof} shows that the pair annihilation deposition rate is always most significant in the vicinity of the BH, and the profile of the pair annihilation deposition rate is similar to each other across the models with different setups. The funnel region of the positive net heating extends with the opening angle of $30$--$45^\circ$, and the model with a geometrically thin torus (see $\left<H/R\right>$ in Table~\ref{tb:model}) tends to have a wide opening angle. The high deposition rate region is extended outside the funnel region regardless of the extent of the low-density regions ($<10^{6}\,{\rm g/cm^3}$). This leads to a low energy conversion efficiency of the neutrino pair annihilation to the relativistic outflow, as discussed below. 

The right panel of Fig.~\ref{fig:qpair-prof} shows the radial profile of the pair annihilation deposition rate within $45^\circ$ from the pole for the fiducial model. The deposition rate approximately follows $r^{-3}$ for $r\lesssim40\,{\rm km}$, and decreases more rapidly for larger radii approximately following $r^{-7}$. This radial dependence of the deposition rate agrees with results found in~\cite{Ruffert:1998qg,Aloy:2004nh,Birkl:2006mu,Zalamea:2010ax,Just:2015dba}. We find that the radial profile of the deposition rate remains approximately the same over time irrespective of the total deposition rate. We also find that the radial profile is similar for the other models, but the radius at which the power-law index changes is approximately proportional to $r_{\rm ms}$.

\begin{figure*}[t]
 	 \includegraphics[width=0.49\linewidth]{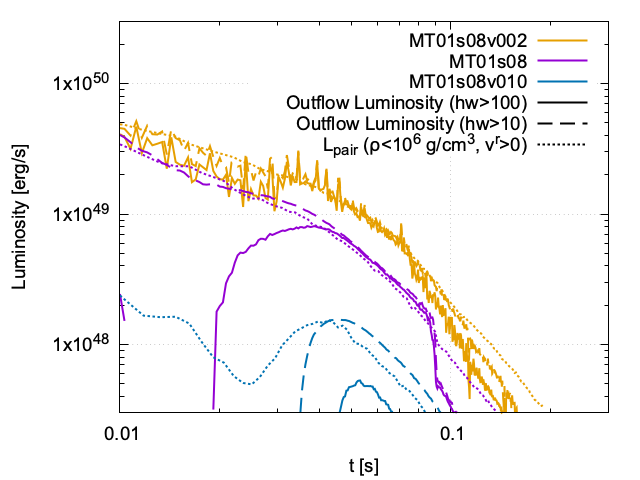}
 	 \includegraphics[width=0.49\linewidth]{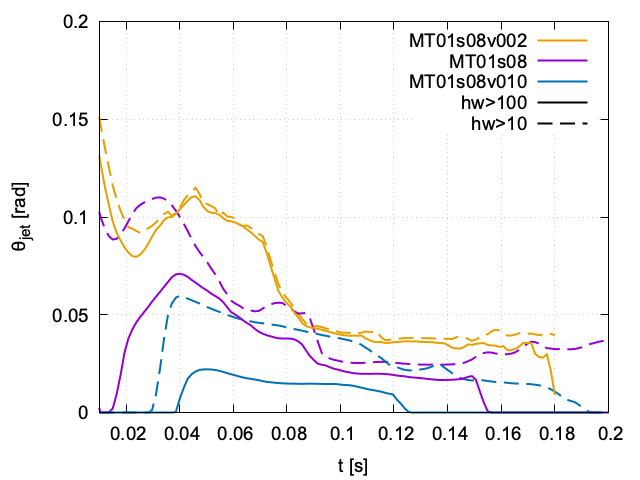}
 	 \caption{(Left panel) Luminosity of the relativistic outflow for the fiducial model. The results for the models with different $\alpha_{\rm vis}$ values are also shown. The solid and dashed curves denote the outflow luminosity for $hw>100$ and $hw>10$, respectively. The total pair annihilation deposition rate of the region with $\rho>10^6\,{\rm g/cm^3}$ and positive radial velocity are shown by the dotted curves. (Right panel) Opening angle of the relativistic outflow as a function of time. The solid and dashed curves denote the outflow luminosity for $hw>100$ and $hw>10$, respectively.}
	 \label{fig:relout_avis}
\end{figure*}

\subsection{Relativistic outflow properties}

The left panel of Fig.~\ref{fig:relout_avis} shows the outflow luminosity for the matter components with $hw>10$ and $hw>100$, $L_{\rm jet,>10}$ and $L_{\rm jet,>100}$, respectively, as functions of time for the fiducial model and its variants with different values of $\alpha_\mathrm{vis}$. We first pay attention to the results of the fiducial setup (MT01s08; the purple curves in Fig.~\ref{fig:relout_avis}). Initially, the matter with $hw>100$ is absent and that with $hw>10$ is only present in the outflow. This reflects the presence of the relatively high rest-mass density in the polar region above the BH, which prevents the matter from having a high Lorentz factor due to the baryon loading effect~\citep{Aloy:2004nh}. 

After $t=20\,{\rm ms}$, the outflow component with $hw>100$ starts appearing as the rest-mass density in the polar region drops due to the expansion of the fireball created in the region above the BH. $L_{\rm jet,>100}$ increases until $t=40\,{\rm ms}$ and reaches a value comparable to $L_{\rm jet,>10}$. After $t=40\,{\rm ms}$, both values of $L_{\rm jet,>100}$ and $L_{\rm jet,>10}$ decrease approximately as $\propto t^{-2}$ until $\approx 100\,{\rm ms}$, and exhibit shut down thereafter.

The outflow luminosity of the relativistic components is much smaller than the total pair annihilation deposition rate (see Fig.~\ref{fig:lnu}). This reflects the fact that only a fraction of the pair annihilation deposition energy is actually deposited in the low-density region, contributing to the launch of a relativistic outflow (see Fig.~\ref{fig:qpair-prof}). In fact, the evolution of $L_{\rm jet,>10}$ and  $L_{\rm jet,>100}$ after reaching the peaks agree approximately with the total pair annihilation deposition rate in the region where the rest-mass density is below
\begin{align}
\rho_{\rm c}&\sim Q_{\rm pair}t_{\rm dyn}/c^2 \\\nonumber &\sim 10^6\,{\rm g/cm^3}  \left(\frac{Q_{\rm pair}}{10^{30}\,{\rm erg/cm^3/s}}\right)\left(\frac{t_{\rm dyn}}{1\,{\rm ms}}\right)^{-1} 
\end{align}
and the radial velocity, $v^r$, is positive. We note that $Q_{\rm pair}\sim 10^{30}\,{\rm erg/cm^3/s}$ and $t_{\rm dyn}\sim 1\,{\rm ms}$ correspond to the typical pair annihilation energy deposition rate for $t\lesssim 50\,{\rm ms}$ and dynamical time scale in the vicinity ($r\approx 30\,{\rm km}$) of the BH, respectively. 

The right panel Fig.~\ref{fig:relout_avis} shows the opening angle of the relativistic outflow for the fiducial model and its variants with different values of $\alpha_\mathrm{vis}$ as a function of time. For the fiducial model (MT01s08), after $t=20\,{\rm ms}$, the opening angles of the relativistic outflow with $hw>100$ and $hw>10$ increase in time up to $t=40\,{\rm ms}$ reaching $0.07\,{\rm rad}$ and $0.11\,{\rm rad}$, respectively. This increase in the opening angle as well as the increase in the value of $L_{\rm jet,>100}$ reflects the development of the fireball due to pair annihilation energy deposition. Thereafter, the opening angles decrease over time. This is due to the decreasing outflow luminosity, which follows the decline in the pair annihilation deposition rate: the lateral pressure of the outflow decreases and this results in narrowing of the outflow in the lateral direction due to the pressure from the torus matter (see~Fig.~\ref{fig:prof_fid}).

\begin{table*}[t]
\caption{Total mass of the matter accreted onto the BH from $t=20\,{\rm ms}$ to $200\,{\rm ms}$ ($M_{\rm fall}$), total emitted neutrino energy ($E_{\rm \nu}$), total pair annihilation deposition energy ($E_{\rm pair}$), total relativistic outflow energy ($E_{\rm jet,>\Gamma_\infty}$), total isotropic relativistic outflow energy ($E^{\rm iso}_{\rm jet,>\Gamma_\infty}$), averaged opening angle ($\left<\theta_{\rm jet,>\Gamma_\infty}\right>$), and duration over which 90\% of the relativistic outflow energy is emitted ($t_{90}^{\rm iso,>\Gamma_\infty}$). The items with $\Gamma_\infty=100$ and $10$ denote the values for the matter component with $hw>100$ and $hw>10$, respectively. Note that the total emitted neutrino energy and total pair annihilation deposition energy are calculated over the time interval of $t=20$--$200\,{\rm ms}$.}
\centering
\begin{tabular}{c|c|c|c|c|c|c|c}\hline
Model	& $M_{\rm fall}$&  $E_{\rm \nu}$& $E_{\rm pair}$& $E_{\rm jet,>100~(10)}$& $E_{\rm jet,>100~(10)}^{\rm iso}$ & $\left<\theta_{\rm jet,>100~(10)}\right>$& $t_{90}^{\rm iso,>100~(10)}$ \\& $[10^{-2}M_\odot]$ & $[10^{51}{\rm erg}]$&$[10^{49}{\rm erg}]$& $[10^{47}\,{\rm erg}]$& $[10^{50}{\rm erg}]$ & [rad]& [s]\\\hline\hline

MT01s08     &3.2&   3.3&   2.4&   3.4~(4.8) &   3.1~(2.0) &0.055~(0.086)&0.12~(0.13)\\
MT01s08LR   &3.2&   3.3&   2.5&   2.5~(4.3) &   3.0~(2.1) &0.047~(0.077)&0.13~(0.13)\\
MT01s08HR   &3.2&   3.3&   2.2&   5.2~(5.5) &   3.1~(2.4) &0.071~(0.079)&0.13~(0.13)\\
MT01s08DF01  &3.2&   3.3&   2.4&   3.4~(4.8) &   2.4~(1.6) &0.060~(0.091)&0.11~(0.12)\\\hline
MT01s08v002  &3.7&   3.3&   2.0&   7.8~(9.2)  &   2.6~(2.9) &0.090~(0.092)&0.12~(0.13)\\
MT01s08v010  &2.6&   2.9&   1.6&   0.17 (0.58)  &   1.1 (0.58) &0.018 (0.049)& 0.11 (0.14)\\
MT01s08v015  &2.4&   2.4&   0.94&   -  &   - &-&-\\\hline
MT01s08ni10 &2.9&   2.8&   2.1&   2.7~(4.1) &   2.6~(1.8) &0.052~(0.081)&0.11~(0.11)\\
MT01s08ni5  &3.0&   3.3&   1.5&   4.3~(5.0) &   1.3~(0.85) &0.12~(0.17)&0.15~(0.17)\\
MT03s08     &9.1&   9.3&   17&   7.1~(13)  &   17~(20) &0.031~(0.043)&0.23~(0.19)\\
\hline
MT01s0      &3.0&   1.0&   0.025&   -  &   - &-&-\\
MT01s095    &2.3&   3.7&   6.4&   2.7~(4.2) &   6.5~(4.5) &0.034~(0.053)&0.12~(0.12)\\
MT03s095    &7.5&   12&   44&   1.6~(4.0) &   9.3~(11) &0.021~(0.032)&0.19~(0.18)\\
\hline
MT02s08MB6  &6.9&   6.2&   5.0&   9.0~(11)  &   1.7~(1.2)  &0.11~(0.14) &0.12~(0.11)\\
MT02s08MB6v0025  &7.7&   6.1&   4.2&   14~(16)  &   1.1~(1.2)  &0.18~(0.18) &0.11~(0.13)\\
\hline
\end{tabular}
\label{tb:jetene}
\end{table*}

Table~\ref{tb:jetene} summarizes several key quantities: total mass of the matter accreted onto the BH from $t=20\,{\rm ms}$ to $200\,{\rm ms}$, total emitted neutrino energy, total pair annihilation deposition energy, total relativistic outflow energy, total isotropic relativistic outflow energy, averaged opening angle, and duration over which 90\% of the relativistic outflow energy is emitted. From $t=20\,{\rm ms}$ to $200\,{\rm ms}$, approximately $0.03\,M_\odot$ ($\approx 30\%$ of the torus mass) has been accreted onto the BH for the fiducial model.
Thus, $\approx$ 6\% of the rest-mass energy of the torus is released in neutrinos. The efficiency is broadly comparable among the other models. The total pair annihilation deposition energy is $\approx$0.4--4\% of the total emitted neutrino energy, and the total relativistic outflow energy is a further $\approx$0.1--5\% of the total pair annihilation deposition energy. In total, the total relativistic outflow energy is $\sim 10^{-5}$--$10^{-6}$ of the rest-mass energy of the accreted torus. Note that our conversion efficiency of pair annihilation from the neutrino emission is a factor of $\approx$ 6--8 higher than those of the similar models in~\cite{Just:2015dba} (SFHo models). We found that a factor of $\approx 4$ of this discrepancy primarily arises because~\cite{Just:2015dba} measured the pair annihilation energy deposition excluding the region with $\theta \geq 45^\circ$ and the innermost area near the black hole ($\lesssim 10\,{\rm km}$), where the energy deposition is strongest. The remaining factor of $\approx 2$ may attribute to the differences in the physical setups.


Figure~\ref{fig:relout_avis} shows that the model with the lower viscous parameter has larger relativistic outflow luminosity. This is primarily because the initial blow-up of the torus matter is less significant and the rest-mass density in the polar region is kept relatively low for the lower viscous parameter models, suppressing the baryon loading effect. In fact, we find that the total pair annihilation deposition rate in the low-density region ($\rho\leq 10^{6}\,{\rm g/cm^3}$) has larger values for the lower viscous parameter models. The opening angle of the relativistic component also increases as the viscous parameter decreases, reflecting the reduced matter pollution in the polar region. On the other hand, we do not find significant launching of relativistic outflows for $\alpha_{\rm vis}=0.15$ (MT01s08v015) due to the heavy matter pollution in the polar region. 



\begin{figure*}
 	 \includegraphics[width=0.49\linewidth]{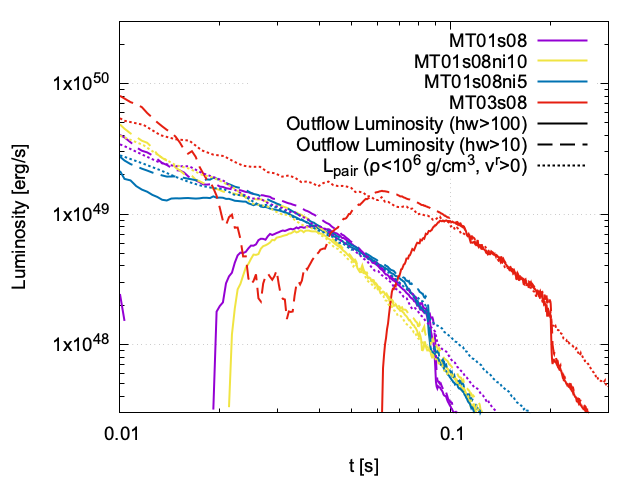}
 	 \includegraphics[width=0.49\linewidth]{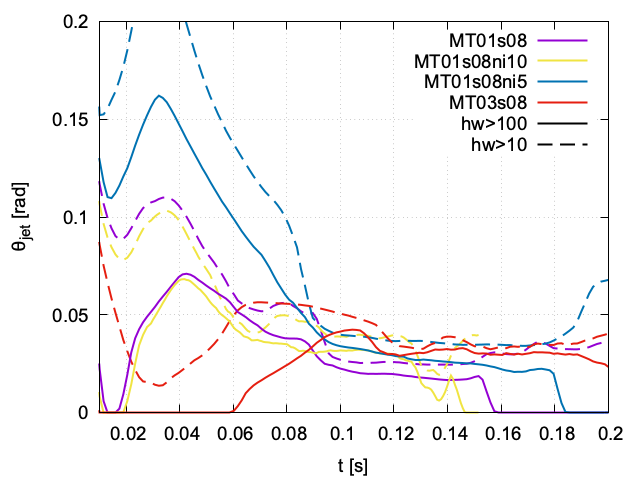}
 	 \caption{The same as Fig.~\ref{fig:relout_avis} but for the models with various angular momentum profiles and torus mass.}
	 \label{fig:relout}
\end{figure*}

The left panel of Fig.~\ref{fig:relout} shows the relativistic outflow luminosity depends only weakly on the initial angular momentum profiles (compare MT01s08ni10 and MT01s08ni5 with MT01s08). 
On the other hand, the model with larger torus mass (MT03s08) results in more energetic and long-lasting relativistic outflow than the fiducial model (MT01s08). The delay and long duration for the peak of MT03s08 are simply due to its large mass.

\begin{figure*}
 	 \includegraphics[width=0.49\linewidth]{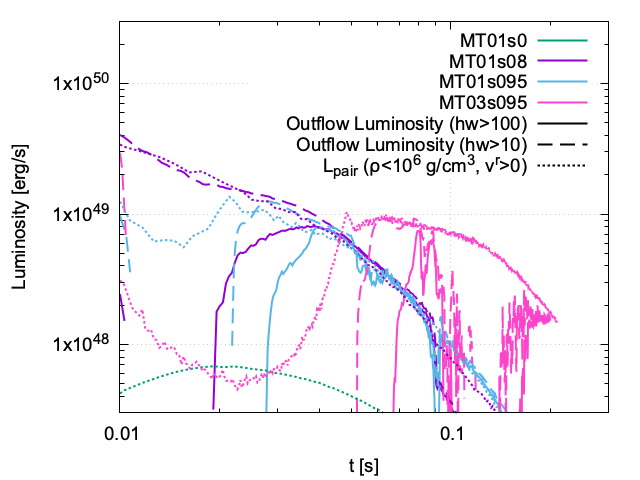}
 	 \includegraphics[width=0.49\linewidth]{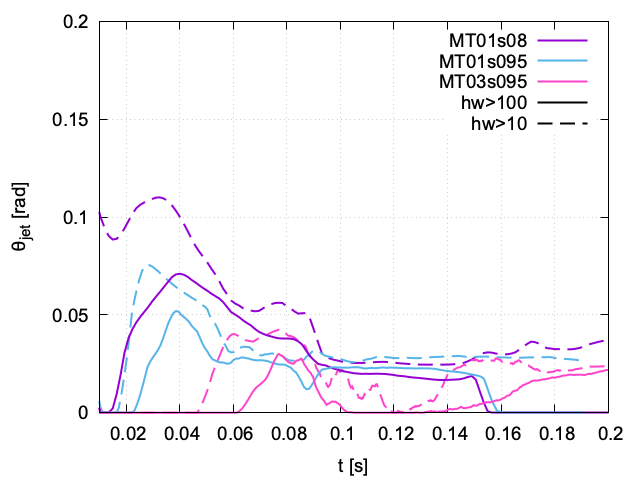}
 	 \caption{The same as Fig.~\ref{fig:relout_avis} but for the models with different $\chi$ values.}
	 \label{fig:relout_chi}
\end{figure*}

Figure~\ref{fig:relout_chi} compares the outflow luminosity and opening angle of the relativistic outflow for the models with different BH spins. We do not find the significant relativistic outflow to be launched for the model with zero BH spin (MT01s0). This reflects that the pair annihilation is inefficient for the zero-spin model. In fact, the total pair annihilation deposition rate and energy are by 1--2 orders of magnitude smaller than those of the fiducial model (see Fig.~\ref{fig:lnu} and Table~\ref{tb:jetene}). 

The model with the larger BH spin ($\chi_{\rm BH}=0.95$; MT01s095) shows only comparable or even slightly lower relativistic outflow luminosity than the fiducial model, although the total pair annihilation deposition rate and energy are more than factor of 2 larger than these of the fiducial model. This reflects the fact that the blow-up of the torus matter and resulting baryon loading in the polar of the BH are more significant for model MT01s095. This is also reflected in the delayed peak time of the outflow luminosity and smaller opening angle of the relativistic outflow (see also Table~\ref{tb:jetene}). 

For MT03s095, as both BH spin and torus mass are large, the initial torus expansion is even more violent than in MT03s08 or MT01s095 enhancing the effects of baryon loading. As a result the relativistic outflow luminosity exhibits intermittent behavior as seen in the left panel of Fig.~\ref{fig:relout_chi}. In particular, the outflow luminosity is smaller than the pair annihilation energy deposition rate in the low-density region. This indicates that the relativistic outflow is not well resolved in this model due to the very narrow opening angle (see the right panel of Fig.~\ref{fig:relout_chi}). Nevertheless, the comparable or larger pair annihilation energy deposition rate with the narrower opening angle suggests that, if the relativistic outflow is well resolved, the isotropic outflow luminosity of MT03s095 could potentially exceed those of MT03s08 and MT01s095.

\begin{figure*}
 	 \includegraphics[width=0.49\linewidth]{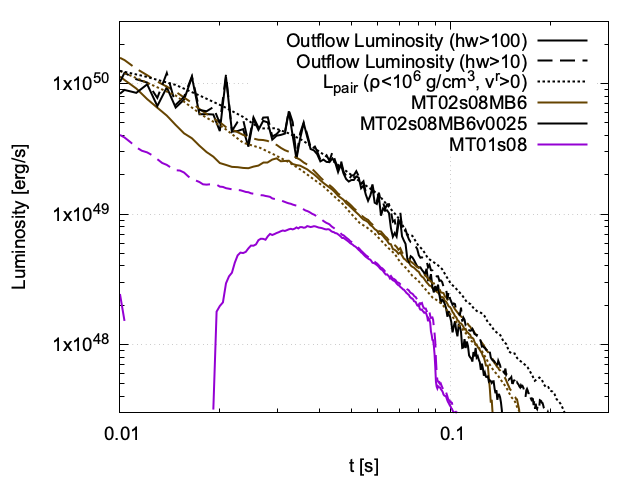}
 	 \includegraphics[width=0.49\linewidth]{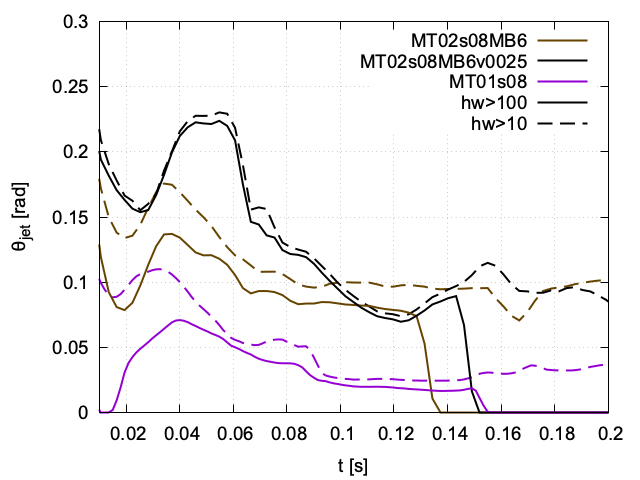}
 	 \caption{The same as Fig.~\ref{fig:relout_avis} but for the models with the BH mass of $6\,M_\odot$ and the torus mass of $0.2\,M_\odot$.}
	 \label{fig:relout_mb6}
\end{figure*}

Figure~\ref{fig:relout_mb6} shows the outflow luminosity and opening angle for the models with the larger BH and torus masses (MT02s08MB6 and MT02s08MB6v0025). The outflow luminosity is by more than a factor of $\approx2$ larger than that of the fiducial model. As a result, the total outflow energy for the components with $hw>100$ is larger by a factor of $\gtrsim 3$ than the fiducial model. On the other hand, the total pair annihilation deposition rate is larger only by a factor of $\approx 2$. These results suggest that the deposited energy due to pair annihilation is more efficiently converted to the outflow energy in these models than the fiducial model. 
This high conversion efficiency of the relativistic outflow is due to the geometrically thin structure of the torus and smaller viscous coefficient (see Table~\ref{tb:model}). By this, the polar region with the low rest-mass density is more extended, and the relativistic fireball is more developed. This is also clearly reflected in the large opening angle, which is approximately twice as large as that of the fiducial model.

\subsection{Isotropic relativistic outflow luminosity and energy}

\begin{figure}
        \includegraphics[width=\linewidth]{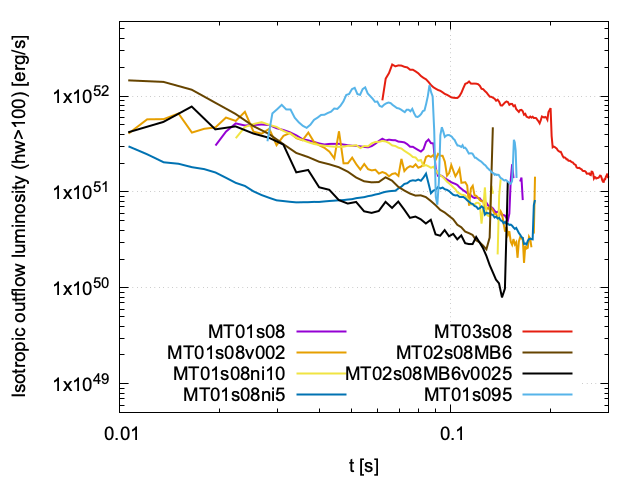}
 	\caption{Isotropic luminosity of the relativistic outflow with $hw>100$ as a function of time for the models exhibiting significant relativistic outflow launching.}
	 \label{fig:reloutiso}
\end{figure}

Figure~\ref{fig:reloutiso} shows the results of the isotropic relativistic outflow luminosity for $hw>100$ as a function of time (see also Table~\ref{tb:jetene}). Here, we only show the results for the models in which significant energy ($E_{\rm jet>100}>10^{47}\,{\rm erg}$) of the relativistic outflow is launched. We also exclude the result of MT03s095 here because the relativistic outflow is likely to be not well resolved. In general, the peak isotropic luminosity and the time scale, $t_\mathrm{iso}$ sustaining the luminosity more than 10\% of the peak value are in the ranges of $\sim 10^{51}$--$10^{52}\,{\rm erg/s}$ and $\sim 0.05$--$0.2\,{\rm s}$, respectively. The isotropic peak luminosity is $\sim 100$--$1000$ times larger than the actual peak luminosity reflecting the opening angle and resulting beaming correction. The time scale, $t_\mathrm{iso}$, is also longer by up to $50\%$ than that of the actual outflow luminosity, because the decrease in the opening angle in the late epoch helps sustaining the isotropic luminosity.

Notably, a high isotropic outflow luminosity does not necessarily reflect a high intrinsic outflow luminosity. Among the models studied, those with a large torus mass (MT03s08) and high BH spin (MT01s095) have the largest isotropic luminosity with peak values exceeding $10^{52}\,{\rm erg/s}$. However, the model with a large torus mass only ranks second or fourth in the intrinsic relativistic outflow luminosity, and the model with a high BH spin only has moderate intrinsic luminosity. In contrast, the isotropic relativistic outflow luminosity of the models with a large BH mass and torus mass (MT02s08MB6 and MT02s08MB6v0025) is always smaller than that of the fiducial model (MT01s08) for $>25\,{\rm ms}$, although the intrinsic luminosity is the largest among the models studied in this paper.

Similar to the isotropic luminosity, the total isotropic energy of the relativistic outflow is also $\sim 10^2$--$10^3$ times larger than the actual energy of the relativistic outflow. The models with the relatively geometrically thin torus, such as MT01s08ni5, MT02s08MB6, and MT02s08MB6v0025 as well as the model with a small viscous parameter ($\alpha_{\rm vis}=0.02$; MT01s08v002) have a smaller enhancement factor compared to others, reflecting its relatively large opening angle.

While the actual relativistic outflow energy varies by more than two orders of magnitudes depending on the properties of the torus and central BH, the total isotropic energy shows relatively less variation among the models. This is because the relativistic outflow energy anti-correlates with the opening angle of the jets and this suppresses the variation in the isotropic energy. In other words, our results suggest that the isotropic energy of the relativistic outflow is basically determined by the isotropic equivalent deposition energy of pair annihilation (see also~\cite{Just:2015dba}). In fact, the total isotropic energy of the relativistic outflow is always by a factor of 3--10 larger than the total pair annihilation energy. This factor broadly agrees with the isotropic correction due to the opening angle of the deposition profile ($\approx 30^\circ$--$45^\circ$). 

Note however that for the models with $\alpha_{\rm vis}=0.15$ (MT01s08v015) and non-spinning BH (MT01s0), in which the pair annihilation deposition rate is too low to sufficiently accelerate the outflow to relativistic speeds ($hw>10$), the approximate correlation between the isotropic relativistic outflow energy and isotropic equivalent deposition energy of pair annihilation is not seen. 
Also, the ratio between the total isotropic relativistic outflow energy and the pair annihilation energy is small ($\approx 3$) compared to the others for the models with a large BH mass (MT02s08MB6 and MT02s08MB6v0025). This is because the region where the pair annihilation heating is significant has a relatively wide opening angle due to the geometrically thin torus.

\begin{figure*}
    \includegraphics[width=0.49\linewidth]{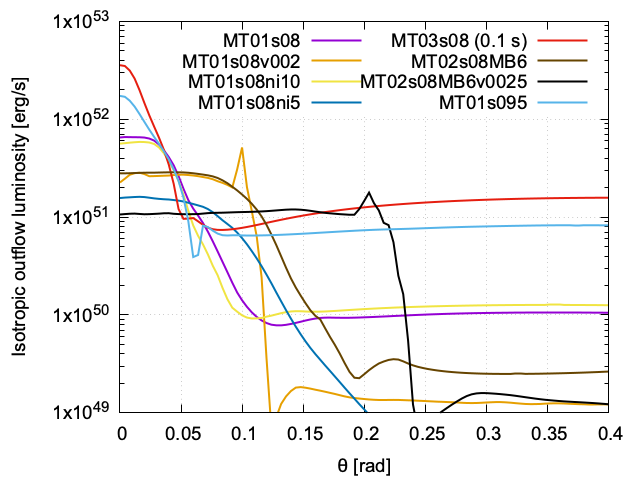}
    \includegraphics[width=0.49\linewidth]{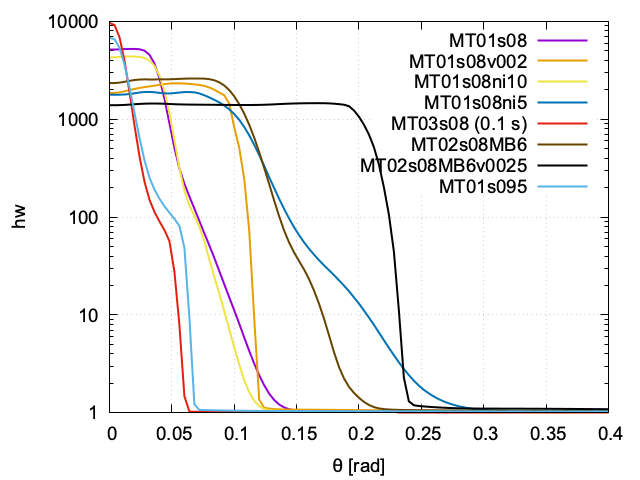}
 	 \caption{(Left panel) Latitudinal profile of the isotropic outflow luminosity averaged over $r=1500$--$3000\,{\rm km}$ obtained for the snapshot at $t=50\,{\rm ms}$ ($t=0.1\,{\rm s}$ for MT03s08). (Right panel) the same as the left panel but for the terminal Lorentz factor.}
	 \label{fig:jet_prof}
\end{figure*}

The left panel of Fig.~\ref{fig:jet_prof} shows the isotropic outflow luminosity at each latitudinal angle, $\theta$, averaged over $r=1500$--$3000\,{\rm km}$. The results are obtained for the snapshot at $t=50\,{\rm ms}$ ($t=0.1\,{\rm s}$ for MT03s08) at which the contribution to the total relativistic outflow energy is approximately the largest. Generally, the isotropic outflow luminosity consists of two parts: one raises in a mountainous shape for $\theta \rightarrow 0$, and the other is approximately constant for large values of $\theta$. The former part corresponds to the relativistic outflow components, while the latter corresponds to the non-relativistic ejecta ($\lesssim 0.1\,c$) which are launched primarily in the beginning of the simulations. The relativistic outflow components are more confined within the small opening angle as the non-relativistic ejecta component becomes energetic. This indicates that the ejecta formed in the initial transient phase, which is not necessarily present in reality, plays a dominant role for the relativistic outflow collimation in our simulations. 


The right panel of Fig.~\ref{fig:jet_prof} shows the latitudinal profile of the terminal Lorentz factor, which is composed of approximately a flat core part with an exponential cutoff on the side. The shape of the profile agrees broadly with the results of~\cite{Aloy:2004nh}, while our results show a steeper structure for a larger latitudinal angle. In particular, the results of the models with small viscous coefficients (MT01s08v002 and MT02s08MB6v0025) show a step-function-like cutoff at the edge of the relativistic outflow. 

In our model, the matter entrainment and mass loading due to the Kelvin-Helmholtz (KH) instability at the lateral edge of the relativistic outflow is less significant, which is explained as the cause of the shallower extension of the Lorentz factor profile in~\cite{Aloy:2004nh}. In fact, in our model, the relativistic outflow is likely to be collimated by the presence of the non-relativistic outflow with the typical radial velocity of $\sim 0.1\,c$. Then, the growth of the KH instability is relatively suppressed compared to the situation in the models in~\cite{Aloy:2004nh} because the relative velocity between the relativistic outflow and the ambient matter is smaller. However, we note that the terminal Lorentz factors in our simulations are significantly affected by errors arising from the imposed floor density and finite grid resolution (see App.~\ref{app:err})

\section{Discussion}\label{sec:dis}

\subsection{Observational implication}\label{sec:dis-obs}
Here, we discuss the implication to the observation of sGRBs, under the assumption that our results on the isotropic luminosity and energy of the relativistic outflow are reliable within an order of magnitude.

The isotropic energy of the relativistic outflow in our models is too small to explain the typical prompt emission of sGRBs~\citep{Nakar:2007yr,Berger:2013jza}. The isotropic gamma-ray energy of sGRB prompt emissions ranges from $10^{48}\,{\rm erg}$ to $10^{52}\,{\rm erg}$~\cite{2021RAA....21..254L}. Assuming a conversion efficiency of $\sim 10\%$ for the relativistic outflow energy into gamma-rays, this indicates that the isotropic outflow energy for some of the observed sGRB prompt emissions has to be larger than $10^{52}\,{\rm erg}$. On the other hand, the isotropic outflow energy of the models studied in this work is typically $\sim 10^{50}\,{\rm erg}$, and at most $10^{51}\,{\rm erg}$. Hence, the relativistic outflow, purely driven by neutrino pair annihilation, is not energetic enough to explain all the typical sGRB prompt emissions, unless the torus mass is much larger than $0.3\,M_\odot$ and the BH spin is close to 1.

Nevertheless, some of the total isotropic energy found in our models is large enough to explain a faint class of sGRBs with the isotropic gamma-ray energy $\alt 10^{50}\,{\rm erg}$. As a point of concern, the duration of jet launch in our models is at most $\approx 0.2\,{\rm s}$. If this duration determines the maximum duration of a GRB, the pair annihilation mechanism can explain only a subclass of sGRBs with the duration $\alt 0.2\,{\rm s}$. However, as is discussed in~\cite{Aloy:2004nh,Janka:2005yh}, the duration of the GRB emission may be approximately by an order of magnitude longer than the intrinsic jet launch duration due to the radial stretching effect of the shell. If this is the case, our models may be still consistent with the observation. It is also worthy to note that the time scale and energy of the relativistic outflow found in this work are consistent with the precursors of GRBs~\cite{2010ApJ...723.1711T,Xiao:2022quv}.

Among the properties of the BH-torus system studied in this paper, the variation in the total pair annihilation deposition energy as well as in the total isotropic energy of the relativistic outflow is found to be only within a factor of $\approx 2$ for the models with the same torus mass, BH mass, and BH spin. On the other hand, the torus mass significantly affects the total emitted neutrino energy, pair annihilation deposition energy, and isotropic relativistic outflow energy. While the total emitted neutrino energy increases approximately linearly with the torus mass, the total pair annihilation deposition energy as well as the isotropic relativistic outflow energy nonlinearly increases with the power index of $\sim 2$. The BH spin also has a strong effect particularly on the total pair annihilation deposition energy and isotropic relativistic outflow energy: a factor of $\approx 2$--3 increase is found by the increase of the BH spin from 0.8 to 0.95.

The dimensionless spin of the remnant BH formed as a result of BNS mergers are typically 0.7--$0.8$~\cite{Kiuchi:2009jt,Dietrich:2016hky,Schianchi:2024vvi} unless the NS spins are extremely large~\citep{Karakas:2024avr}. On the other hand, the torus mass can have diversity depending on the NS masses and NS equation of state, and it can be up to $0.3\,M_\odot$ for the scenario that the remnant central object eventually becomes a BH ~\cite{Kiuchi:2009jt,Hotokezaka:2013iia,Dietrich:2016hky}. However, for the case that the BH is formed within a short time scale ($\lesssim 10\,{\rm ms}$) after the onset of the merger, which we focus on in this paper, the torus mass is typically small ($\lesssim 0.01\,M_\odot$) unless the mass ratio is substantially different from the unity~\citep{Kiuchi:2019lls}. Hence, our results suggest that the energetic relativistic outflow driven by neutrino pair annihilation may not be launched from the BH-torus system formed immediately after the merger of equal-mass BNSs.

Initial profiles we employed in this paper are idealized because only a BH and a torus are present. In reality, the remnant torus formed in a BNS merger is likely surrounded by matter ejected during the merger and post-merger phases, forming a high-density envelope along the axis that suppresses the formation of the relativistic outflow~\cite{Gottlieb:2022sis,Kiuchi:2023obe}. On the other hand, if a remnant massive NS formed after a BNS merger survives for a long time scale ($\gtrsim 100\,{\rm ms}$), it is a bright source of neutrinos, which could help driving relativistic outflows by neutrino irradiation and energy injection from neutrino pair annihilation. To date, the quantitative understanding of such effects is still incomplete because they require a detailed study of neutrino transfer. Specifically, to solve the high opacity region in the current work, we need to follow a large number of the MC packets, and hence, the computational cost becomes quite demanding. For studying such systems with high-opacity regions, we need to develop a method to reduce the computational cost in the high opacity region~\citep{Foucart:2017mbt,Izquierdo:2023fub}, which we leave as the future task.

In BH-NS mergers, the polar region of the torus will be less polluted by matter~\citep{Kyutoku:2015gda,Hayashi:2021oxy}. This gives an optimal condition for the relativistic fireball to be well developed. Also, if the initial BH is rapidly spinning, the remnant BH can have the BH spin larger than the case of BNSs, for example, with the dimensionless value larger than 0.9~\citep{Kyutoku:2011vz}. However, our result also suggests that the increase in the BH mass for the fixed torus mass decreases the efficiency of converting the neutrino radiation energy into the matter internal energy by pair annihilation. The matter remains outside the remnant BH after the merger also decreases as the initial BH mass increases and initial BH spin decreases~\cite{Rosswog:2005su,Shibata:2007zm,Etienne:2008re,Lovelace:2013vma,Kyutoku:2015gda,Foucart:2018rjc}. Therefore, energetic relativistic outflows driven by pair annihilation are more likely in BH-NS mergers with a low-mass, rapidly spinning BH.

Nevertheless, the pair annihilation-driven relativistic outflow can be a central engine of sGRBs if a BH-NS binary is composed of a low-mass and high-spin BH. For example, a merger of $2.7\,M_\odot$ BH and $1.35\,M_\odot$ NS with the dimensionless BH spin of 0.75 and the NS radius of $13.6\,{\rm km}$ results in a remnant BH of $3.7\,M_\odot$ with the dimensionless BH spin of $\approx 0.9$ surrounded by matter of $\approx 0.3\,M_\odot$~\citep{Kyutoku:2011vz}. Our results suggest that the pair annihilation rate is approximately $\propto {\hat r}_{\rm isco}^{-5}$ and $\propto M_{\rm BH}^{-3/2}$, and thus, the enhancement due to a large spin approximately compensates with the suppression due to the increases in the BH mass. Hence, the total pair annihilation deposition energy and total isotropic relativistic outflow energy are likely to be comparable to those of the model with $M_{\rm torus}=0.3\,M_\odot$ (MT03s08) for this system.

Our results may imply that jets driven by the other mechanism, namely magnetically driven relativistic outflows, are needed to interpret at least the bright class of observed sGRBs. MHD effects will also be important for determining the property of pair annihilation driven outflows as they drive the angular momentum transport and release of gravitational binding energy in the BH-torus system. In fact, we found that the results on the pair annihilation rates and relativistic outflow property strongly depend on the effective viscous parameter, which should be in reality determined by the magnetically driven turbulence~\cite{Balbus:1998ja,Fernandez:2018kax,Christie:2019lim,Kiuchi:2022nin,Hayashi:2022cdq}. On the other hand, the energy deposition from pair annihilation could aid in launching and accelerating a magnetically driven jet by blowing up matter in the polar region and entraining poloidal magnetic field lines, creating a globally coherent structure. To clarify the effects that can be induced by the interplay of MHD effects and pair annihilation energy deposition, we need to perform a radiation hydrodynamics simulations consistently following the pair annihilation process, also taking into account MHD effects.

\subsection{Limitation of our numerical simulation results}
Our simulation results in this paper can be affected by the error due to the finite grid resolution, presence of the floor density, artifact of the initial setups, simplified physics, and assumption of axisymmetry. Here, we broadly evaluate which part of the results can be relatively robust and which cannot, and how they affect the implication of our results to the observables of sGRB discussed above. A detailed summary of the finite grid resolution and density floor effects is provided in App.~\ref{app:err}.

The total emitted neutrino luminosity and energy are among the quantities that are relatively reliable in our results. In fact, we find that they depend only weakly on the grid resolutions and the floor density setup (see Table~\ref{tb:jetene}). The broad agreement in the neutrino luminosity among the models with different angular momentum profiles as well as with the results of the previous studies including the 3D simulation~\citep{Just:2015dba,Fujibayashi:2020qda,Fujibayashi:2020jfr,Kiuchi:2022nin} also suggests that the results are relatively robust within a factor of 2--3 except for the very early epoch ($\lesssim 10\,{\rm ms}$) and late time ($\gtrsim 1\,{\rm s}$). 
On the other hand, the variation in the luminosity among the models with different viscous parameters indicates that the uncertainty in the angular momentum transport process has a large impact on the results.

Our results of the total pair annihilation deposition rate and energy can suffer more strongly from the systematic errors compared to the total emitted neutrino luminosity and energy. In fact, we find that the difference in the total pair annihilation deposition energy among the runs with different grid resolutions is larger than that in the total emitted neutrino energy (see Table~\ref{tb:jetene}). The difference in the total pair annihilation deposition energy among the runs with different viscous parameters shows also larger variety than that for the total emitted neutrino energy, suggesting that the results are more sensitive to the uncertainty in the angular momentum transport mechanism. Nevertheless, the results are still in agreement with a factor of $\approx 2$, and this may indicate that the results are reliable within an order of magnitude. Additionally, the fact that the total pair annihilation deposition rate as a function of the mass accretion rate is largely insensitive to the initial mass and angular momentum profile of the torus suggests that this relationship is semi-quantitatively reliable.

By contrast, the results of the relativistic outflow in our simulation should be interpreted with caution. The error of the relativistic outflow energy as well as the opening angle of the outflow in our simulations due to the finite grid resolution can be at least more than a factor of 2 (see Table~\ref{tb:jetene}). The presence of the floor density in the simulations also has a large impact on the estimation of the terminal Lorentz factor of the outflow, while it does not strongly impact the relativistic outflow luminosity (see App.~\ref{app:err}). Furthermore, in our models, the density structure in the polar region is influenced by the initial blow-up of the torus and mass ejection during the relaxation of the initial profile. Since the density profile around the polar region is crucial for determining jet properties, the relativistic outflow energy and opening angle may depend strongly on the initial data setup. As mentioned earlier, we employ idealized initial profiles in which the matter around the BH-tours is absent. From this point of view, our initial settings of the simulations represent optimal conditions for the formation of relativistic outflows of no ram pressure by infalling matter and no baryon loading effect~\citep{Aloy:2004nh,Just:2015dba}.

Nevertheless, the total isotropic relativistic outflow energy may be less affected by these errors and uncertainty. The total isotropic energy of the relativistic outflow shows less variation among the models, while the intrinsic relativistic outflow energy varies by more than two orders of magnitude depending on the properties of the torus and central BH. In particular, the total isotropic relativistic outflow energy shows only the variation within $\sim 20\%$ among the different resolution runs and different floor density runs (Table~\ref{tb:jetene}). These reflect that the isotropic energy of the relativistic outflow is likely to be determined primarily by the isotropic equivalent deposition energy of pair annihilation (see also~\cite{Just:2015dba}). If this is the case, our results of the isotropic relativistic outflow energy may be reliable within an order of magnitude. Thus, it is reasonable to conclude that pair annihilation is a possible explanation for sGRB prompt emissions with $\lesssim 10^{50}\,{\rm erg}$.

\section{Summary}\label{sec:sum}

In this paper, we studied the evolution of BH-torus systems with various setups which can be formed as a result of NS mergers by performing axisymmetric MC-based full-Boltzmann neutrino radiation viscous hydrodynamics simulations. We focused on their properties of neutrino emission, energy deposition due to pair annihilation, and relativistic outflow. In particular, neutrino-antineutrino pair annihilation is taken into account for both electron and heavy lepton types, considering non-thermal distribution effects. This is the first study to dynamically model neutrino-antineutrino pair annihilation-driven relativistic outflows, taking into account full Boltzmann transport in various setups of BH-torus systems.

We obtained the results for the dependence of the emitted neutrino luminosity and energy on the viscous parameter and property of BH-torus systems consistent with those in the previous studies~\citep{Just:2015dba,Fujibayashi:2020qda,Fujibayashi:2020jfr}. The pair annihilation deposition rate and energy follow a similar dependence as the neutrino luminosity and energy, while the relative difference among the different models is more pronounced. In our models, $\approx$2--10\% percent of the rest-mass energy of the accreted torus is released in neutrinos, and a further $\approx$0.4--4\% of the total emitted neutrino energy is deposited to matter by pair annihilation. The pair annihilation deposition rate shows a steeper decline than the neutrino luminosity in contrast to~\cite{Setiawan:2005ah}, which is likely due to the higher initial $Y_e$ profiles of the torus in our models. 


We found a clear relationship between the pair-annihilation energy deposition rate and mass accretion rate, which is consistent with the previous work~\cite{Zalamea:2010ax}. This relation is largely insensitive to the initial torus angular momentum profile and mass. On the other hand, the efficiency depends strongly on the BH spin; it is by a factor of $\approx 30$ larger for $\chi_\mathrm{BH}=0.95$ (MT01s095 and MT03s095) than for $\chi_\mathrm{BH}=0$ (MT01s0). This is again in broad agreement with that in~\citep{Zalamea:2010ax}. The efficiency also decreases with the increase of the BH mass, approximately following $M_{\rm BH}^{-3/2}$, and with the increase of the viscous parameter, particularly in the regime in which mass accretion rate is lower than $0.1\,M_\odot/{\rm s}$. 

We derived a fitting formula that reproduces the total pair-annihilation energy deposition rate as a function of the mass accretion rate with an accuracy of $\sim 30\%$ across our model set. The spatial profile of the deposition rate shows a quasi-universal power-law structure with a radial index of $-3$ to $-4$ within an opening angle of $30^\circ$--$45^\circ$, which is broadly consistent with the findings in the previous studies~\cite{Ruffert:1998qg,Aloy:2004nh,Birkl:2006mu,Zalamea:2010ax,Just:2015dba}. This indicates a potential strategy for approximating pair-annihilation heating without explicitly solving it in simulations. Although such an approach violates strict energy-momentum conservation, it provides a practical way to capture pair annihilation heating effects based on accretion rate and a parameterized spatial distribution, particularly useful for exploratory 3D MHD simulations for which full neutrino transport is still computationally prohibited.

In most of the models studied, except those with the largest viscous parameter or a non-spinning BH, relativistic outflows are driven by pair annihilation with the terminal Lorentz factors exceeding 100. The actual relativistic outflow energy and opening angle are typically in the ranges of $10^{46}$--$10^{48}\,{\rm erg}$ and $0.05$--$0.2\,{\rm rad}$, respectively, with the typical duration being $\sim 0.1\,{\rm s}$. This corresponds to $\sim 10^{-5}$--$10^{-6}$ of the accreted torus rest-mass energy. The low efficiency arises because only the pair annihilation energy deposition in low-density regions contributes to the outflow launching. The opening angle of the relativistic outflow is determined by the matter density structure around the symmetric axis, which is in our models determined by the initial blow-up of the torus and mass ejection due to the relaxation of the initial profile. 

The outflow energy, opening angle, and terminal Lorentz factor in our simulations depend on grid resolution and floor density settings. Since matter density near the symmetry axis is sensitive to initial conditions, our results should be interpreted qualitatively as indicative of possible realizations. In contrast, the isotropic relativistic outflow energy and launching duration appear more reliable, as they are less sensitive to grid resolution and floor density. Their strong correlation with the isotropic-equivalent pair annihilation deposition rate suggests less dependence on outflow collimation, which is mainly influenced by the initial simulation setup.

Under the assumption that our results of the isotropic relativistic outflow energy as well as the launching duration are reliable within a factor of 2--3, our results suggest that a faint class of the sGRB prompt emission of which isotropic gamma-ray energy is less than $10^{49}\,{\rm erg}$ or GRB precursors can be explained by the pair annihilation driven relativistic outflows, if the gamma-ray efficiency is as large as $\sim 10\%$. The sGRB prompt emission with the isotropic gamma-ray energy of $10^{50}\,{\rm erg}$ can be also explained if the torus mass is sufficiently large ($>0.3\,M_\odot$), the dimensionless BH spin is sufficiently high ($>0.95$), and the BH mass is sufficiently small ($\lesssim 3\,M_\odot$).

In order to robustly conclude these findings, we need further investigation by performing a simulation with more realistic initial setups, with a higher grid resolution, and with lower floor density. It is also important to consistently follow the propagation of the relativistic outflow for longer time scale and in more extended computational domain to determine the observational property. A radiation-MHD simulation consistently following the pair annihilation process is also needed to clarify the effects of the interplay between pair annihilation driven fireballs and MHD effects on launching the relativistic outflows.

\acknowledgments We thank Kota Hayashi, Hiroki Nagakura, Ioka Kunihito, and Kenta Hotokezaka for the valuable discussion. Numerical computation was performed on Yukawa21 at Yukawa Institute for Theoretical Physics, Kyoto University and the Sakura and Momiji clusters at Max Planck Computing and Data Facility. This work was supported by Grant-in-Aid for Scientific Research (23H04900) of JSPS/MEXT.

\bibliographystyle{apsrev4-2}

\begin{thebibliography}{123}%
\makeatletter
\providecommand \@ifxundefined [1]{%
 \@ifx{#1\undefined}
}%
\providecommand \@ifnum [1]{%
 \ifnum #1\expandafter \@firstoftwo
 \else \expandafter \@secondoftwo
 \fi
}%
\providecommand \@ifx [1]{%
 \ifx #1\expandafter \@firstoftwo
 \else \expandafter \@secondoftwo
 \fi
}%
\providecommand \natexlab [1]{#1}%
\providecommand \enquote  [1]{``#1''}%
\providecommand \bibnamefont  [1]{#1}%
\providecommand \bibfnamefont [1]{#1}%
\providecommand \citenamefont [1]{#1}%
\providecommand \href@noop [0]{\@secondoftwo}%
\providecommand \href [0]{\begingroup \@sanitize@url \@href}%
\providecommand \@href[1]{\@@startlink{#1}\@@href}%
\providecommand \@@href[1]{\endgroup#1\@@endlink}%
\providecommand \@sanitize@url [0]{\catcode `\\12\catcode `\$12\catcode
  `\&12\catcode `\#12\catcode `\^12\catcode `\_12\catcode `\%12\relax}%
\providecommand \@@startlink[1]{}%
\providecommand \@@endlink[0]{}%
\providecommand \url  [0]{\begingroup\@sanitize@url \@url }%
\providecommand \@url [1]{\endgroup\@href {#1}{\urlprefix }}%
\providecommand \urlprefix  [0]{URL }%
\providecommand \Eprint [0]{\href }%
\providecommand \doibase [0]{https://doi.org/}%
\providecommand \selectlanguage [0]{\@gobble}%
\providecommand \bibinfo  [0]{\@secondoftwo}%
\providecommand \bibfield  [0]{\@secondoftwo}%
\providecommand \translation [1]{[#1]}%
\providecommand \BibitemOpen [0]{}%
\providecommand \bibitemStop [0]{}%
\providecommand \bibitemNoStop [0]{.\EOS\space}%
\providecommand \EOS [0]{\spacefactor3000\relax}%
\providecommand \BibitemShut  [1]{\csname bibitem#1\endcsname}%
\let\auto@bib@innerbib\@empty
\bibitem [{\citenamefont {{Paczynski}}(1986)}]{1986ApJ...308L..43P}%
  \BibitemOpen
  \bibfield  {author} {\bibinfo {author} {\bibfnamefont {B.}~\bibnamefont
  {{Paczynski}}},\ }\href {https://doi.org/10.1086/184740} {\bibfield
  {journal} {\bibinfo  {journal} {\apjl}\ }\textbf {\bibinfo {volume} {308}},\
  \bibinfo {pages} {L43} (\bibinfo {year} {1986})}\BibitemShut {NoStop}%
\bibitem [{\citenamefont {{Goodman}}(1986)}]{1986ApJ...308L..47G}%
  \BibitemOpen
  \bibfield  {author} {\bibinfo {author} {\bibfnamefont {J.}~\bibnamefont
  {{Goodman}}},\ }\href {https://doi.org/10.1086/184741} {\bibfield  {journal}
  {\bibinfo  {journal} {\apjl}\ }\textbf {\bibinfo {volume} {308}},\ \bibinfo
  {pages} {L47} (\bibinfo {year} {1986})}\BibitemShut {NoStop}%
\bibitem [{\citenamefont {{Fong}}\ \emph {et~al.}(2010)\citenamefont {{Fong}},
  \citenamefont {{Berger}},\ and\ \citenamefont {{Fox}}}]{2010ApJ...708....9F}%
  \BibitemOpen
  \bibfield  {author} {\bibinfo {author} {\bibfnamefont {W.}~\bibnamefont
  {{Fong}}}, \bibinfo {author} {\bibfnamefont {E.}~\bibnamefont {{Berger}}},\
  and\ \bibinfo {author} {\bibfnamefont {D.~B.}\ \bibnamefont {{Fox}}},\ }\href
  {https://doi.org/10.1088/0004-637X/708/1/9} {\bibfield  {journal} {\bibinfo
  {journal} {\apj}\ }\textbf {\bibinfo {volume} {708}},\ \bibinfo {pages} {9}
  (\bibinfo {year} {2010})},\ \Eprint {https://arxiv.org/abs/0909.1804}
  {arXiv:0909.1804 [astro-ph.HE]} \BibitemShut {NoStop}%
\bibitem [{\citenamefont {{Fong}}\ \emph {et~al.}(2022)\citenamefont {{Fong}},
  \citenamefont {{Nugent}}, \citenamefont {{Dong}}, \citenamefont {{Berger}},
  \citenamefont {{Paterson}}, \citenamefont {{Chornock}}, \citenamefont
  {{Levan}}, \citenamefont {{Blanchard}}, \citenamefont {{Alexander}},
  \citenamefont {{Andrews}}, \citenamefont {{Cobb}}, \citenamefont
  {{Cucchiara}}, \citenamefont {{Fox}}, \citenamefont {{Fryer}}, \citenamefont
  {{Gordon}}, \citenamefont {{Kilpatrick}}, \citenamefont {{Lunnan}},
  \citenamefont {{Margutti}}, \citenamefont {{Miller}}, \citenamefont
  {{Milne}}, \citenamefont {{Nicholl}}, \citenamefont {{Perley}}, \citenamefont
  {{Rastinejad}}, \citenamefont {{Escorial}}, \citenamefont {{Schroeder}},
  \citenamefont {{Smith}}, \citenamefont {{Tanvir}},\ and\ \citenamefont
  {{Terreran}}}]{2022ApJ...940...56F}%
  \BibitemOpen
  \bibfield  {author} {\bibinfo {author} {\bibfnamefont {W.-f.}\ \bibnamefont
  {{Fong}}}, \bibinfo {author} {\bibfnamefont {A.~E.}\ \bibnamefont
  {{Nugent}}}, \bibinfo {author} {\bibfnamefont {Y.}~\bibnamefont {{Dong}}},
  \bibinfo {author} {\bibfnamefont {E.}~\bibnamefont {{Berger}}}, \bibinfo
  {author} {\bibfnamefont {K.}~\bibnamefont {{Paterson}}}, \bibinfo {author}
  {\bibfnamefont {R.}~\bibnamefont {{Chornock}}}, \bibinfo {author}
  {\bibfnamefont {A.}~\bibnamefont {{Levan}}}, \bibinfo {author} {\bibfnamefont
  {P.}~\bibnamefont {{Blanchard}}}, \bibinfo {author} {\bibfnamefont {K.~D.}\
  \bibnamefont {{Alexander}}}, \bibinfo {author} {\bibfnamefont
  {J.}~\bibnamefont {{Andrews}}}, \bibinfo {author} {\bibfnamefont {B.~E.}\
  \bibnamefont {{Cobb}}}, \bibinfo {author} {\bibfnamefont {A.}~\bibnamefont
  {{Cucchiara}}}, \bibinfo {author} {\bibfnamefont {D.}~\bibnamefont {{Fox}}},
  \bibinfo {author} {\bibfnamefont {C.~L.}\ \bibnamefont {{Fryer}}}, \bibinfo
  {author} {\bibfnamefont {A.~C.}\ \bibnamefont {{Gordon}}}, \bibinfo {author}
  {\bibfnamefont {C.~D.}\ \bibnamefont {{Kilpatrick}}}, \bibinfo {author}
  {\bibfnamefont {R.}~\bibnamefont {{Lunnan}}}, \bibinfo {author}
  {\bibfnamefont {R.}~\bibnamefont {{Margutti}}}, \bibinfo {author}
  {\bibfnamefont {A.}~\bibnamefont {{Miller}}}, \bibinfo {author}
  {\bibfnamefont {P.}~\bibnamefont {{Milne}}}, \bibinfo {author} {\bibfnamefont
  {M.}~\bibnamefont {{Nicholl}}}, \bibinfo {author} {\bibfnamefont
  {D.}~\bibnamefont {{Perley}}}, \bibinfo {author} {\bibfnamefont
  {J.}~\bibnamefont {{Rastinejad}}}, \bibinfo {author} {\bibfnamefont {A.~R.}\
  \bibnamefont {{Escorial}}}, \bibinfo {author} {\bibfnamefont
  {G.}~\bibnamefont {{Schroeder}}}, \bibinfo {author} {\bibfnamefont
  {N.}~\bibnamefont {{Smith}}}, \bibinfo {author} {\bibfnamefont
  {N.}~\bibnamefont {{Tanvir}}},\ and\ \bibinfo {author} {\bibfnamefont
  {G.}~\bibnamefont {{Terreran}}},\ }\href
  {https://doi.org/10.3847/1538-4357/ac91d0} {\bibfield  {journal} {\bibinfo
  {journal} {\apj}\ }\textbf {\bibinfo {volume} {940}},\ \bibinfo {eid} {56}
  (\bibinfo {year} {2022})},\ \Eprint {https://arxiv.org/abs/2206.01763}
  {arXiv:2206.01763 [astro-ph.GA]} \BibitemShut {NoStop}%
\bibitem [{\citenamefont {{Nugent}}\ \emph {et~al.}(2022)\citenamefont
  {{Nugent}}, \citenamefont {{Fong}}, \citenamefont {{Dong}}, \citenamefont
  {{Leja}}, \citenamefont {{Berger}}, \citenamefont {{Zevin}}, \citenamefont
  {{Chornock}}, \citenamefont {{Cobb}}, \citenamefont {{Kelley}}, \citenamefont
  {{Kilpatrick}}, \citenamefont {{Levan}}, \citenamefont {{Margutti}},
  \citenamefont {{Paterson}}, \citenamefont {{Perley}}, \citenamefont
  {{Escorial}}, \citenamefont {{Smith}},\ and\ \citenamefont
  {{Tanvir}}}]{2022ApJ...940...57N}%
  \BibitemOpen
  \bibfield  {author} {\bibinfo {author} {\bibfnamefont {A.~E.}\ \bibnamefont
  {{Nugent}}}, \bibinfo {author} {\bibfnamefont {W.-F.}\ \bibnamefont
  {{Fong}}}, \bibinfo {author} {\bibfnamefont {Y.}~\bibnamefont {{Dong}}},
  \bibinfo {author} {\bibfnamefont {J.}~\bibnamefont {{Leja}}}, \bibinfo
  {author} {\bibfnamefont {E.}~\bibnamefont {{Berger}}}, \bibinfo {author}
  {\bibfnamefont {M.}~\bibnamefont {{Zevin}}}, \bibinfo {author} {\bibfnamefont
  {R.}~\bibnamefont {{Chornock}}}, \bibinfo {author} {\bibfnamefont {B.~E.}\
  \bibnamefont {{Cobb}}}, \bibinfo {author} {\bibfnamefont {L.~Z.}\
  \bibnamefont {{Kelley}}}, \bibinfo {author} {\bibfnamefont {C.~D.}\
  \bibnamefont {{Kilpatrick}}}, \bibinfo {author} {\bibfnamefont
  {A.}~\bibnamefont {{Levan}}}, \bibinfo {author} {\bibfnamefont
  {R.}~\bibnamefont {{Margutti}}}, \bibinfo {author} {\bibfnamefont
  {K.}~\bibnamefont {{Paterson}}}, \bibinfo {author} {\bibfnamefont
  {D.}~\bibnamefont {{Perley}}}, \bibinfo {author} {\bibfnamefont {A.~R.}\
  \bibnamefont {{Escorial}}}, \bibinfo {author} {\bibfnamefont
  {N.}~\bibnamefont {{Smith}}},\ and\ \bibinfo {author} {\bibfnamefont
  {N.}~\bibnamefont {{Tanvir}}},\ }\href
  {https://doi.org/10.3847/1538-4357/ac91d1} {\bibfield  {journal} {\bibinfo
  {journal} {\apj}\ }\textbf {\bibinfo {volume} {940}},\ \bibinfo {eid} {57}
  (\bibinfo {year} {2022})},\ \Eprint {https://arxiv.org/abs/2206.01764}
  {arXiv:2206.01764 [astro-ph.GA]} \BibitemShut {NoStop}%
\bibitem [{\citenamefont {{O'Connor}}\ \emph {et~al.}(2022)\citenamefont
  {{O'Connor}}, \citenamefont {{Troja}}, \citenamefont {{Dichiara}},
  \citenamefont {{Beniamini}}, \citenamefont {{Cenko}}, \citenamefont
  {{Kouveliotou}}, \citenamefont {{Gonz{\'a}lez}}, \citenamefont {{Durbak}},
  \citenamefont {{Gatkine}}, \citenamefont {{Kutyrev}}, \citenamefont
  {{Sakamoto}}, \citenamefont {{S{\'a}nchez-Ram{\'\i}rez}},\ and\ \citenamefont
  {{Veilleux}}}]{2022MNRAS.515.4890O}%
  \BibitemOpen
  \bibfield  {author} {\bibinfo {author} {\bibfnamefont {B.}~\bibnamefont
  {{O'Connor}}}, \bibinfo {author} {\bibfnamefont {E.}~\bibnamefont {{Troja}}},
  \bibinfo {author} {\bibfnamefont {S.}~\bibnamefont {{Dichiara}}}, \bibinfo
  {author} {\bibfnamefont {P.}~\bibnamefont {{Beniamini}}}, \bibinfo {author}
  {\bibfnamefont {S.~B.}\ \bibnamefont {{Cenko}}}, \bibinfo {author}
  {\bibfnamefont {C.}~\bibnamefont {{Kouveliotou}}}, \bibinfo {author}
  {\bibfnamefont {J.~B.}\ \bibnamefont {{Gonz{\'a}lez}}}, \bibinfo {author}
  {\bibfnamefont {J.}~\bibnamefont {{Durbak}}}, \bibinfo {author}
  {\bibfnamefont {P.}~\bibnamefont {{Gatkine}}}, \bibinfo {author}
  {\bibfnamefont {A.}~\bibnamefont {{Kutyrev}}}, \bibinfo {author}
  {\bibfnamefont {T.}~\bibnamefont {{Sakamoto}}}, \bibinfo {author}
  {\bibfnamefont {R.}~\bibnamefont {{S{\'a}nchez-Ram{\'\i}rez}}},\ and\
  \bibinfo {author} {\bibfnamefont {S.}~\bibnamefont {{Veilleux}}},\ }\href
  {https://doi.org/10.1093/mnras/stac1982} {\bibfield  {journal} {\bibinfo
  {journal} {\mnras}\ }\textbf {\bibinfo {volume} {515}},\ \bibinfo {pages}
  {4890} (\bibinfo {year} {2022})},\ \Eprint {https://arxiv.org/abs/2204.09059}
  {arXiv:2204.09059 [astro-ph.HE]} \BibitemShut {NoStop}%
\bibitem [{\citenamefont {{Iwamoto}}\ \emph {et~al.}(1998)\citenamefont
  {{Iwamoto}}, \citenamefont {{Mazzali}}, \citenamefont {{Nomoto}},
  \citenamefont {{Umeda}}, \citenamefont {{Nakamura}}, \citenamefont {{Patat}},
  \citenamefont {{Danziger}}, \citenamefont {{Young}}, \citenamefont
  {{Suzuki}}, \citenamefont {{Shigeyama}}, \citenamefont {{Augusteijn}},
  \citenamefont {{Doublier}}, \citenamefont {{Gonzalez}}, \citenamefont
  {{Boehnhardt}}, \citenamefont {{Brewer}}, \citenamefont {{Hainaut}},
  \citenamefont {{Lidman}}, \citenamefont {{Leibundgut}}, \citenamefont
  {{Cappellaro}}, \citenamefont {{Turatto}}, \citenamefont {{Galama}},
  \citenamefont {{Vreeswijk}}, \citenamefont {{Kouveliotou}}, \citenamefont
  {{van Paradijs}}, \citenamefont {{Pian}}, \citenamefont {{Palazzi}},\ and\
  \citenamefont {{Frontera}}}]{1998Natur.395..672I}%
  \BibitemOpen
  \bibfield  {author} {\bibinfo {author} {\bibfnamefont {K.}~\bibnamefont
  {{Iwamoto}}}, \bibinfo {author} {\bibfnamefont {P.~A.}\ \bibnamefont
  {{Mazzali}}}, \bibinfo {author} {\bibfnamefont {K.}~\bibnamefont {{Nomoto}}},
  \bibinfo {author} {\bibfnamefont {H.}~\bibnamefont {{Umeda}}}, \bibinfo
  {author} {\bibfnamefont {T.}~\bibnamefont {{Nakamura}}}, \bibinfo {author}
  {\bibfnamefont {F.}~\bibnamefont {{Patat}}}, \bibinfo {author} {\bibfnamefont
  {I.~J.}\ \bibnamefont {{Danziger}}}, \bibinfo {author} {\bibfnamefont
  {T.~R.}\ \bibnamefont {{Young}}}, \bibinfo {author} {\bibfnamefont
  {T.}~\bibnamefont {{Suzuki}}}, \bibinfo {author} {\bibfnamefont
  {T.}~\bibnamefont {{Shigeyama}}}, \bibinfo {author} {\bibfnamefont
  {T.}~\bibnamefont {{Augusteijn}}}, \bibinfo {author} {\bibfnamefont
  {V.}~\bibnamefont {{Doublier}}}, \bibinfo {author} {\bibfnamefont {J.-F.}\
  \bibnamefont {{Gonzalez}}}, \bibinfo {author} {\bibfnamefont
  {H.}~\bibnamefont {{Boehnhardt}}}, \bibinfo {author} {\bibfnamefont
  {J.}~\bibnamefont {{Brewer}}}, \bibinfo {author} {\bibfnamefont {O.~R.}\
  \bibnamefont {{Hainaut}}}, \bibinfo {author} {\bibfnamefont {C.}~\bibnamefont
  {{Lidman}}}, \bibinfo {author} {\bibfnamefont {B.}~\bibnamefont
  {{Leibundgut}}}, \bibinfo {author} {\bibfnamefont {E.}~\bibnamefont
  {{Cappellaro}}}, \bibinfo {author} {\bibfnamefont {M.}~\bibnamefont
  {{Turatto}}}, \bibinfo {author} {\bibfnamefont {T.~J.}\ \bibnamefont
  {{Galama}}}, \bibinfo {author} {\bibfnamefont {P.~M.}\ \bibnamefont
  {{Vreeswijk}}}, \bibinfo {author} {\bibfnamefont {C.}~\bibnamefont
  {{Kouveliotou}}}, \bibinfo {author} {\bibfnamefont {J.}~\bibnamefont {{van
  Paradijs}}}, \bibinfo {author} {\bibfnamefont {E.}~\bibnamefont {{Pian}}},
  \bibinfo {author} {\bibfnamefont {E.}~\bibnamefont {{Palazzi}}},\ and\
  \bibinfo {author} {\bibfnamefont {F.}~\bibnamefont {{Frontera}}},\ }\href
  {https://doi.org/10.1038/27155} {\bibfield  {journal} {\bibinfo  {journal}
  {\nat}\ }\textbf {\bibinfo {volume} {395}},\ \bibinfo {pages} {672} (\bibinfo
  {year} {1998})},\ \Eprint {https://arxiv.org/abs/astro-ph/9806382}
  {astro-ph/9806382} \BibitemShut {NoStop}%
\bibitem [{\citenamefont {{Stanek}}\ \emph {et~al.}(2003)\citenamefont
  {{Stanek}}, \citenamefont {{Matheson}}, \citenamefont {{Garnavich}},
  \citenamefont {{Martini}}, \citenamefont {{Berlind}}, \citenamefont
  {{Caldwell}}, \citenamefont {{Challis}}, \citenamefont {{Brown}},
  \citenamefont {{Schild}}, \citenamefont {{Krisciunas}}, \citenamefont
  {{Calkins}}, \citenamefont {{Lee}}, \citenamefont {{Hathi}}, \citenamefont
  {{Jansen}}, \citenamefont {{Windhorst}}, \citenamefont {{Echevarria}},
  \citenamefont {{Eisenstein}}, \citenamefont {{Pindor}}, \citenamefont
  {{Olszewski}}, \citenamefont {{Harding}}, \citenamefont {{Holland}},\ and\
  \citenamefont {{Bersier}}}]{2003ApJ...591L..17S}%
  \BibitemOpen
  \bibfield  {author} {\bibinfo {author} {\bibfnamefont {K.~Z.}\ \bibnamefont
  {{Stanek}}}, \bibinfo {author} {\bibfnamefont {T.}~\bibnamefont
  {{Matheson}}}, \bibinfo {author} {\bibfnamefont {P.~M.}\ \bibnamefont
  {{Garnavich}}}, \bibinfo {author} {\bibfnamefont {P.}~\bibnamefont
  {{Martini}}}, \bibinfo {author} {\bibfnamefont {P.}~\bibnamefont
  {{Berlind}}}, \bibinfo {author} {\bibfnamefont {N.}~\bibnamefont
  {{Caldwell}}}, \bibinfo {author} {\bibfnamefont {P.}~\bibnamefont
  {{Challis}}}, \bibinfo {author} {\bibfnamefont {W.~R.}\ \bibnamefont
  {{Brown}}}, \bibinfo {author} {\bibfnamefont {R.}~\bibnamefont {{Schild}}},
  \bibinfo {author} {\bibfnamefont {K.}~\bibnamefont {{Krisciunas}}}, \bibinfo
  {author} {\bibfnamefont {M.~L.}\ \bibnamefont {{Calkins}}}, \bibinfo {author}
  {\bibfnamefont {J.~C.}\ \bibnamefont {{Lee}}}, \bibinfo {author}
  {\bibfnamefont {N.}~\bibnamefont {{Hathi}}}, \bibinfo {author} {\bibfnamefont
  {R.~A.}\ \bibnamefont {{Jansen}}}, \bibinfo {author} {\bibfnamefont
  {R.}~\bibnamefont {{Windhorst}}}, \bibinfo {author} {\bibfnamefont
  {L.}~\bibnamefont {{Echevarria}}}, \bibinfo {author} {\bibfnamefont {D.~J.}\
  \bibnamefont {{Eisenstein}}}, \bibinfo {author} {\bibfnamefont
  {B.}~\bibnamefont {{Pindor}}}, \bibinfo {author} {\bibfnamefont {E.~W.}\
  \bibnamefont {{Olszewski}}}, \bibinfo {author} {\bibfnamefont
  {P.}~\bibnamefont {{Harding}}}, \bibinfo {author} {\bibfnamefont {S.~T.}\
  \bibnamefont {{Holland}}},\ and\ \bibinfo {author} {\bibfnamefont
  {D.}~\bibnamefont {{Bersier}}},\ }\href {https://doi.org/10.1086/376976}
  {\bibfield  {journal} {\bibinfo  {journal} {\apjl}\ }\textbf {\bibinfo
  {volume} {591}},\ \bibinfo {pages} {L17} (\bibinfo {year} {2003})},\ \Eprint
  {https://arxiv.org/abs/astro-ph/0304173} {arXiv:astro-ph/0304173 [astro-ph]}
  \BibitemShut {NoStop}%
\bibitem [{\citenamefont {{MacFadyen}}\ and\ \citenamefont
  {{Woosley}}(1999)}]{1999ApJ...524..262M}%
  \BibitemOpen
  \bibfield  {author} {\bibinfo {author} {\bibfnamefont {A.~I.}\ \bibnamefont
  {{MacFadyen}}}\ and\ \bibinfo {author} {\bibfnamefont {S.~E.}\ \bibnamefont
  {{Woosley}}},\ }\href {https://doi.org/10.1086/307790} {\bibfield  {journal}
  {\bibinfo  {journal} {\apj}\ }\textbf {\bibinfo {volume} {524}},\ \bibinfo
  {pages} {262} (\bibinfo {year} {1999})},\ \Eprint
  {https://arxiv.org/abs/astro-ph/9810274} {astro-ph/9810274} \BibitemShut
  {NoStop}%
\bibitem [{\citenamefont {Shibata}\ \emph {et~al.}(2025)\citenamefont
  {Shibata}, \citenamefont {Fujibayashi}, \citenamefont {Wanajo}, \citenamefont
  {Ioka}, \citenamefont {Lam},\ and\ \citenamefont
  {Sekiguchi}}]{Shibata:2025gix}%
  \BibitemOpen
  \bibfield  {author} {\bibinfo {author} {\bibfnamefont {M.}~\bibnamefont
  {Shibata}}, \bibinfo {author} {\bibfnamefont {S.}~\bibnamefont
  {Fujibayashi}}, \bibinfo {author} {\bibfnamefont {S.}~\bibnamefont {Wanajo}},
  \bibinfo {author} {\bibfnamefont {K.}~\bibnamefont {Ioka}}, \bibinfo {author}
  {\bibfnamefont {A.~T.-L.}\ \bibnamefont {Lam}},\ and\ \bibinfo {author}
  {\bibfnamefont {Y.}~\bibnamefont {Sekiguchi}},\ }\href@noop {} {\bibfield
  {journal} {\bibinfo  {journal} {arXiv}\ } (\bibinfo {year} {2025})},\ \Eprint
  {https://arxiv.org/abs/2502.02077} {arXiv:2502.02077 [astro-ph.HE]}
  \BibitemShut {NoStop}%
\bibitem [{\citenamefont {Nakar}(2007)}]{Nakar:2007yr}%
  \BibitemOpen
  \bibfield  {author} {\bibinfo {author} {\bibfnamefont {E.}~\bibnamefont
  {Nakar}},\ }\href {https://doi.org/10.1016/j.physrep.2007.02.005} {\bibfield
  {journal} {\bibinfo  {journal} {Phys. Rept.}\ }\textbf {\bibinfo {volume}
  {442}},\ \bibinfo {pages} {166} (\bibinfo {year} {2007})},\ \Eprint
  {https://arxiv.org/abs/astro-ph/0701748} {arXiv:astro-ph/0701748}
  \BibitemShut {NoStop}%
\bibitem [{\citenamefont {Berger}(2014)}]{Berger:2013jza}%
  \BibitemOpen
  \bibfield  {author} {\bibinfo {author} {\bibfnamefont {E.}~\bibnamefont
  {Berger}},\ }\href {https://doi.org/10.1146/annurev-astro-081913-035926}
  {\bibfield  {journal} {\bibinfo  {journal} {Ann. Rev. Astron. Astrophys.}\
  }\textbf {\bibinfo {volume} {52}},\ \bibinfo {pages} {43} (\bibinfo {year}
  {2014})},\ \Eprint {https://arxiv.org/abs/1311.2603} {arXiv:1311.2603
  [astro-ph.HE]} \BibitemShut {NoStop}%
\bibitem [{\citenamefont {Gompertz}\ \emph {et~al.}(2023)\citenamefont
  {Gompertz} \emph {et~al.}}]{GRB211211AG}%
  \BibitemOpen
  \bibfield  {author} {\bibinfo {author} {\bibfnamefont {B.~P.}\ \bibnamefont
  {Gompertz}} \emph {et~al.},\ }\href
  {https://doi.org/10.1038/s41550-022-01819-4} {\bibfield  {journal} {\bibinfo
  {journal} {Nature Astron.}\ }\textbf {\bibinfo {volume} {7}},\ \bibinfo
  {pages} {67} (\bibinfo {year} {2023})},\ \Eprint
  {https://arxiv.org/abs/2205.05008} {arXiv:2205.05008 [astro-ph.HE]}
  \BibitemShut {NoStop}%
\bibitem [{\citenamefont {Rastinejad}\ \emph {et~al.}(2022)\citenamefont
  {Rastinejad} \emph {et~al.}}]{GRB211211AR}%
  \BibitemOpen
  \bibfield  {author} {\bibinfo {author} {\bibfnamefont {J.~C.}\ \bibnamefont
  {Rastinejad}} \emph {et~al.},\ }\href
  {https://doi.org/10.1038/s41586-022-05390-w} {\bibfield  {journal} {\bibinfo
  {journal} {Nature}\ }\textbf {\bibinfo {volume} {612}},\ \bibinfo {pages}
  {223} (\bibinfo {year} {2022})},\ \Eprint {https://arxiv.org/abs/2204.10864}
  {arXiv:2204.10864 [astro-ph.HE]} \BibitemShut {NoStop}%
\bibitem [{\citenamefont {Troja}\ \emph {et~al.}(2022)\citenamefont {Troja}
  \emph {et~al.}}]{GRB211211AT}%
  \BibitemOpen
  \bibfield  {author} {\bibinfo {author} {\bibfnamefont {E.}~\bibnamefont
  {Troja}} \emph {et~al.},\ }\href {https://doi.org/10.1038/s41586-022-05327-3}
  {\bibfield  {journal} {\bibinfo  {journal} {Nature}\ }\textbf {\bibinfo
  {volume} {612}},\ \bibinfo {pages} {228} (\bibinfo {year} {2022})},\ \Eprint
  {https://arxiv.org/abs/2209.03363} {arXiv:2209.03363 [astro-ph.HE]}
  \BibitemShut {NoStop}%
\bibitem [{\citenamefont {Levan}\ \emph {et~al.}(2024)\citenamefont {Levan}
  \emph {et~al.}}]{GRB230307AL}%
  \BibitemOpen
  \bibfield  {author} {\bibinfo {author} {\bibfnamefont {A.~J.}\ \bibnamefont
  {Levan}} \emph {et~al.} (\bibinfo {collaboration} {JWST}),\ }\href
  {https://doi.org/10.1038/s41586-023-06759-1} {\bibfield  {journal} {\bibinfo
  {journal} {Nature}\ }\textbf {\bibinfo {volume} {626}},\ \bibinfo {pages}
  {737} (\bibinfo {year} {2024})},\ \Eprint {https://arxiv.org/abs/2307.02098}
  {arXiv:2307.02098 [astro-ph.HE]} \BibitemShut {NoStop}%
\bibitem [{\citenamefont {Blandford}\ and\ \citenamefont
  {Payne}(1982)}]{Blandford:1982di}%
  \BibitemOpen
  \bibfield  {author} {\bibinfo {author} {\bibfnamefont {R.~D.}\ \bibnamefont
  {Blandford}}\ and\ \bibinfo {author} {\bibfnamefont {D.~G.}\ \bibnamefont
  {Payne}},\ }\href@noop {} {\bibfield  {journal} {\bibinfo  {journal} {Mon.
  Not. Roy. Astron. Soc.}\ }\textbf {\bibinfo {volume} {199}},\ \bibinfo
  {pages} {883} (\bibinfo {year} {1982})}\BibitemShut {NoStop}%
\bibitem [{\citenamefont {Komissarov}(2005)}]{Komissarov:2005wj}%
  \BibitemOpen
  \bibfield  {author} {\bibinfo {author} {\bibfnamefont {S.~S.}\ \bibnamefont
  {Komissarov}},\ }\href {https://doi.org/10.1111/j.1365-2966.2005.08974.x}
  {\bibfield  {journal} {\bibinfo  {journal} {Mon. Not. Roy. Astron. Soc.}\
  }\textbf {\bibinfo {volume} {359}},\ \bibinfo {pages} {801} (\bibinfo {year}
  {2005})},\ \Eprint {https://arxiv.org/abs/astro-ph/0501599}
  {arXiv:astro-ph/0501599} \BibitemShut {NoStop}%
\bibitem [{\citenamefont {McKinney}(2006)}]{McKinney:2006tf}%
  \BibitemOpen
  \bibfield  {author} {\bibinfo {author} {\bibfnamefont {J.~C.}\ \bibnamefont
  {McKinney}},\ }\href {https://doi.org/10.1111/j.1365-2966.2006.10256.x}
  {\bibfield  {journal} {\bibinfo  {journal} {Mon. Not. Roy. Astron. Soc.}\
  }\textbf {\bibinfo {volume} {368}},\ \bibinfo {pages} {1561} (\bibinfo {year}
  {2006})},\ \Eprint {https://arxiv.org/abs/astro-ph/0603045}
  {arXiv:astro-ph/0603045} \BibitemShut {NoStop}%
\bibitem [{\citenamefont {Barkov}\ and\ \citenamefont
  {Komissarov}(2008)}]{Barkov:2007us}%
  \BibitemOpen
  \bibfield  {author} {\bibinfo {author} {\bibfnamefont {M.~V.}\ \bibnamefont
  {Barkov}}\ and\ \bibinfo {author} {\bibfnamefont {S.~S.}\ \bibnamefont
  {Komissarov}},\ }\href {https://doi.org/10.1111/j.1745-3933.2008.00427.x}
  {\bibfield  {journal} {\bibinfo  {journal} {Mon. Not. Roy. Astron. Soc.}\
  }\textbf {\bibinfo {volume} {385}},\ \bibinfo {pages} {28} (\bibinfo {year}
  {2008})},\ \Eprint {https://arxiv.org/abs/0710.2654} {arXiv:0710.2654
  [astro-ph]} \BibitemShut {NoStop}%
\bibitem [{\citenamefont {Komissarov}\ and\ \citenamefont
  {Barkov}(2009)}]{Komissarov_2009}%
  \BibitemOpen
  \bibfield  {author} {\bibinfo {author} {\bibfnamefont {S.~S.}\ \bibnamefont
  {Komissarov}}\ and\ \bibinfo {author} {\bibfnamefont {M.~V.}\ \bibnamefont
  {Barkov}},\ }\href {https://doi.org/10.1111/j.1365-2966.2009.14831.x}
  {\bibfield  {journal} {\bibinfo  {journal} {Monthly Notices of the Royal
  Astronomical Society}\ }\textbf {\bibinfo {volume} {397}},\ \bibinfo {pages}
  {1153–1168} (\bibinfo {year} {2009})}\BibitemShut {NoStop}%
\bibitem [{\citenamefont {Tchekhovskoy}\ \emph {et~al.}(2011)\citenamefont
  {Tchekhovskoy}, \citenamefont {Narayan},\ and\ \citenamefont
  {McKinney}}]{Tchekhovskoy:2011zx}%
  \BibitemOpen
  \bibfield  {author} {\bibinfo {author} {\bibfnamefont {A.}~\bibnamefont
  {Tchekhovskoy}}, \bibinfo {author} {\bibfnamefont {R.}~\bibnamefont
  {Narayan}},\ and\ \bibinfo {author} {\bibfnamefont {J.~C.}\ \bibnamefont
  {McKinney}},\ }\href {https://doi.org/10.1111/j.1745-3933.2011.01147.x}
  {\bibfield  {journal} {\bibinfo  {journal} {Mon. Not. Roy. Astron. Soc.}\
  }\textbf {\bibinfo {volume} {418}},\ \bibinfo {pages} {L79} (\bibinfo {year}
  {2011})},\ \Eprint {https://arxiv.org/abs/1108.0412} {arXiv:1108.0412
  [astro-ph.HE]} \BibitemShut {NoStop}%
\bibitem [{\citenamefont {Bromberg}\ and\ \citenamefont
  {Tchekhovskoy}(2016)}]{Bromberg:2015wra}%
  \BibitemOpen
  \bibfield  {author} {\bibinfo {author} {\bibfnamefont {O.}~\bibnamefont
  {Bromberg}}\ and\ \bibinfo {author} {\bibfnamefont {A.}~\bibnamefont
  {Tchekhovskoy}},\ }\href {https://doi.org/10.1093/mnras/stv2591} {\bibfield
  {journal} {\bibinfo  {journal} {Mon. Not. Roy. Astron. Soc.}\ }\textbf
  {\bibinfo {volume} {456}},\ \bibinfo {pages} {1739} (\bibinfo {year}
  {2016})},\ \Eprint {https://arxiv.org/abs/1508.02721} {arXiv:1508.02721
  [astro-ph.HE]} \BibitemShut {NoStop}%
\bibitem [{\citenamefont {Gottlieb}\ \emph
  {et~al.}(2022{\natexlab{a}})\citenamefont {Gottlieb}, \citenamefont
  {Lalakos}, \citenamefont {Bromberg}, \citenamefont {Liska},\ and\
  \citenamefont {Tchekhovskoy}}]{Gottlieb:2021srg}%
  \BibitemOpen
  \bibfield  {author} {\bibinfo {author} {\bibfnamefont {O.}~\bibnamefont
  {Gottlieb}}, \bibinfo {author} {\bibfnamefont {A.}~\bibnamefont {Lalakos}},
  \bibinfo {author} {\bibfnamefont {O.}~\bibnamefont {Bromberg}}, \bibinfo
  {author} {\bibfnamefont {M.}~\bibnamefont {Liska}},\ and\ \bibinfo {author}
  {\bibfnamefont {A.}~\bibnamefont {Tchekhovskoy}},\ }\href
  {https://doi.org/10.1093/mnras/stab3784} {\bibfield  {journal} {\bibinfo
  {journal} {Mon. Not. Roy. Astron. Soc.}\ }\textbf {\bibinfo {volume} {510}},\
  \bibinfo {pages} {4962} (\bibinfo {year} {2022}{\natexlab{a}})},\ \Eprint
  {https://arxiv.org/abs/2109.14619} {arXiv:2109.14619 [astro-ph.HE]}
  \BibitemShut {NoStop}%
\bibitem [{\citenamefont {Hayashi}\ \emph {et~al.}(2024)\citenamefont
  {Hayashi}, \citenamefont {Kiuchi}, \citenamefont {Kyutoku}, \citenamefont
  {Sekiguchi},\ and\ \citenamefont {Shibata}}]{Hayashi:2024jwt}%
  \BibitemOpen
  \bibfield  {author} {\bibinfo {author} {\bibfnamefont {K.}~\bibnamefont
  {Hayashi}}, \bibinfo {author} {\bibfnamefont {K.}~\bibnamefont {Kiuchi}},
  \bibinfo {author} {\bibfnamefont {K.}~\bibnamefont {Kyutoku}}, \bibinfo
  {author} {\bibfnamefont {Y.}~\bibnamefont {Sekiguchi}},\ and\ \bibinfo
  {author} {\bibfnamefont {M.}~\bibnamefont {Shibata}},\ }\href@noop {} {\
  (\bibinfo {year} {2024})},\ \Eprint {https://arxiv.org/abs/2410.10958}
  {arXiv:2410.10958 [astro-ph.HE]} \BibitemShut {NoStop}%
\bibitem [{\citenamefont {{Metzger}}\ \emph {et~al.}(2011)\citenamefont
  {{Metzger}}, \citenamefont {{Giannios}}, \citenamefont {{Thompson}},
  \citenamefont {{Bucciantini}},\ and\ \citenamefont
  {{Quataert}}}]{Metzger:2011}%
  \BibitemOpen
  \bibfield  {author} {\bibinfo {author} {\bibfnamefont {B.~D.}\ \bibnamefont
  {{Metzger}}}, \bibinfo {author} {\bibfnamefont {D.}~\bibnamefont
  {{Giannios}}}, \bibinfo {author} {\bibfnamefont {T.~A.}\ \bibnamefont
  {{Thompson}}}, \bibinfo {author} {\bibfnamefont {N.}~\bibnamefont
  {{Bucciantini}}},\ and\ \bibinfo {author} {\bibfnamefont {E.}~\bibnamefont
  {{Quataert}}},\ }\href {https://doi.org/10.1111/j.1365-2966.2011.18280.x}
  {\bibfield  {journal} {\bibinfo  {journal} {\mnras}\ }\textbf {\bibinfo
  {volume} {413}},\ \bibinfo {pages} {2031} (\bibinfo {year} {2011})},\ \Eprint
  {https://arxiv.org/abs/1012.0001} {arXiv:1012.0001 [astro-ph.HE]}
  \BibitemShut {NoStop}%
\bibitem [{\citenamefont {M\"osta}\ \emph {et~al.}(2020)\citenamefont
  {M\"osta}, \citenamefont {Radice}, \citenamefont {Haas}, \citenamefont
  {Schnetter},\ and\ \citenamefont {Bernuzzi}}]{Mosta:2020hlh}%
  \BibitemOpen
  \bibfield  {author} {\bibinfo {author} {\bibfnamefont {P.}~\bibnamefont
  {M\"osta}}, \bibinfo {author} {\bibfnamefont {D.}~\bibnamefont {Radice}},
  \bibinfo {author} {\bibfnamefont {R.}~\bibnamefont {Haas}}, \bibinfo {author}
  {\bibfnamefont {E.}~\bibnamefont {Schnetter}},\ and\ \bibinfo {author}
  {\bibfnamefont {S.}~\bibnamefont {Bernuzzi}},\ }\href
  {https://doi.org/10.3847/2041-8213/abb6ef} {\bibfield  {journal} {\bibinfo
  {journal} {Astrophys. J. Lett.}\ }\textbf {\bibinfo {volume} {901}},\
  \bibinfo {pages} {L37} (\bibinfo {year} {2020})},\ \Eprint
  {https://arxiv.org/abs/2003.06043} {arXiv:2003.06043 [astro-ph.HE]}
  \BibitemShut {NoStop}%
\bibitem [{\citenamefont {Most}\ and\ \citenamefont
  {Quataert}(2023)}]{Most:2023sft}%
  \BibitemOpen
  \bibfield  {author} {\bibinfo {author} {\bibfnamefont {E.~R.}\ \bibnamefont
  {Most}}\ and\ \bibinfo {author} {\bibfnamefont {E.}~\bibnamefont
  {Quataert}},\ }\href {https://doi.org/10.3847/2041-8213/acca84} {\bibfield
  {journal} {\bibinfo  {journal} {Astrophys. J. Lett.}\ }\textbf {\bibinfo
  {volume} {947}},\ \bibinfo {pages} {L15} (\bibinfo {year} {2023})},\ \Eprint
  {https://arxiv.org/abs/2303.08062} {arXiv:2303.08062 [astro-ph.HE]}
  \BibitemShut {NoStop}%
\bibitem [{\citenamefont {Kiuchi}\ \emph {et~al.}(2024)\citenamefont {Kiuchi},
  \citenamefont {Reboul-Salze}, \citenamefont {Shibata},\ and\ \citenamefont
  {Sekiguchi}}]{Kiuchi:2023obe}%
  \BibitemOpen
  \bibfield  {author} {\bibinfo {author} {\bibfnamefont {K.}~\bibnamefont
  {Kiuchi}}, \bibinfo {author} {\bibfnamefont {A.}~\bibnamefont
  {Reboul-Salze}}, \bibinfo {author} {\bibfnamefont {M.}~\bibnamefont
  {Shibata}},\ and\ \bibinfo {author} {\bibfnamefont {Y.}~\bibnamefont
  {Sekiguchi}},\ }\href {https://doi.org/10.1038/s41550-024-02194-y} {\bibfield
   {journal} {\bibinfo  {journal} {Nature Astron.}\ }\textbf {\bibinfo {volume}
  {8}},\ \bibinfo {pages} {298} (\bibinfo {year} {2024})},\ \Eprint
  {https://arxiv.org/abs/2306.15721} {arXiv:2306.15721 [astro-ph.HE]}
  \BibitemShut {NoStop}%
\bibitem [{\citenamefont {{Meszaros}}\ and\ \citenamefont
  {{Rees}}(1993)}]{1993ApJ...405..278M}%
  \BibitemOpen
  \bibfield  {author} {\bibinfo {author} {\bibfnamefont {P.}~\bibnamefont
  {{Meszaros}}}\ and\ \bibinfo {author} {\bibfnamefont {M.~J.}\ \bibnamefont
  {{Rees}}},\ }\href {https://doi.org/10.1086/172360} {\bibfield  {journal}
  {\bibinfo  {journal} {\apj}\ }\textbf {\bibinfo {volume} {405}},\ \bibinfo
  {pages} {278} (\bibinfo {year} {1993})}\BibitemShut {NoStop}%
\bibitem [{\citenamefont {{Jaroszynski}}(1996)}]{1996A&A...305..839J}%
  \BibitemOpen
  \bibfield  {author} {\bibinfo {author} {\bibfnamefont {M.}~\bibnamefont
  {{Jaroszynski}}},\ }\href@noop {} {\bibfield  {journal} {\bibinfo  {journal}
  {\aap}\ }\textbf {\bibinfo {volume} {305}},\ \bibinfo {pages} {839} (\bibinfo
  {year} {1996})},\ \Eprint {https://arxiv.org/abs/astro-ph/9506062}
  {arXiv:astro-ph/9506062 [astro-ph]} \BibitemShut {NoStop}%
\bibitem [{\citenamefont {{Popham}}\ \emph {et~al.}(1999)\citenamefont
  {{Popham}}, \citenamefont {{Woosley}},\ and\ \citenamefont
  {{Fryer}}}]{1999ApJ...518..356P}%
  \BibitemOpen
  \bibfield  {author} {\bibinfo {author} {\bibfnamefont {R.}~\bibnamefont
  {{Popham}}}, \bibinfo {author} {\bibfnamefont {S.~E.}\ \bibnamefont
  {{Woosley}}},\ and\ \bibinfo {author} {\bibfnamefont {C.}~\bibnamefont
  {{Fryer}}},\ }\href {https://doi.org/10.1086/307259} {\bibfield  {journal}
  {\bibinfo  {journal} {\apj}\ }\textbf {\bibinfo {volume} {518}},\ \bibinfo
  {pages} {356} (\bibinfo {year} {1999})},\ \Eprint
  {https://arxiv.org/abs/astro-ph/9807028} {arXiv:astro-ph/9807028 [astro-ph]}
  \BibitemShut {NoStop}%
\bibitem [{\citenamefont {Shibata}\ and\ \citenamefont
  {Hotokezaka}(2019)}]{Shibata:2019wef}%
  \BibitemOpen
  \bibfield  {author} {\bibinfo {author} {\bibfnamefont {M.}~\bibnamefont
  {Shibata}}\ and\ \bibinfo {author} {\bibfnamefont {K.}~\bibnamefont
  {Hotokezaka}},\ }\href {https://doi.org/10.1146/annurev-nucl-101918-023625}
  {\bibfield  {journal} {\bibinfo  {journal} {Ann. Rev. Nucl. Part. Sci.}\
  }\textbf {\bibinfo {volume} {69}},\ \bibinfo {pages} {41} (\bibinfo {year}
  {2019})},\ \Eprint {https://arxiv.org/abs/1908.02350} {arXiv:1908.02350
  [astro-ph.HE]} \BibitemShut {NoStop}%
\bibitem [{\citenamefont {Kyutoku}\ \emph {et~al.}(2021)\citenamefont
  {Kyutoku}, \citenamefont {Shibata},\ and\ \citenamefont
  {Taniguchi}}]{Kyutoku:2021icp}%
  \BibitemOpen
  \bibfield  {author} {\bibinfo {author} {\bibfnamefont {K.}~\bibnamefont
  {Kyutoku}}, \bibinfo {author} {\bibfnamefont {M.}~\bibnamefont {Shibata}},\
  and\ \bibinfo {author} {\bibfnamefont {K.}~\bibnamefont {Taniguchi}},\ }\href
  {https://doi.org/10.1007/s41114-021-00033-4} {\bibfield  {journal} {\bibinfo
  {journal} {Living Rev. Rel.}\ }\textbf {\bibinfo {volume} {24}},\ \bibinfo
  {pages} {5} (\bibinfo {year} {2021})},\ \Eprint
  {https://arxiv.org/abs/2110.06218} {arXiv:2110.06218 [astro-ph.HE]}
  \BibitemShut {NoStop}%
\bibitem [{\citenamefont {Shibata}\ \emph
  {et~al.}(2017{\natexlab{a}})\citenamefont {Shibata}, \citenamefont
  {Fujibayashi}, \citenamefont {Hotokezaka}, \citenamefont {Kiuchi},
  \citenamefont {Kyutoku}, \citenamefont {Sekiguchi},\ and\ \citenamefont
  {Tanaka}}]{Shibata:2017xdx}%
  \BibitemOpen
  \bibfield  {author} {\bibinfo {author} {\bibfnamefont {M.}~\bibnamefont
  {Shibata}}, \bibinfo {author} {\bibfnamefont {S.}~\bibnamefont
  {Fujibayashi}}, \bibinfo {author} {\bibfnamefont {K.}~\bibnamefont
  {Hotokezaka}}, \bibinfo {author} {\bibfnamefont {K.}~\bibnamefont {Kiuchi}},
  \bibinfo {author} {\bibfnamefont {K.}~\bibnamefont {Kyutoku}}, \bibinfo
  {author} {\bibfnamefont {Y.}~\bibnamefont {Sekiguchi}},\ and\ \bibinfo
  {author} {\bibfnamefont {M.}~\bibnamefont {Tanaka}},\ }\href
  {https://doi.org/10.1103/PhysRevD.96.123012} {\bibfield  {journal} {\bibinfo
  {journal} {Phys. Rev.}\ }\textbf {\bibinfo {volume} {D96}},\ \bibinfo {pages}
  {123012} (\bibinfo {year} {2017}{\natexlab{a}})},\ \Eprint
  {https://arxiv.org/abs/1710.07579} {arXiv:1710.07579 [astro-ph.HE]}
  \BibitemShut {NoStop}%
\bibitem [{\citenamefont {Fujibayashi}\ \emph {et~al.}(2018)\citenamefont
  {Fujibayashi}, \citenamefont {Kiuchi}, \citenamefont {Nishimura},
  \citenamefont {Sekiguchi},\ and\ \citenamefont
  {Shibata}}]{Fujibayashi:2017puw}%
  \BibitemOpen
  \bibfield  {author} {\bibinfo {author} {\bibfnamefont {S.}~\bibnamefont
  {Fujibayashi}}, \bibinfo {author} {\bibfnamefont {K.}~\bibnamefont {Kiuchi}},
  \bibinfo {author} {\bibfnamefont {N.}~\bibnamefont {Nishimura}}, \bibinfo
  {author} {\bibfnamefont {Y.}~\bibnamefont {Sekiguchi}},\ and\ \bibinfo
  {author} {\bibfnamefont {M.}~\bibnamefont {Shibata}},\ }\href
  {https://doi.org/10.3847/1538-4357/aabafd} {\bibfield  {journal} {\bibinfo
  {journal} {Astrophys. J.}\ }\textbf {\bibinfo {volume} {860}},\ \bibinfo
  {pages} {64} (\bibinfo {year} {2018})},\ \Eprint
  {https://arxiv.org/abs/1711.02093} {arXiv:1711.02093 [astro-ph.HE]}
  \BibitemShut {NoStop}%
\bibitem [{\citenamefont {Fujibayashi}\ \emph
  {et~al.}(2020{\natexlab{a}})\citenamefont {Fujibayashi}, \citenamefont
  {Shibata}, \citenamefont {Wanajo}, \citenamefont {Kiuchi}, \citenamefont
  {Kyutoku},\ and\ \citenamefont {Sekiguchi}}]{Fujibayashi:2020qda}%
  \BibitemOpen
  \bibfield  {author} {\bibinfo {author} {\bibfnamefont {S.}~\bibnamefont
  {Fujibayashi}}, \bibinfo {author} {\bibfnamefont {M.}~\bibnamefont
  {Shibata}}, \bibinfo {author} {\bibfnamefont {S.}~\bibnamefont {Wanajo}},
  \bibinfo {author} {\bibfnamefont {K.}~\bibnamefont {Kiuchi}}, \bibinfo
  {author} {\bibfnamefont {K.}~\bibnamefont {Kyutoku}},\ and\ \bibinfo {author}
  {\bibfnamefont {Y.}~\bibnamefont {Sekiguchi}},\ }\href
  {https://doi.org/10.1103/PhysRevD.101.083029} {\bibfield  {journal} {\bibinfo
   {journal} {Phys. Rev. D}\ }\textbf {\bibinfo {volume} {101}},\ \bibinfo
  {pages} {083029} (\bibinfo {year} {2020}{\natexlab{a}})},\ \Eprint
  {https://arxiv.org/abs/2001.04467} {arXiv:2001.04467 [astro-ph.HE]}
  \BibitemShut {NoStop}%
\bibitem [{\citenamefont {Hayashi}\ \emph {et~al.}(2022)\citenamefont
  {Hayashi}, \citenamefont {Fujibayashi}, \citenamefont {Kiuchi}, \citenamefont
  {Kyutoku}, \citenamefont {Sekiguchi},\ and\ \citenamefont
  {Shibata}}]{Hayashi:2021oxy}%
  \BibitemOpen
  \bibfield  {author} {\bibinfo {author} {\bibfnamefont {K.}~\bibnamefont
  {Hayashi}}, \bibinfo {author} {\bibfnamefont {S.}~\bibnamefont
  {Fujibayashi}}, \bibinfo {author} {\bibfnamefont {K.}~\bibnamefont {Kiuchi}},
  \bibinfo {author} {\bibfnamefont {K.}~\bibnamefont {Kyutoku}}, \bibinfo
  {author} {\bibfnamefont {Y.}~\bibnamefont {Sekiguchi}},\ and\ \bibinfo
  {author} {\bibfnamefont {M.}~\bibnamefont {Shibata}},\ }\href
  {https://doi.org/10.1103/PhysRevD.106.023008} {\bibfield  {journal} {\bibinfo
   {journal} {Phys. Rev. D}\ }\textbf {\bibinfo {volume} {106}},\ \bibinfo
  {pages} {023008} (\bibinfo {year} {2022})},\ \Eprint
  {https://arxiv.org/abs/2111.04621} {arXiv:2111.04621 [astro-ph.HE]}
  \BibitemShut {NoStop}%
\bibitem [{\citenamefont {Hayashi}\ \emph {et~al.}(2023)\citenamefont
  {Hayashi}, \citenamefont {Kiuchi}, \citenamefont {Kyutoku}, \citenamefont
  {Sekiguchi},\ and\ \citenamefont {Shibata}}]{Hayashi:2022cdq}%
  \BibitemOpen
  \bibfield  {author} {\bibinfo {author} {\bibfnamefont {K.}~\bibnamefont
  {Hayashi}}, \bibinfo {author} {\bibfnamefont {K.}~\bibnamefont {Kiuchi}},
  \bibinfo {author} {\bibfnamefont {K.}~\bibnamefont {Kyutoku}}, \bibinfo
  {author} {\bibfnamefont {Y.}~\bibnamefont {Sekiguchi}},\ and\ \bibinfo
  {author} {\bibfnamefont {M.}~\bibnamefont {Shibata}},\ }\href
  {https://doi.org/10.1103/PhysRevD.107.123001} {\bibfield  {journal} {\bibinfo
   {journal} {Phys. Rev. D}\ }\textbf {\bibinfo {volume} {107}},\ \bibinfo
  {pages} {123001} (\bibinfo {year} {2023})},\ \Eprint
  {https://arxiv.org/abs/2211.07158} {arXiv:2211.07158 [astro-ph.HE]}
  \BibitemShut {NoStop}%
\bibitem [{\citenamefont {Kiuchi}\ \emph {et~al.}(2023)\citenamefont {Kiuchi},
  \citenamefont {Fujibayashi}, \citenamefont {Hayashi}, \citenamefont
  {Kyutoku}, \citenamefont {Sekiguchi},\ and\ \citenamefont
  {Shibata}}]{Kiuchi:2022nin}%
  \BibitemOpen
  \bibfield  {author} {\bibinfo {author} {\bibfnamefont {K.}~\bibnamefont
  {Kiuchi}}, \bibinfo {author} {\bibfnamefont {S.}~\bibnamefont {Fujibayashi}},
  \bibinfo {author} {\bibfnamefont {K.}~\bibnamefont {Hayashi}}, \bibinfo
  {author} {\bibfnamefont {K.}~\bibnamefont {Kyutoku}}, \bibinfo {author}
  {\bibfnamefont {Y.}~\bibnamefont {Sekiguchi}},\ and\ \bibinfo {author}
  {\bibfnamefont {M.}~\bibnamefont {Shibata}},\ }\href
  {https://doi.org/10.1103/PhysRevLett.131.011401} {\bibfield  {journal}
  {\bibinfo  {journal} {Phys. Rev. Lett.}\ }\textbf {\bibinfo {volume} {131}},\
  \bibinfo {pages} {011401} (\bibinfo {year} {2023})},\ \Eprint
  {https://arxiv.org/abs/2211.07637} {arXiv:2211.07637 [astro-ph.HE]}
  \BibitemShut {NoStop}%
\bibitem [{\citenamefont {Just}\ \emph {et~al.}(2023)\citenamefont {Just},
  \citenamefont {Vijayan}, \citenamefont {Xiong}, \citenamefont {Goriely},
  \citenamefont {Soultanis}, \citenamefont {Bauswein}, \citenamefont {Guilet},
  \citenamefont {Janka}, \citenamefont {Janka},\ and\ \citenamefont
  {Mart\'\i{}nez-Pinedo}}]{Just:2023wtj}%
  \BibitemOpen
  \bibfield  {author} {\bibinfo {author} {\bibfnamefont {O.}~\bibnamefont
  {Just}}, \bibinfo {author} {\bibfnamefont {V.}~\bibnamefont {Vijayan}},
  \bibinfo {author} {\bibfnamefont {Z.}~\bibnamefont {Xiong}}, \bibinfo
  {author} {\bibfnamefont {S.}~\bibnamefont {Goriely}}, \bibinfo {author}
  {\bibfnamefont {T.}~\bibnamefont {Soultanis}}, \bibinfo {author}
  {\bibfnamefont {A.}~\bibnamefont {Bauswein}}, \bibinfo {author}
  {\bibfnamefont {J.}~\bibnamefont {Guilet}}, \bibinfo {author} {\bibfnamefont
  {H.~T.}\ \bibnamefont {Janka}}, \bibinfo {author} {\bibfnamefont {H.-T.}\
  \bibnamefont {Janka}},\ and\ \bibinfo {author} {\bibfnamefont
  {G.}~\bibnamefont {Mart\'\i{}nez-Pinedo}},\ }\href
  {https://doi.org/10.3847/2041-8213/acdad2} {\bibfield  {journal} {\bibinfo
  {journal} {Astrophys. J. Lett.}\ }\textbf {\bibinfo {volume} {951}},\
  \bibinfo {pages} {L12} (\bibinfo {year} {2023})},\ \Eprint
  {https://arxiv.org/abs/2302.10928} {arXiv:2302.10928 [astro-ph.HE]}
  \BibitemShut {NoStop}%
\bibitem [{\citenamefont {Shibata}\ \emph
  {et~al.}(2021{\natexlab{a}})\citenamefont {Shibata}, \citenamefont
  {Fujibayashi},\ and\ \citenamefont {Sekiguchi}}]{Shibata:2021xmo}%
  \BibitemOpen
  \bibfield  {author} {\bibinfo {author} {\bibfnamefont {M.}~\bibnamefont
  {Shibata}}, \bibinfo {author} {\bibfnamefont {S.}~\bibnamefont
  {Fujibayashi}},\ and\ \bibinfo {author} {\bibfnamefont {Y.}~\bibnamefont
  {Sekiguchi}},\ }\href {https://doi.org/10.1103/PhysRevD.104.063026}
  {\bibfield  {journal} {\bibinfo  {journal} {Phys. Rev. D}\ }\textbf {\bibinfo
  {volume} {104}},\ \bibinfo {pages} {063026} (\bibinfo {year}
  {2021}{\natexlab{a}})},\ \Eprint {https://arxiv.org/abs/2109.08732}
  {arXiv:2109.08732 [astro-ph.HE]} \BibitemShut {NoStop}%
\bibitem [{\citenamefont {Shibata}\ \emph
  {et~al.}(2021{\natexlab{b}})\citenamefont {Shibata}, \citenamefont
  {Fujibayashi},\ and\ \citenamefont {Sekiguchi}}]{Shibata:2021bbj}%
  \BibitemOpen
  \bibfield  {author} {\bibinfo {author} {\bibfnamefont {M.}~\bibnamefont
  {Shibata}}, \bibinfo {author} {\bibfnamefont {S.}~\bibnamefont
  {Fujibayashi}},\ and\ \bibinfo {author} {\bibfnamefont {Y.}~\bibnamefont
  {Sekiguchi}},\ }\href {https://doi.org/10.1103/PhysRevD.103.043022}
  {\bibfield  {journal} {\bibinfo  {journal} {Phys. Rev. D}\ }\textbf {\bibinfo
  {volume} {103}},\ \bibinfo {pages} {043022} (\bibinfo {year}
  {2021}{\natexlab{b}})},\ \Eprint {https://arxiv.org/abs/2102.01346}
  {arXiv:2102.01346 [astro-ph.HE]} \BibitemShut {NoStop}%
\bibitem [{\citenamefont {Muller}\ \emph {et~al.}(2010)\citenamefont {Muller},
  \citenamefont {Janka},\ and\ \citenamefont {Dimmelmeier}}]{Muller:2010ymw}%
  \BibitemOpen
  \bibfield  {author} {\bibinfo {author} {\bibfnamefont {B.}~\bibnamefont
  {Muller}}, \bibinfo {author} {\bibfnamefont {H.-T.}\ \bibnamefont {Janka}},\
  and\ \bibinfo {author} {\bibfnamefont {H.}~\bibnamefont {Dimmelmeier}},\
  }\href {https://doi.org/10.1088/0067-0049/189/1/104} {\bibfield  {journal}
  {\bibinfo  {journal} {Astrophys. J. Suppl.}\ }\textbf {\bibinfo {volume}
  {189}},\ \bibinfo {pages} {104} (\bibinfo {year} {2010})},\ \Eprint
  {https://arxiv.org/abs/1001.4841} {arXiv:1001.4841 [astro-ph.SR]}
  \BibitemShut {NoStop}%
\bibitem [{\citenamefont {Fujibayashi}\ \emph {et~al.}(2021)\citenamefont
  {Fujibayashi}, \citenamefont {Takahashi}, \citenamefont {Sekiguchi},\ and\
  \citenamefont {Shibata}}]{Fujibayashi:2021wvv}%
  \BibitemOpen
  \bibfield  {author} {\bibinfo {author} {\bibfnamefont {S.}~\bibnamefont
  {Fujibayashi}}, \bibinfo {author} {\bibfnamefont {K.}~\bibnamefont
  {Takahashi}}, \bibinfo {author} {\bibfnamefont {Y.}~\bibnamefont
  {Sekiguchi}},\ and\ \bibinfo {author} {\bibfnamefont {M.}~\bibnamefont
  {Shibata}},\ }\href {https://doi.org/10.3847/1538-4357/ac10cb} {\bibfield
  {journal} {\bibinfo  {journal} {Astrophys. J.}\ }\textbf {\bibinfo {volume}
  {919}},\ \bibinfo {pages} {80} (\bibinfo {year} {2021})},\ \Eprint
  {https://arxiv.org/abs/2102.04467} {arXiv:2102.04467 [astro-ph.HE]}
  \BibitemShut {NoStop}%
\bibitem [{\citenamefont {Fujibayashi}\ \emph {et~al.}(2023)\citenamefont
  {Fujibayashi}, \citenamefont {Sekiguchi}, \citenamefont {Shibata},\ and\
  \citenamefont {Wanajo}}]{Fujibayashi:2022xsm}%
  \BibitemOpen
  \bibfield  {author} {\bibinfo {author} {\bibfnamefont {S.}~\bibnamefont
  {Fujibayashi}}, \bibinfo {author} {\bibfnamefont {Y.}~\bibnamefont
  {Sekiguchi}}, \bibinfo {author} {\bibfnamefont {M.}~\bibnamefont {Shibata}},\
  and\ \bibinfo {author} {\bibfnamefont {S.}~\bibnamefont {Wanajo}},\ }\href
  {https://doi.org/10.3847/1538-4357/acf5e5} {\bibfield  {journal} {\bibinfo
  {journal} {Astrophys. J.}\ }\textbf {\bibinfo {volume} {956}},\ \bibinfo
  {pages} {100} (\bibinfo {year} {2023})},\ \Eprint
  {https://arxiv.org/abs/2212.03958} {arXiv:2212.03958 [astro-ph.HE]}
  \BibitemShut {NoStop}%
\bibitem [{\citenamefont {Fujibayashi}\ \emph {et~al.}(2024)\citenamefont
  {Fujibayashi}, \citenamefont {Lam}, \citenamefont {Shibata},\ and\
  \citenamefont {Sekiguchi}}]{Fujibayashi:2023oyt}%
  \BibitemOpen
  \bibfield  {author} {\bibinfo {author} {\bibfnamefont {S.}~\bibnamefont
  {Fujibayashi}}, \bibinfo {author} {\bibfnamefont {A.~T.-L.}\ \bibnamefont
  {Lam}}, \bibinfo {author} {\bibfnamefont {M.}~\bibnamefont {Shibata}},\ and\
  \bibinfo {author} {\bibfnamefont {Y.}~\bibnamefont {Sekiguchi}},\ }\href
  {https://doi.org/10.1103/PhysRevD.109.023031} {\bibfield  {journal} {\bibinfo
   {journal} {Phys. Rev. D}\ }\textbf {\bibinfo {volume} {109}},\ \bibinfo
  {pages} {023031} (\bibinfo {year} {2024})},\ \Eprint
  {https://arxiv.org/abs/2309.02161} {arXiv:2309.02161 [astro-ph.HE]}
  \BibitemShut {NoStop}%
\bibitem [{\citenamefont {Shibata}\ \emph {et~al.}(2024)\citenamefont
  {Shibata}, \citenamefont {Fujibayashi}, \citenamefont {Lam}, \citenamefont
  {Ioka},\ and\ \citenamefont {Sekiguchi}}]{Shibata:2023tho}%
  \BibitemOpen
  \bibfield  {author} {\bibinfo {author} {\bibfnamefont {M.}~\bibnamefont
  {Shibata}}, \bibinfo {author} {\bibfnamefont {S.}~\bibnamefont
  {Fujibayashi}}, \bibinfo {author} {\bibfnamefont {A.~T.-L.}\ \bibnamefont
  {Lam}}, \bibinfo {author} {\bibfnamefont {K.}~\bibnamefont {Ioka}},\ and\
  \bibinfo {author} {\bibfnamefont {Y.}~\bibnamefont {Sekiguchi}},\ }\href
  {https://doi.org/10.1103/PhysRevD.109.043051} {\bibfield  {journal} {\bibinfo
   {journal} {Phys. Rev. D}\ }\textbf {\bibinfo {volume} {109}},\ \bibinfo
  {pages} {043051} (\bibinfo {year} {2024})},\ \Eprint
  {https://arxiv.org/abs/2309.12086} {arXiv:2309.12086 [astro-ph.HE]}
  \BibitemShut {NoStop}%
\bibitem [{\citenamefont {Kuroda}\ and\ \citenamefont
  {Shibata}(2023)}]{Kuroda:2023mzi}%
  \BibitemOpen
  \bibfield  {author} {\bibinfo {author} {\bibfnamefont {T.}~\bibnamefont
  {Kuroda}}\ and\ \bibinfo {author} {\bibfnamefont {M.}~\bibnamefont
  {Shibata}},\ }\href {https://doi.org/10.1093/mnras/stad2710} {\bibfield
  {journal} {\bibinfo  {journal} {Mon. Not. Roy. Astron. Soc.}\ }\textbf
  {\bibinfo {volume} {526}},\ \bibinfo {pages} {152} (\bibinfo {year}
  {2023})},\ \Eprint {https://arxiv.org/abs/2307.06192} {arXiv:2307.06192
  [astro-ph.HE]} \BibitemShut {NoStop}%
\bibitem [{\citenamefont {Shibagaki}\ \emph {et~al.}(2024)\citenamefont
  {Shibagaki}, \citenamefont {Kuroda}, \citenamefont {Kotake}, \citenamefont
  {Takiwaki},\ and\ \citenamefont {Fischer}}]{Shibagaki:2023tmh}%
  \BibitemOpen
  \bibfield  {author} {\bibinfo {author} {\bibfnamefont {S.}~\bibnamefont
  {Shibagaki}}, \bibinfo {author} {\bibfnamefont {T.}~\bibnamefont {Kuroda}},
  \bibinfo {author} {\bibfnamefont {K.}~\bibnamefont {Kotake}}, \bibinfo
  {author} {\bibfnamefont {T.}~\bibnamefont {Takiwaki}},\ and\ \bibinfo
  {author} {\bibfnamefont {T.}~\bibnamefont {Fischer}},\ }\href
  {https://doi.org/10.1093/mnras/stae1361} {\bibfield  {journal} {\bibinfo
  {journal} {Mon. Not. Roy. Astron. Soc.}\ }\textbf {\bibinfo {volume} {531}},\
  \bibinfo {pages} {3732} (\bibinfo {year} {2024})},\ \Eprint
  {https://arxiv.org/abs/2309.05161} {arXiv:2309.05161 [astro-ph.HE]}
  \BibitemShut {NoStop}%
\bibitem [{\citenamefont {Kuroda}\ and\ \citenamefont
  {Shibata}(2024)}]{Kuroda:2024xbe}%
  \BibitemOpen
  \bibfield  {author} {\bibinfo {author} {\bibfnamefont {T.}~\bibnamefont
  {Kuroda}}\ and\ \bibinfo {author} {\bibfnamefont {M.}~\bibnamefont
  {Shibata}},\ }\href {https://doi.org/10.1093/mnrasl/slae069} {\bibfield
  {journal} {\bibinfo  {journal} {Mon. Not. Roy. Astron. Soc.}\ }\textbf
  {\bibinfo {volume} {533}},\ \bibinfo {pages} {L107} (\bibinfo {year}
  {2024})},\ \Eprint {https://arxiv.org/abs/2404.02792} {arXiv:2404.02792
  [astro-ph.HE]} \BibitemShut {NoStop}%
\bibitem [{\citenamefont {Issa}\ \emph {et~al.}(2024)\citenamefont {Issa},
  \citenamefont {Gottlieb}, \citenamefont {Metzger}, \citenamefont
  {Jacquemin-Ide}, \citenamefont {Liska}, \citenamefont {Foucart},
  \citenamefont {Halevi},\ and\ \citenamefont {Tchekhovskoy}}]{Issa:2024sts}%
  \BibitemOpen
  \bibfield  {author} {\bibinfo {author} {\bibfnamefont {D.}~\bibnamefont
  {Issa}}, \bibinfo {author} {\bibfnamefont {O.}~\bibnamefont {Gottlieb}},
  \bibinfo {author} {\bibfnamefont {B.}~\bibnamefont {Metzger}}, \bibinfo
  {author} {\bibfnamefont {J.}~\bibnamefont {Jacquemin-Ide}}, \bibinfo {author}
  {\bibfnamefont {M.}~\bibnamefont {Liska}}, \bibinfo {author} {\bibfnamefont
  {F.}~\bibnamefont {Foucart}}, \bibinfo {author} {\bibfnamefont
  {G.}~\bibnamefont {Halevi}},\ and\ \bibinfo {author} {\bibfnamefont
  {A.}~\bibnamefont {Tchekhovskoy}},\ }\href@noop {} {\  (\bibinfo {year}
  {2024})},\ \Eprint {https://arxiv.org/abs/2410.02852} {arXiv:2410.02852
  [astro-ph.HE]} \BibitemShut {NoStop}%
\bibitem [{\citenamefont {Bruenn}(1985)}]{Bruenn:1985en}%
  \BibitemOpen
  \bibfield  {author} {\bibinfo {author} {\bibfnamefont {S.~W.}\ \bibnamefont
  {Bruenn}},\ }\href {https://doi.org/10.1086/191056} {\bibfield  {journal}
  {\bibinfo  {journal} {Astrophys. J. Suppl.}\ }\textbf {\bibinfo {volume}
  {58}},\ \bibinfo {pages} {771} (\bibinfo {year} {1985})}\BibitemShut
  {NoStop}%
\bibitem [{200(2004)}]{2004StellarCollapse}%
  \BibitemOpen
  \href {https://doi.org/10.1007/978-0-306-48599-2} {\emph {\bibinfo {title}
  {Stellar Collapse}}}\ (\bibinfo  {publisher} {Springer Netherlands},\
  \bibinfo {year} {2004})\BibitemShut {NoStop}%
\bibitem [{\citenamefont {Galeazzi}\ \emph {et~al.}(2013)\citenamefont
  {Galeazzi}, \citenamefont {Kastaun}, \citenamefont {Rezzolla},\ and\
  \citenamefont {Font}}]{Galeazzi:2013mia}%
  \BibitemOpen
  \bibfield  {author} {\bibinfo {author} {\bibfnamefont {F.}~\bibnamefont
  {Galeazzi}}, \bibinfo {author} {\bibfnamefont {W.}~\bibnamefont {Kastaun}},
  \bibinfo {author} {\bibfnamefont {L.}~\bibnamefont {Rezzolla}},\ and\
  \bibinfo {author} {\bibfnamefont {J.~A.}\ \bibnamefont {Font}},\ }\href
  {https://doi.org/10.1103/PhysRevD.88.064009} {\bibfield  {journal} {\bibinfo
  {journal} {Phys. Rev. D}\ }\textbf {\bibinfo {volume} {88}},\ \bibinfo
  {pages} {064009} (\bibinfo {year} {2013})},\ \Eprint
  {https://arxiv.org/abs/1306.4953} {arXiv:1306.4953 [gr-qc]} \BibitemShut
  {NoStop}%
\bibitem [{\citenamefont {Neilsen}\ \emph {et~al.}(2014)\citenamefont
  {Neilsen}, \citenamefont {Liebling}, \citenamefont {Anderson}, \citenamefont
  {Lehner}, \citenamefont {O'Connor},\ and\ \citenamefont
  {Palenzuela}}]{Neilsen:2014hha}%
  \BibitemOpen
  \bibfield  {author} {\bibinfo {author} {\bibfnamefont {D.}~\bibnamefont
  {Neilsen}}, \bibinfo {author} {\bibfnamefont {S.~L.}\ \bibnamefont
  {Liebling}}, \bibinfo {author} {\bibfnamefont {M.}~\bibnamefont {Anderson}},
  \bibinfo {author} {\bibfnamefont {L.}~\bibnamefont {Lehner}}, \bibinfo
  {author} {\bibfnamefont {E.}~\bibnamefont {O'Connor}},\ and\ \bibinfo
  {author} {\bibfnamefont {C.}~\bibnamefont {Palenzuela}},\ }\href
  {https://doi.org/10.1103/PhysRevD.89.104029} {\bibfield  {journal} {\bibinfo
  {journal} {Phys. Rev. D}\ }\textbf {\bibinfo {volume} {89}},\ \bibinfo
  {pages} {104029} (\bibinfo {year} {2014})},\ \Eprint
  {https://arxiv.org/abs/1403.3680} {arXiv:1403.3680 [gr-qc]} \BibitemShut
  {NoStop}%
\bibitem [{\citenamefont {{Takahashi}}\ and\ \citenamefont
  {{Ohsuga}}(2013)}]{2013ApJ...772..127T}%
  \BibitemOpen
  \bibfield  {author} {\bibinfo {author} {\bibfnamefont {H.~R.}\ \bibnamefont
  {{Takahashi}}}\ and\ \bibinfo {author} {\bibfnamefont {K.}~\bibnamefont
  {{Ohsuga}}},\ }\href {https://doi.org/10.1088/0004-637X/772/2/127} {\bibfield
   {journal} {\bibinfo  {journal} {\apj}\ }\textbf {\bibinfo {volume} {772}},\
  \bibinfo {eid} {127} (\bibinfo {year} {2013})},\ \Eprint
  {https://arxiv.org/abs/1306.0049} {arXiv:1306.0049 [astro-ph.HE]}
  \BibitemShut {NoStop}%
\bibitem [{\citenamefont {{Sadowski}}\ \emph {et~al.}(2014)\citenamefont
  {{Sadowski}}, \citenamefont {{Narayan}}, \citenamefont {{McKinney}},\ and\
  \citenamefont {{Tchekhovskoy}}}]{2014MNRAS.439..503S}%
  \BibitemOpen
  \bibfield  {author} {\bibinfo {author} {\bibfnamefont {A.}~\bibnamefont
  {{Sadowski}}}, \bibinfo {author} {\bibfnamefont {R.}~\bibnamefont
  {{Narayan}}}, \bibinfo {author} {\bibfnamefont {J.~C.}\ \bibnamefont
  {{McKinney}}},\ and\ \bibinfo {author} {\bibfnamefont {A.}~\bibnamefont
  {{Tchekhovskoy}}},\ }\href {https://doi.org/10.1093/mnras/stt2479} {\bibfield
   {journal} {\bibinfo  {journal} {\mnras}\ }\textbf {\bibinfo {volume}
  {439}},\ \bibinfo {pages} {503} (\bibinfo {year} {2014})},\ \Eprint
  {https://arxiv.org/abs/1311.5900} {arXiv:1311.5900 [astro-ph.HE]}
  \BibitemShut {NoStop}%
\bibitem [{\citenamefont {{McKinney}}\ \emph {et~al.}(2014)\citenamefont
  {{McKinney}}, \citenamefont {{Tchekhovskoy}}, \citenamefont {{Sadowski}},\
  and\ \citenamefont {{Narayan}}}]{2014MNRAS.441.3177M}%
  \BibitemOpen
  \bibfield  {author} {\bibinfo {author} {\bibfnamefont {J.~C.}\ \bibnamefont
  {{McKinney}}}, \bibinfo {author} {\bibfnamefont {A.}~\bibnamefont
  {{Tchekhovskoy}}}, \bibinfo {author} {\bibfnamefont {A.}~\bibnamefont
  {{Sadowski}}},\ and\ \bibinfo {author} {\bibfnamefont {R.}~\bibnamefont
  {{Narayan}}},\ }\href {https://doi.org/10.1093/mnras/stu762} {\bibfield
  {journal} {\bibinfo  {journal} {\mnras}\ }\textbf {\bibinfo {volume} {441}},\
  \bibinfo {pages} {3177} (\bibinfo {year} {2014})},\ \Eprint
  {https://arxiv.org/abs/1312.6127} {arXiv:1312.6127 [astro-ph.CO]}
  \BibitemShut {NoStop}%
\bibitem [{\citenamefont {{Sadowski}}\ \emph {et~al.}(2015)\citenamefont
  {{Sadowski}}, \citenamefont {{Narayan}}, \citenamefont {{Tchekhovskoy}},
  \citenamefont {{Abarca}}, \citenamefont {{Zhu}},\ and\ \citenamefont
  {{McKinney}}}]{2015MNRAS.447...49S}%
  \BibitemOpen
  \bibfield  {author} {\bibinfo {author} {\bibfnamefont {A.}~\bibnamefont
  {{Sadowski}}}, \bibinfo {author} {\bibfnamefont {R.}~\bibnamefont
  {{Narayan}}}, \bibinfo {author} {\bibfnamefont {A.}~\bibnamefont
  {{Tchekhovskoy}}}, \bibinfo {author} {\bibfnamefont {D.}~\bibnamefont
  {{Abarca}}}, \bibinfo {author} {\bibfnamefont {Y.}~\bibnamefont {{Zhu}}},\
  and\ \bibinfo {author} {\bibfnamefont {J.~C.}\ \bibnamefont {{McKinney}}},\
  }\href {https://doi.org/10.1093/mnras/stu2387} {\bibfield  {journal}
  {\bibinfo  {journal} {\mnras}\ }\textbf {\bibinfo {volume} {447}},\ \bibinfo
  {pages} {49} (\bibinfo {year} {2015})},\ \Eprint
  {https://arxiv.org/abs/1407.4421} {arXiv:1407.4421 [astro-ph.HE]}
  \BibitemShut {NoStop}%
\bibitem [{\citenamefont {Sekiguchi}\ \emph {et~al.}(2015)\citenamefont
  {Sekiguchi}, \citenamefont {Kiuchi}, \citenamefont {Kyutoku},\ and\
  \citenamefont {Shibata}}]{Sekiguchi:2015dma}%
  \BibitemOpen
  \bibfield  {author} {\bibinfo {author} {\bibfnamefont {Y.}~\bibnamefont
  {Sekiguchi}}, \bibinfo {author} {\bibfnamefont {K.}~\bibnamefont {Kiuchi}},
  \bibinfo {author} {\bibfnamefont {K.}~\bibnamefont {Kyutoku}},\ and\ \bibinfo
  {author} {\bibfnamefont {M.}~\bibnamefont {Shibata}},\ }\href
  {https://doi.org/10.1103/PhysRevD.91.064059} {\bibfield  {journal} {\bibinfo
  {journal} {Phys. Rev.}\ }\textbf {\bibinfo {volume} {D91}},\ \bibinfo {pages}
  {064059} (\bibinfo {year} {2015})},\ \Eprint
  {https://arxiv.org/abs/1502.06660} {arXiv:1502.06660 [astro-ph.HE]}
  \BibitemShut {NoStop}%
\bibitem [{\citenamefont {{Foucart}}\ \emph {et~al.}(2015)\citenamefont
  {{Foucart}}, \citenamefont {{O'Connor}}, \citenamefont {{Roberts}},
  \citenamefont {{Duez}}, \citenamefont {{Haas}}, \citenamefont {{Kidder}},
  \citenamefont {{Ott}}, \citenamefont {{Pfeiffer}}, \citenamefont {{Scheel}},\
  and\ \citenamefont {{Szilagyi}}}]{2015PhRvD..91l4021F}%
  \BibitemOpen
  \bibfield  {author} {\bibinfo {author} {\bibfnamefont {F.}~\bibnamefont
  {{Foucart}}}, \bibinfo {author} {\bibfnamefont {E.}~\bibnamefont
  {{O'Connor}}}, \bibinfo {author} {\bibfnamefont {L.}~\bibnamefont
  {{Roberts}}}, \bibinfo {author} {\bibfnamefont {M.~D.}\ \bibnamefont
  {{Duez}}}, \bibinfo {author} {\bibfnamefont {R.}~\bibnamefont {{Haas}}},
  \bibinfo {author} {\bibfnamefont {L.~E.}\ \bibnamefont {{Kidder}}}, \bibinfo
  {author} {\bibfnamefont {C.~D.}\ \bibnamefont {{Ott}}}, \bibinfo {author}
  {\bibfnamefont {H.~P.}\ \bibnamefont {{Pfeiffer}}}, \bibinfo {author}
  {\bibfnamefont {M.~A.}\ \bibnamefont {{Scheel}}},\ and\ \bibinfo {author}
  {\bibfnamefont {B.}~\bibnamefont {{Szilagyi}}},\ }\href
  {https://doi.org/10.1103/PhysRevD.91.124021} {\bibfield  {journal} {\bibinfo
  {journal} {\prd}\ }\textbf {\bibinfo {volume} {91}},\ \bibinfo {eid} {124021}
  (\bibinfo {year} {2015})},\ \Eprint {https://arxiv.org/abs/1502.04146}
  {arXiv:1502.04146 [astro-ph.HE]} \BibitemShut {NoStop}%
\bibitem [{\citenamefont {Sekiguchi}\ \emph {et~al.}(2016)\citenamefont
  {Sekiguchi}, \citenamefont {Kiuchi}, \citenamefont {Kyutoku}, \citenamefont
  {Shibata},\ and\ \citenamefont {Taniguchi}}]{Sekiguchi:2016bjd}%
  \BibitemOpen
  \bibfield  {author} {\bibinfo {author} {\bibfnamefont {Y.}~\bibnamefont
  {Sekiguchi}}, \bibinfo {author} {\bibfnamefont {K.}~\bibnamefont {Kiuchi}},
  \bibinfo {author} {\bibfnamefont {K.}~\bibnamefont {Kyutoku}}, \bibinfo
  {author} {\bibfnamefont {M.}~\bibnamefont {Shibata}},\ and\ \bibinfo {author}
  {\bibfnamefont {K.}~\bibnamefont {Taniguchi}},\ }\href
  {https://doi.org/10.1103/PhysRevD.93.124046} {\bibfield  {journal} {\bibinfo
  {journal} {Phys. Rev.}\ }\textbf {\bibinfo {volume} {D93}},\ \bibinfo {pages}
  {124046} (\bibinfo {year} {2016})},\ \Eprint
  {https://arxiv.org/abs/1603.01918} {arXiv:1603.01918 [astro-ph.HE]}
  \BibitemShut {NoStop}%
\bibitem [{\citenamefont {Radice}\ \emph {et~al.}(2016)\citenamefont {Radice},
  \citenamefont {Galeazzi}, \citenamefont {Lippuner}, \citenamefont {Roberts},
  \citenamefont {Ott},\ and\ \citenamefont {Rezzolla}}]{Radice:2016dwd}%
  \BibitemOpen
  \bibfield  {author} {\bibinfo {author} {\bibfnamefont {D.}~\bibnamefont
  {Radice}}, \bibinfo {author} {\bibfnamefont {F.}~\bibnamefont {Galeazzi}},
  \bibinfo {author} {\bibfnamefont {J.}~\bibnamefont {Lippuner}}, \bibinfo
  {author} {\bibfnamefont {L.~F.}\ \bibnamefont {Roberts}}, \bibinfo {author}
  {\bibfnamefont {C.~D.}\ \bibnamefont {Ott}},\ and\ \bibinfo {author}
  {\bibfnamefont {L.}~\bibnamefont {Rezzolla}},\ }\href
  {https://doi.org/10.1093/mnras/stw1227} {\bibfield  {journal} {\bibinfo
  {journal} {Mon. Not. Roy. Astron. Soc.}\ }\textbf {\bibinfo {volume} {460}},\
  \bibinfo {pages} {3255} (\bibinfo {year} {2016})},\ \Eprint
  {https://arxiv.org/abs/1601.02426} {arXiv:1601.02426 [astro-ph.HE]}
  \BibitemShut {NoStop}%
\bibitem [{\citenamefont {Radice}\ \emph {et~al.}(2022)\citenamefont {Radice},
  \citenamefont {Bernuzzi}, \citenamefont {Perego},\ and\ \citenamefont
  {Haas}}]{Radice:2021jtw}%
  \BibitemOpen
  \bibfield  {author} {\bibinfo {author} {\bibfnamefont {D.}~\bibnamefont
  {Radice}}, \bibinfo {author} {\bibfnamefont {S.}~\bibnamefont {Bernuzzi}},
  \bibinfo {author} {\bibfnamefont {A.}~\bibnamefont {Perego}},\ and\ \bibinfo
  {author} {\bibfnamefont {R.}~\bibnamefont {Haas}},\ }\href
  {https://doi.org/10.1093/mnras/stac589} {\bibfield  {journal} {\bibinfo
  {journal} {Mon. Not. Roy. Astron. Soc.}\ }\textbf {\bibinfo {volume} {512}},\
  \bibinfo {pages} {1499} (\bibinfo {year} {2022})},\ \Eprint
  {https://arxiv.org/abs/2111.14858} {arXiv:2111.14858 [astro-ph.HE]}
  \BibitemShut {NoStop}%
\bibitem [{\citenamefont {{Kiuchi}}\ \emph {et~al.}(2022)\citenamefont
  {{Kiuchi}}, \citenamefont {{Held}}, \citenamefont {{Sekiguchi}},\ and\
  \citenamefont {{Shibata}}}]{2022arXiv220504487K}%
  \BibitemOpen
  \bibfield  {author} {\bibinfo {author} {\bibfnamefont {K.}~\bibnamefont
  {{Kiuchi}}}, \bibinfo {author} {\bibfnamefont {L.~E.}\ \bibnamefont
  {{Held}}}, \bibinfo {author} {\bibfnamefont {Y.}~\bibnamefont
  {{Sekiguchi}}},\ and\ \bibinfo {author} {\bibfnamefont {M.}~\bibnamefont
  {{Shibata}}},\ }\href@noop {} {\bibfield  {journal} {\bibinfo  {journal}
  {arXiv e-prints}\ ,\ \bibinfo {eid} {arXiv:2205.04487}} (\bibinfo {year}
  {2022})},\ \Eprint {https://arxiv.org/abs/2205.04487} {arXiv:2205.04487
  [astro-ph.HE]} \BibitemShut {NoStop}%
\bibitem [{\citenamefont {Sun}\ \emph {et~al.}(2022)\citenamefont {Sun},
  \citenamefont {Ruiz}, \citenamefont {Shapiro},\ and\ \citenamefont
  {Tsokaros}}]{Sun:2022vri}%
  \BibitemOpen
  \bibfield  {author} {\bibinfo {author} {\bibfnamefont {L.}~\bibnamefont
  {Sun}}, \bibinfo {author} {\bibfnamefont {M.}~\bibnamefont {Ruiz}}, \bibinfo
  {author} {\bibfnamefont {S.~L.}\ \bibnamefont {Shapiro}},\ and\ \bibinfo
  {author} {\bibfnamefont {A.}~\bibnamefont {Tsokaros}},\ }\href
  {https://doi.org/10.1103/PhysRevD.105.104028} {\bibfield  {journal} {\bibinfo
   {journal} {Phys. Rev. D}\ }\textbf {\bibinfo {volume} {105}},\ \bibinfo
  {pages} {104028} (\bibinfo {year} {2022})},\ \Eprint
  {https://arxiv.org/abs/2202.12901} {arXiv:2202.12901 [astro-ph.HE]}
  \BibitemShut {NoStop}%
\bibitem [{\citenamefont {Werneck}\ \emph {et~al.}(2023)\citenamefont {Werneck}
  \emph {et~al.}}]{Werneck:2022exo}%
  \BibitemOpen
  \bibfield  {author} {\bibinfo {author} {\bibfnamefont {L.~R.}\ \bibnamefont
  {Werneck}} \emph {et~al.},\ }\href
  {https://doi.org/10.1103/PhysRevD.107.044037} {\bibfield  {journal} {\bibinfo
   {journal} {Phys. Rev. D}\ }\textbf {\bibinfo {volume} {107}},\ \bibinfo
  {pages} {044037} (\bibinfo {year} {2023})},\ \Eprint
  {https://arxiv.org/abs/2208.14487} {arXiv:2208.14487 [gr-qc]} \BibitemShut
  {NoStop}%
\bibitem [{\citenamefont {Foucart}\ \emph {et~al.}(2024)\citenamefont
  {Foucart}, \citenamefont {Cheong}, \citenamefont {Duez}, \citenamefont
  {Kidder}, \citenamefont {Pfeiffer},\ and\ \citenamefont
  {Scheel}}]{Foucart:2024npn}%
  \BibitemOpen
  \bibfield  {author} {\bibinfo {author} {\bibfnamefont {F.}~\bibnamefont
  {Foucart}}, \bibinfo {author} {\bibfnamefont {P.~C.-K.}\ \bibnamefont
  {Cheong}}, \bibinfo {author} {\bibfnamefont {M.~D.}\ \bibnamefont {Duez}},
  \bibinfo {author} {\bibfnamefont {L.~E.}\ \bibnamefont {Kidder}}, \bibinfo
  {author} {\bibfnamefont {H.~P.}\ \bibnamefont {Pfeiffer}},\ and\ \bibinfo
  {author} {\bibfnamefont {M.~A.}\ \bibnamefont {Scheel}},\ }\href@noop {}
  {\bibfield  {journal} {\bibinfo  {journal} {arXiv}\ } (\bibinfo {year}
  {2024})},\ \Eprint {https://arxiv.org/abs/2407.15989} {arXiv:2407.15989
  [astro-ph.HE]} \BibitemShut {NoStop}%
\bibitem [{\citenamefont {Ruffert}\ and\ \citenamefont
  {Janka}(1999)}]{Ruffert:1998qg}%
  \BibitemOpen
  \bibfield  {author} {\bibinfo {author} {\bibfnamefont {M.}~\bibnamefont
  {Ruffert}}\ and\ \bibinfo {author} {\bibfnamefont {H.~T.}\ \bibnamefont
  {Janka}},\ }\href@noop {} {\bibfield  {journal} {\bibinfo  {journal} {Astron.
  Astrophys.}\ }\textbf {\bibinfo {volume} {344}},\ \bibinfo {pages} {573}
  (\bibinfo {year} {1999})},\ \Eprint {https://arxiv.org/abs/astro-ph/9809280}
  {arXiv:astro-ph/9809280} \BibitemShut {NoStop}%
\bibitem [{\citenamefont {Setiawan}\ \emph {et~al.}(2004)\citenamefont
  {Setiawan}, \citenamefont {Ruffert},\ and\ \citenamefont
  {Janka}}]{Setiawan:2004xy}%
  \BibitemOpen
  \bibfield  {author} {\bibinfo {author} {\bibfnamefont {S.}~\bibnamefont
  {Setiawan}}, \bibinfo {author} {\bibfnamefont {M.}~\bibnamefont {Ruffert}},\
  and\ \bibinfo {author} {\bibfnamefont {H.-T.}\ \bibnamefont {Janka}},\ }\href
  {https://doi.org/10.1111/j.1365-2966.2004.07974.x} {\bibfield  {journal}
  {\bibinfo  {journal} {Mon. Not. Roy. Astron. Soc.}\ }\textbf {\bibinfo
  {volume} {352}},\ \bibinfo {pages} {753} (\bibinfo {year} {2004})},\ \Eprint
  {https://arxiv.org/abs/astro-ph/0402481} {arXiv:astro-ph/0402481}
  \BibitemShut {NoStop}%
\bibitem [{\citenamefont {Setiawan}\ \emph {et~al.}(2006)\citenamefont
  {Setiawan}, \citenamefont {Ruffert},\ and\ \citenamefont
  {Janka}}]{Setiawan:2005ah}%
  \BibitemOpen
  \bibfield  {author} {\bibinfo {author} {\bibfnamefont {S.}~\bibnamefont
  {Setiawan}}, \bibinfo {author} {\bibfnamefont {M.}~\bibnamefont {Ruffert}},\
  and\ \bibinfo {author} {\bibfnamefont {H.~T.}\ \bibnamefont {Janka}},\ }\href
  {https://doi.org/10.1051/0004-6361:20054193} {\bibfield  {journal} {\bibinfo
  {journal} {Astron. Astrophys.}\ }\textbf {\bibinfo {volume} {458}},\ \bibinfo
  {pages} {553} (\bibinfo {year} {2006})},\ \Eprint
  {https://arxiv.org/abs/astro-ph/0509300} {arXiv:astro-ph/0509300}
  \BibitemShut {NoStop}%
\bibitem [{\citenamefont {Richers}\ \emph {et~al.}(2015)\citenamefont
  {Richers}, \citenamefont {Kasen}, \citenamefont {O'Connor}, \citenamefont
  {Fern\'andez},\ and\ \citenamefont {Ott}}]{Richers:2015lma}%
  \BibitemOpen
  \bibfield  {author} {\bibinfo {author} {\bibfnamefont {S.}~\bibnamefont
  {Richers}}, \bibinfo {author} {\bibfnamefont {D.}~\bibnamefont {Kasen}},
  \bibinfo {author} {\bibfnamefont {E.}~\bibnamefont {O'Connor}}, \bibinfo
  {author} {\bibfnamefont {R.}~\bibnamefont {Fern\'andez}},\ and\ \bibinfo
  {author} {\bibfnamefont {C.~D.}\ \bibnamefont {Ott}},\ }\href
  {https://doi.org/10.1088/0004-637X/813/1/38} {\bibfield  {journal} {\bibinfo
  {journal} {Astrophys. J.}\ }\textbf {\bibinfo {volume} {813}},\ \bibinfo
  {pages} {38} (\bibinfo {year} {2015})},\ \Eprint
  {https://arxiv.org/abs/1507.03606} {arXiv:1507.03606 [astro-ph.HE]}
  \BibitemShut {NoStop}%
\bibitem [{\citenamefont {Perego}\ \emph {et~al.}(2017)\citenamefont {Perego},
  \citenamefont {Yasin},\ and\ \citenamefont {Arcones}}]{Perego:2017fho}%
  \BibitemOpen
  \bibfield  {author} {\bibinfo {author} {\bibfnamefont {A.}~\bibnamefont
  {Perego}}, \bibinfo {author} {\bibfnamefont {H.}~\bibnamefont {Yasin}},\ and\
  \bibinfo {author} {\bibfnamefont {A.}~\bibnamefont {Arcones}},\ }\href
  {https://doi.org/10.1088/1361-6471/aa7bdc} {\bibfield  {journal} {\bibinfo
  {journal} {J. Phys. G}\ }\textbf {\bibinfo {volume} {44}},\ \bibinfo {pages}
  {084007} (\bibinfo {year} {2017})},\ \Eprint
  {https://arxiv.org/abs/1701.02017} {arXiv:1701.02017 [astro-ph.HE]}
  \BibitemShut {NoStop}%
\bibitem [{\citenamefont {Sumiyoshi}\ \emph {et~al.}(2021)\citenamefont
  {Sumiyoshi}, \citenamefont {Fujibayashi}, \citenamefont {Sekiguchi},\ and\
  \citenamefont {Shibata}}]{Sumiyoshi:2020bdh}%
  \BibitemOpen
  \bibfield  {author} {\bibinfo {author} {\bibfnamefont {K.}~\bibnamefont
  {Sumiyoshi}}, \bibinfo {author} {\bibfnamefont {S.}~\bibnamefont
  {Fujibayashi}}, \bibinfo {author} {\bibfnamefont {Y.}~\bibnamefont
  {Sekiguchi}},\ and\ \bibinfo {author} {\bibfnamefont {M.}~\bibnamefont
  {Shibata}},\ }\href {https://doi.org/10.3847/1538-4357/abce63} {\bibfield
  {journal} {\bibinfo  {journal} {Astrophys. J.}\ }\textbf {\bibinfo {volume}
  {907}},\ \bibinfo {pages} {92} (\bibinfo {year} {2021})},\ \Eprint
  {https://arxiv.org/abs/2010.10865} {arXiv:2010.10865 [astro-ph.HE]}
  \BibitemShut {NoStop}%
\bibitem [{\citenamefont {Just}\ \emph {et~al.}(2016)\citenamefont {Just},
  \citenamefont {Obergaulinger}, \citenamefont {Janka}, \citenamefont
  {Bauswein},\ and\ \citenamefont {Schwarz}}]{Just:2015dba}%
  \BibitemOpen
  \bibfield  {author} {\bibinfo {author} {\bibfnamefont {O.}~\bibnamefont
  {Just}}, \bibinfo {author} {\bibfnamefont {M.}~\bibnamefont {Obergaulinger}},
  \bibinfo {author} {\bibfnamefont {H.~T.}\ \bibnamefont {Janka}}, \bibinfo
  {author} {\bibfnamefont {A.}~\bibnamefont {Bauswein}},\ and\ \bibinfo
  {author} {\bibfnamefont {N.}~\bibnamefont {Schwarz}},\ }\href
  {https://doi.org/10.3847/2041-8205/816/2/L30} {\bibfield  {journal} {\bibinfo
   {journal} {Astrophys. J. Lett.}\ }\textbf {\bibinfo {volume} {816}},\
  \bibinfo {pages} {L30} (\bibinfo {year} {2016})},\ \Eprint
  {https://arxiv.org/abs/1510.04288} {arXiv:1510.04288 [astro-ph.HE]}
  \BibitemShut {NoStop}%
\bibitem [{\citenamefont {Fujibayashi}\ \emph {et~al.}(2017)\citenamefont
  {Fujibayashi}, \citenamefont {Sekiguchi}, \citenamefont {Kiuchi},\ and\
  \citenamefont {Shibata}}]{Fujibayashi:2017xsz}%
  \BibitemOpen
  \bibfield  {author} {\bibinfo {author} {\bibfnamefont {S.}~\bibnamefont
  {Fujibayashi}}, \bibinfo {author} {\bibfnamefont {Y.}~\bibnamefont
  {Sekiguchi}}, \bibinfo {author} {\bibfnamefont {K.}~\bibnamefont {Kiuchi}},\
  and\ \bibinfo {author} {\bibfnamefont {M.}~\bibnamefont {Shibata}},\ }\href
  {https://doi.org/10.3847/1538-4357/aa8039} {\bibfield  {journal} {\bibinfo
  {journal} {Astrophys. J.}\ }\textbf {\bibinfo {volume} {846}},\ \bibinfo
  {pages} {114} (\bibinfo {year} {2017})},\ \Eprint
  {https://arxiv.org/abs/1703.10191} {arXiv:1703.10191 [astro-ph.HE]}
  \BibitemShut {NoStop}%
\bibitem [{\citenamefont {{Abdikamalov}}\ \emph {et~al.}(2012)\citenamefont
  {{Abdikamalov}}, \citenamefont {{Burrows}}, \citenamefont {{Ott}},
  \citenamefont {{L{\"o}ffler}}, \citenamefont {{O'Connor}}, \citenamefont
  {{Dolence}},\ and\ \citenamefont {{Schnetter}}}]{2012ApJ...755..111A}%
  \BibitemOpen
  \bibfield  {author} {\bibinfo {author} {\bibfnamefont {E.}~\bibnamefont
  {{Abdikamalov}}}, \bibinfo {author} {\bibfnamefont {A.}~\bibnamefont
  {{Burrows}}}, \bibinfo {author} {\bibfnamefont {C.~D.}\ \bibnamefont
  {{Ott}}}, \bibinfo {author} {\bibfnamefont {F.}~\bibnamefont
  {{L{\"o}ffler}}}, \bibinfo {author} {\bibfnamefont {E.}~\bibnamefont
  {{O'Connor}}}, \bibinfo {author} {\bibfnamefont {J.~C.}\ \bibnamefont
  {{Dolence}}},\ and\ \bibinfo {author} {\bibfnamefont {E.}~\bibnamefont
  {{Schnetter}}},\ }\href {https://doi.org/10.1088/0004-637X/755/2/111}
  {\bibfield  {journal} {\bibinfo  {journal} {\apj}\ }\textbf {\bibinfo
  {volume} {755}},\ \bibinfo {eid} {111} (\bibinfo {year} {2012})},\ \Eprint
  {https://arxiv.org/abs/1203.2915} {arXiv:1203.2915 [astro-ph.SR]}
  \BibitemShut {NoStop}%
\bibitem [{\citenamefont {{Roth}}\ and\ \citenamefont
  {{Kasen}}(2015)}]{2015ApJS..217....9R}%
  \BibitemOpen
  \bibfield  {author} {\bibinfo {author} {\bibfnamefont {N.}~\bibnamefont
  {{Roth}}}\ and\ \bibinfo {author} {\bibfnamefont {D.}~\bibnamefont
  {{Kasen}}},\ }\href {https://doi.org/10.1088/0067-0049/217/1/9} {\bibfield
  {journal} {\bibinfo  {journal} {\apjs}\ }\textbf {\bibinfo {volume} {217}},\
  \bibinfo {eid} {9} (\bibinfo {year} {2015})},\ \Eprint
  {https://arxiv.org/abs/1404.4652} {arXiv:1404.4652 [astro-ph.IM]}
  \BibitemShut {NoStop}%
\bibitem [{\citenamefont {{Ryan}}\ \emph {et~al.}(2015)\citenamefont {{Ryan}},
  \citenamefont {{Dolence}},\ and\ \citenamefont
  {{Gammie}}}]{2015ApJ...807...31R}%
  \BibitemOpen
  \bibfield  {author} {\bibinfo {author} {\bibfnamefont {B.~R.}\ \bibnamefont
  {{Ryan}}}, \bibinfo {author} {\bibfnamefont {J.~C.}\ \bibnamefont
  {{Dolence}}},\ and\ \bibinfo {author} {\bibfnamefont {C.~F.}\ \bibnamefont
  {{Gammie}}},\ }\href {https://doi.org/10.1088/0004-637X/807/1/31} {\bibfield
  {journal} {\bibinfo  {journal} {\apj}\ }\textbf {\bibinfo {volume} {807}},\
  \bibinfo {eid} {31} (\bibinfo {year} {2015})},\ \Eprint
  {https://arxiv.org/abs/1505.05119} {arXiv:1505.05119 [astro-ph.HE]}
  \BibitemShut {NoStop}%
\bibitem [{\citenamefont {{Foucart}}(2018)}]{2018MNRAS.475.4186F}%
  \BibitemOpen
  \bibfield  {author} {\bibinfo {author} {\bibfnamefont {F.}~\bibnamefont
  {{Foucart}}},\ }\href {https://doi.org/10.1093/mnras/sty108} {\bibfield
  {journal} {\bibinfo  {journal} {\mnras}\ }\textbf {\bibinfo {volume} {475}},\
  \bibinfo {pages} {4186} (\bibinfo {year} {2018})},\ \Eprint
  {https://arxiv.org/abs/1708.08452} {arXiv:1708.08452 [astro-ph.HE]}
  \BibitemShut {NoStop}%
\bibitem [{\citenamefont {{Miller}}\ \emph
  {et~al.}(2019{\natexlab{a}})\citenamefont {{Miller}}, \citenamefont {{Ryan}},
  \citenamefont {{Dolence}}, \citenamefont {{Burrows}}, \citenamefont
  {{Fontes}}, \citenamefont {{Fryer}}, \citenamefont {{Korobkin}},
  \citenamefont {{Lippuner}}, \citenamefont {{Mumpower}},\ and\ \citenamefont
  {{Wollaeger}}}]{2019PhRvD.100b3008M}%
  \BibitemOpen
  \bibfield  {author} {\bibinfo {author} {\bibfnamefont {J.~M.}\ \bibnamefont
  {{Miller}}}, \bibinfo {author} {\bibfnamefont {B.~R.}\ \bibnamefont
  {{Ryan}}}, \bibinfo {author} {\bibfnamefont {J.~C.}\ \bibnamefont
  {{Dolence}}}, \bibinfo {author} {\bibfnamefont {A.}~\bibnamefont
  {{Burrows}}}, \bibinfo {author} {\bibfnamefont {C.~J.}\ \bibnamefont
  {{Fontes}}}, \bibinfo {author} {\bibfnamefont {C.~L.}\ \bibnamefont
  {{Fryer}}}, \bibinfo {author} {\bibfnamefont {O.}~\bibnamefont {{Korobkin}}},
  \bibinfo {author} {\bibfnamefont {J.}~\bibnamefont {{Lippuner}}}, \bibinfo
  {author} {\bibfnamefont {M.~R.}\ \bibnamefont {{Mumpower}}},\ and\ \bibinfo
  {author} {\bibfnamefont {R.~T.}\ \bibnamefont {{Wollaeger}}},\ }\href
  {https://doi.org/10.1103/PhysRevD.100.023008} {\bibfield  {journal} {\bibinfo
   {journal} {\prd}\ }\textbf {\bibinfo {volume} {100}},\ \bibinfo {eid}
  {023008} (\bibinfo {year} {2019}{\natexlab{a}})},\ \Eprint
  {https://arxiv.org/abs/1905.07477} {arXiv:1905.07477 [astro-ph.HE]}
  \BibitemShut {NoStop}%
\bibitem [{\citenamefont {{Miller}}\ \emph
  {et~al.}(2019{\natexlab{b}})\citenamefont {{Miller}}, \citenamefont
  {{Ryan}},\ and\ \citenamefont {{Dolence}}}]{2019ApJS..241...30M}%
  \BibitemOpen
  \bibfield  {author} {\bibinfo {author} {\bibfnamefont {J.~M.}\ \bibnamefont
  {{Miller}}}, \bibinfo {author} {\bibfnamefont {B.~R.}\ \bibnamefont
  {{Ryan}}},\ and\ \bibinfo {author} {\bibfnamefont {J.~C.}\ \bibnamefont
  {{Dolence}}},\ }\href {https://doi.org/10.3847/1538-4365/ab09fc} {\bibfield
  {journal} {\bibinfo  {journal} {\apjs}\ }\textbf {\bibinfo {volume} {241}},\
  \bibinfo {eid} {30} (\bibinfo {year} {2019}{\natexlab{b}})},\ \Eprint
  {https://arxiv.org/abs/1903.09273} {arXiv:1903.09273 [astro-ph.IM]}
  \BibitemShut {NoStop}%
\bibitem [{\citenamefont {Foucart}\ \emph {et~al.}(2020)\citenamefont
  {Foucart}, \citenamefont {Duez}, \citenamefont {Hebert}, \citenamefont
  {Kidder}, \citenamefont {Pfeiffer},\ and\ \citenamefont
  {Scheel}}]{Foucart:2020qjb}%
  \BibitemOpen
  \bibfield  {author} {\bibinfo {author} {\bibfnamefont {F.}~\bibnamefont
  {Foucart}}, \bibinfo {author} {\bibfnamefont {M.~D.}\ \bibnamefont {Duez}},
  \bibinfo {author} {\bibfnamefont {F.}~\bibnamefont {Hebert}}, \bibinfo
  {author} {\bibfnamefont {L.~E.}\ \bibnamefont {Kidder}}, \bibinfo {author}
  {\bibfnamefont {H.~P.}\ \bibnamefont {Pfeiffer}},\ and\ \bibinfo {author}
  {\bibfnamefont {M.~A.}\ \bibnamefont {Scheel}},\ }\href
  {https://doi.org/10.3847/2041-8213/abbb87} {\bibfield  {journal} {\bibinfo
  {journal} {Astrophys. J. Lett.}\ }\textbf {\bibinfo {volume} {902}},\
  \bibinfo {pages} {L27} (\bibinfo {year} {2020})},\ \Eprint
  {https://arxiv.org/abs/2008.08089} {arXiv:2008.08089 [astro-ph.HE]}
  \BibitemShut {NoStop}%
\bibitem [{\citenamefont {{Foucart}}\ \emph {et~al.}(2021)\citenamefont
  {{Foucart}}, \citenamefont {{Duez}}, \citenamefont {{H{\'e}bert}},
  \citenamefont {{Kidder}}, \citenamefont {{Kovarik}}, \citenamefont
  {{Pfeiffer}},\ and\ \citenamefont {{Scheel}}}]{2021ApJ...920...82F}%
  \BibitemOpen
  \bibfield  {author} {\bibinfo {author} {\bibfnamefont {F.}~\bibnamefont
  {{Foucart}}}, \bibinfo {author} {\bibfnamefont {M.~D.}\ \bibnamefont
  {{Duez}}}, \bibinfo {author} {\bibfnamefont {F.}~\bibnamefont
  {{H{\'e}bert}}}, \bibinfo {author} {\bibfnamefont {L.~E.}\ \bibnamefont
  {{Kidder}}}, \bibinfo {author} {\bibfnamefont {P.}~\bibnamefont {{Kovarik}}},
  \bibinfo {author} {\bibfnamefont {H.~P.}\ \bibnamefont {{Pfeiffer}}},\ and\
  \bibinfo {author} {\bibfnamefont {M.~A.}\ \bibnamefont {{Scheel}}},\ }\href
  {https://doi.org/10.3847/1538-4357/ac1737} {\bibfield  {journal} {\bibinfo
  {journal} {\apj}\ }\textbf {\bibinfo {volume} {920}},\ \bibinfo {eid} {82}
  (\bibinfo {year} {2021})},\ \Eprint {https://arxiv.org/abs/2103.16588}
  {arXiv:2103.16588 [astro-ph.HE]} \BibitemShut {NoStop}%
\bibitem [{\citenamefont {{Roth}}\ \emph {et~al.}(2022)\citenamefont {{Roth}},
  \citenamefont {{Anninos}}, \citenamefont {{Robinson}}, \citenamefont
  {{Peterson}}, \citenamefont {{Polak}}, \citenamefont {{Mangan}},\ and\
  \citenamefont {{Beyer}}}]{2022ApJ...933..226R}%
  \BibitemOpen
  \bibfield  {author} {\bibinfo {author} {\bibfnamefont {N.}~\bibnamefont
  {{Roth}}}, \bibinfo {author} {\bibfnamefont {P.}~\bibnamefont {{Anninos}}},
  \bibinfo {author} {\bibfnamefont {P.~B.}\ \bibnamefont {{Robinson}}},
  \bibinfo {author} {\bibfnamefont {J.~L.}\ \bibnamefont {{Peterson}}},
  \bibinfo {author} {\bibfnamefont {B.}~\bibnamefont {{Polak}}}, \bibinfo
  {author} {\bibfnamefont {T.~K.}\ \bibnamefont {{Mangan}}},\ and\ \bibinfo
  {author} {\bibfnamefont {K.}~\bibnamefont {{Beyer}}},\ }\href
  {https://doi.org/10.3847/1538-4357/ac75cb} {\bibfield  {journal} {\bibinfo
  {journal} {\apj}\ }\textbf {\bibinfo {volume} {933}},\ \bibinfo {eid} {226}
  (\bibinfo {year} {2022})},\ \Eprint {https://arxiv.org/abs/2206.01760}
  {arXiv:2206.01760 [astro-ph.IM]} \BibitemShut {NoStop}%
\bibitem [{\citenamefont {Izquierdo}\ \emph {et~al.}(2024)\citenamefont
  {Izquierdo}, \citenamefont {Abalos},\ and\ \citenamefont
  {Palenzuela}}]{Izquierdo:2023fub}%
  \BibitemOpen
  \bibfield  {author} {\bibinfo {author} {\bibfnamefont {M.~R.}\ \bibnamefont
  {Izquierdo}}, \bibinfo {author} {\bibfnamefont {J.~F.}\ \bibnamefont
  {Abalos}},\ and\ \bibinfo {author} {\bibfnamefont {C.}~\bibnamefont
  {Palenzuela}},\ }\href {https://doi.org/10.1103/PhysRevD.109.043044}
  {\bibfield  {journal} {\bibinfo  {journal} {Phys. Rev. D}\ }\textbf {\bibinfo
  {volume} {109}},\ \bibinfo {pages} {043044} (\bibinfo {year} {2024})},\
  \Eprint {https://arxiv.org/abs/2312.09275} {arXiv:2312.09275 [astro-ph.HE]}
  \BibitemShut {NoStop}%
\bibitem [{\citenamefont {Kawaguchi}\ \emph {et~al.}(2023)\citenamefont
  {Kawaguchi}, \citenamefont {Fujibayashi},\ and\ \citenamefont
  {Shibata}}]{Kawaguchi:2022tae}%
  \BibitemOpen
  \bibfield  {author} {\bibinfo {author} {\bibfnamefont {K.}~\bibnamefont
  {Kawaguchi}}, \bibinfo {author} {\bibfnamefont {S.}~\bibnamefont
  {Fujibayashi}},\ and\ \bibinfo {author} {\bibfnamefont {M.}~\bibnamefont
  {Shibata}},\ }\href {https://doi.org/10.1103/PhysRevD.107.023026} {\bibfield
  {journal} {\bibinfo  {journal} {Phys. Rev. D}\ }\textbf {\bibinfo {volume}
  {107}},\ \bibinfo {pages} {023026} (\bibinfo {year} {2023})},\ \Eprint
  {https://arxiv.org/abs/2209.12472} {arXiv:2209.12472 [astro-ph.HE]}
  \BibitemShut {NoStop}%
\bibitem [{\citenamefont {Kawaguchi}\ \emph {et~al.}(2025)\citenamefont
  {Kawaguchi}, \citenamefont {Fujibayashi},\ and\ \citenamefont
  {Shibata}}]{Kawaguchi:2024naa}%
  \BibitemOpen
  \bibfield  {author} {\bibinfo {author} {\bibfnamefont {K.}~\bibnamefont
  {Kawaguchi}}, \bibinfo {author} {\bibfnamefont {S.}~\bibnamefont
  {Fujibayashi}},\ and\ \bibinfo {author} {\bibfnamefont {M.}~\bibnamefont
  {Shibata}},\ }\href {https://doi.org/10.1103/PhysRevD.111.023015} {\bibfield
  {journal} {\bibinfo  {journal} {Phys. Rev. D}\ }\textbf {\bibinfo {volume}
  {111}},\ \bibinfo {pages} {023015} (\bibinfo {year} {2025})},\ \Eprint
  {https://arxiv.org/abs/2410.02380} {arXiv:2410.02380 [astro-ph.HE]}
  \BibitemShut {NoStop}%
\bibitem [{\citenamefont {{Font}}\ \emph {et~al.}(1998)\citenamefont {{Font}},
  \citenamefont {{Ib{\'a}{\~n}ez}},\ and\ \citenamefont
  {{Papadopoulos}}}]{1998ApJ...507L..67F}%
  \BibitemOpen
  \bibfield  {author} {\bibinfo {author} {\bibfnamefont {J.~A.}\ \bibnamefont
  {{Font}}}, \bibinfo {author} {\bibfnamefont {J.~M.}\ \bibnamefont
  {{Ib{\'a}{\~n}ez}}},\ and\ \bibinfo {author} {\bibfnamefont {P.}~\bibnamefont
  {{Papadopoulos}}},\ }\href {https://doi.org/10.1086/311666} {\bibfield
  {journal} {\bibinfo  {journal} {\apjl}\ }\textbf {\bibinfo {volume} {507}},\
  \bibinfo {pages} {L67} (\bibinfo {year} {1998})},\ \Eprint
  {https://arxiv.org/abs/astro-ph/9805269} {arXiv:astro-ph/9805269 [astro-ph]}
  \BibitemShut {NoStop}%
\bibitem [{\citenamefont {Shibata}\ \emph
  {et~al.}(2017{\natexlab{b}})\citenamefont {Shibata}, \citenamefont {Kiuchi},\
  and\ \citenamefont {Sekiguchi}}]{Shibata:2017jyf}%
  \BibitemOpen
  \bibfield  {author} {\bibinfo {author} {\bibfnamefont {M.}~\bibnamefont
  {Shibata}}, \bibinfo {author} {\bibfnamefont {K.}~\bibnamefont {Kiuchi}},\
  and\ \bibinfo {author} {\bibfnamefont {Y.-i.}\ \bibnamefont {Sekiguchi}},\
  }\href {https://doi.org/10.1103/PhysRevD.95.083005} {\bibfield  {journal}
  {\bibinfo  {journal} {Phys. Rev. D}\ }\textbf {\bibinfo {volume} {95}},\
  \bibinfo {pages} {083005} (\bibinfo {year} {2017}{\natexlab{b}})},\ \Eprint
  {https://arxiv.org/abs/1703.10303} {arXiv:1703.10303 [astro-ph.HE]}
  \BibitemShut {NoStop}%
\bibitem [{\citenamefont {Balbus}\ and\ \citenamefont
  {Hawley}(1998)}]{Balbus:1998ja}%
  \BibitemOpen
  \bibfield  {author} {\bibinfo {author} {\bibfnamefont {S.~A.}\ \bibnamefont
  {Balbus}}\ and\ \bibinfo {author} {\bibfnamefont {J.~F.}\ \bibnamefont
  {Hawley}},\ }\href {https://doi.org/10.1103/RevModPhys.70.1} {\bibfield
  {journal} {\bibinfo  {journal} {Rev. Mod. Phys.}\ }\textbf {\bibinfo {volume}
  {70}},\ \bibinfo {pages} {1} (\bibinfo {year} {1998})}\BibitemShut {NoStop}%
\bibitem [{\citenamefont {{Shakura}}\ and\ \citenamefont
  {{Sunyaev}}(1973)}]{1973A&A....24..337S}%
  \BibitemOpen
  \bibfield  {author} {\bibinfo {author} {\bibfnamefont {N.~I.}\ \bibnamefont
  {{Shakura}}}\ and\ \bibinfo {author} {\bibfnamefont {R.~A.}\ \bibnamefont
  {{Sunyaev}}},\ }\href@noop {} {\bibfield  {journal} {\bibinfo  {journal}
  {\aap}\ }\textbf {\bibinfo {volume} {24}},\ \bibinfo {pages} {337} (\bibinfo
  {year} {1973})}\BibitemShut {NoStop}%
\bibitem [{\citenamefont {Horowitz}(2002)}]{Horowitz:2001xf}%
  \BibitemOpen
  \bibfield  {author} {\bibinfo {author} {\bibfnamefont {C.~J.}\ \bibnamefont
  {Horowitz}},\ }\href {https://doi.org/10.1103/PhysRevD.65.043001} {\bibfield
  {journal} {\bibinfo  {journal} {Phys. Rev. D}\ }\textbf {\bibinfo {volume}
  {65}},\ \bibinfo {pages} {043001} (\bibinfo {year} {2002})},\ \Eprint
  {https://arxiv.org/abs/astro-ph/0109209} {arXiv:astro-ph/0109209}
  \BibitemShut {NoStop}%
\bibitem [{\citenamefont {Meszaros}\ and\ \citenamefont
  {Rees}(2000)}]{Meszaros:1999gb}%
  \BibitemOpen
  \bibfield  {author} {\bibinfo {author} {\bibfnamefont {P.}~\bibnamefont
  {Meszaros}}\ and\ \bibinfo {author} {\bibfnamefont {M.~J.}\ \bibnamefont
  {Rees}},\ }\href {https://doi.org/10.1086/308371} {\bibfield  {journal}
  {\bibinfo  {journal} {Astrophys. J.}\ }\textbf {\bibinfo {volume} {530}},\
  \bibinfo {pages} {292} (\bibinfo {year} {2000})},\ \Eprint
  {https://arxiv.org/abs/astro-ph/9908126} {arXiv:astro-ph/9908126}
  \BibitemShut {NoStop}%
\bibitem [{\citenamefont {Metzger}\ \emph {et~al.}(2008)\citenamefont
  {Metzger}, \citenamefont {Piro},\ and\ \citenamefont
  {Quataert}}]{Metzger:2008av}%
  \BibitemOpen
  \bibfield  {author} {\bibinfo {author} {\bibfnamefont {B.~D.}\ \bibnamefont
  {Metzger}}, \bibinfo {author} {\bibfnamefont {A.~L.}\ \bibnamefont {Piro}},\
  and\ \bibinfo {author} {\bibfnamefont {E.}~\bibnamefont {Quataert}},\ }\href
  {https://doi.org/10.1111/j.1365-2966.2008.13789.x} {\bibfield  {journal}
  {\bibinfo  {journal} {Mon. Not. Roy. Astron. Soc.}\ }\textbf {\bibinfo
  {volume} {390}},\ \bibinfo {pages} {781} (\bibinfo {year} {2008})},\ \Eprint
  {https://arxiv.org/abs/0805.4415} {arXiv:0805.4415 [astro-ph]} \BibitemShut
  {NoStop}%
\bibitem [{\citenamefont {Chen}\ and\ \citenamefont
  {Beloborodov}(2007)}]{Chen:2006rra}%
  \BibitemOpen
  \bibfield  {author} {\bibinfo {author} {\bibfnamefont {W.-X.}\ \bibnamefont
  {Chen}}\ and\ \bibinfo {author} {\bibfnamefont {A.~M.}\ \bibnamefont
  {Beloborodov}},\ }\href {https://doi.org/10.1086/508923} {\bibfield
  {journal} {\bibinfo  {journal} {Astrophys. J.}\ }\textbf {\bibinfo {volume}
  {657}},\ \bibinfo {pages} {383} (\bibinfo {year} {2007})},\ \Eprint
  {https://arxiv.org/abs/astro-ph/0607145} {arXiv:astro-ph/0607145}
  \BibitemShut {NoStop}%
\bibitem [{\citenamefont {Zalamea}\ and\ \citenamefont
  {Beloborodov}(2011)}]{Zalamea:2010ax}%
  \BibitemOpen
  \bibfield  {author} {\bibinfo {author} {\bibfnamefont {I.}~\bibnamefont
  {Zalamea}}\ and\ \bibinfo {author} {\bibfnamefont {A.~M.}\ \bibnamefont
  {Beloborodov}},\ }\href {https://doi.org/10.1111/j.1365-2966.2010.17600.x}
  {\bibfield  {journal} {\bibinfo  {journal} {Mon. Not. Roy. Astron. Soc.}\
  }\textbf {\bibinfo {volume} {410}},\ \bibinfo {pages} {2302} (\bibinfo {year}
  {2011})},\ \Eprint {https://arxiv.org/abs/1003.0710} {arXiv:1003.0710
  [astro-ph.HE]} \BibitemShut {NoStop}%
\bibitem [{\citenamefont {Agarwal}\ \emph {et~al.}(2025)\citenamefont
  {Agarwal}, \citenamefont {Siegel}, \citenamefont {Metzger},\ and\
  \citenamefont {Nagele}}]{Agarwal:2025gbw}%
  \BibitemOpen
  \bibfield  {author} {\bibinfo {author} {\bibfnamefont {A.}~\bibnamefont
  {Agarwal}}, \bibinfo {author} {\bibfnamefont {D.~M.}\ \bibnamefont {Siegel}},
  \bibinfo {author} {\bibfnamefont {B.~D.}\ \bibnamefont {Metzger}},\ and\
  \bibinfo {author} {\bibfnamefont {C.}~\bibnamefont {Nagele}},\ }\href@noop {}
  {\bibfield  {journal} {\bibinfo  {journal} {arXiv}\ } (\bibinfo {year}
  {2025})},\ \Eprint {https://arxiv.org/abs/2503.15729} {arXiv:2503.15729
  [astro-ph.HE]} \BibitemShut {NoStop}%
\bibitem [{\citenamefont {Aloy}\ \emph {et~al.}(2004)\citenamefont {Aloy},
  \citenamefont {Janka},\ and\ \citenamefont {Muller}}]{Aloy:2004nh}%
  \BibitemOpen
  \bibfield  {author} {\bibinfo {author} {\bibfnamefont {M.~A.}\ \bibnamefont
  {Aloy}}, \bibinfo {author} {\bibfnamefont {H.~T.}\ \bibnamefont {Janka}},\
  and\ \bibinfo {author} {\bibfnamefont {E.}~\bibnamefont {Muller}},\ }\href
  {https://doi.org/10.1051/0004-6361:20041865} {\bibfield  {journal} {\bibinfo
  {journal} {eConf}\ }\textbf {\bibinfo {volume} {C041213}},\ \bibinfo {pages}
  {0109} (\bibinfo {year} {2004})},\ \Eprint
  {https://arxiv.org/abs/astro-ph/0408291} {arXiv:astro-ph/0408291}
  \BibitemShut {NoStop}%
\bibitem [{\citenamefont {Birkl}\ \emph {et~al.}(2007)\citenamefont {Birkl},
  \citenamefont {Aloy}, \citenamefont {Janka},\ and\ \citenamefont
  {Mueller}}]{Birkl:2006mu}%
  \BibitemOpen
  \bibfield  {author} {\bibinfo {author} {\bibfnamefont {R.}~\bibnamefont
  {Birkl}}, \bibinfo {author} {\bibfnamefont {M.~A.}\ \bibnamefont {Aloy}},
  \bibinfo {author} {\bibfnamefont {H.~T.}\ \bibnamefont {Janka}},\ and\
  \bibinfo {author} {\bibfnamefont {E.}~\bibnamefont {Mueller}},\ }\href
  {https://doi.org/10.1051/0004-6361:20066293} {\bibfield  {journal} {\bibinfo
  {journal} {Astron. Astrophys.}\ }\textbf {\bibinfo {volume} {463}},\ \bibinfo
  {pages} {51} (\bibinfo {year} {2007})},\ \Eprint
  {https://arxiv.org/abs/astro-ph/0608543} {arXiv:astro-ph/0608543}
  \BibitemShut {NoStop}%
\bibitem [{\citenamefont {{Liu}}\ \emph {et~al.}(2021)\citenamefont {{Liu}},
  \citenamefont {{Zhang}},\ and\ \citenamefont {{Zhu}}}]{2021RAA....21..254L}%
  \BibitemOpen
  \bibfield  {author} {\bibinfo {author} {\bibfnamefont {Z.-Y.}\ \bibnamefont
  {{Liu}}}, \bibinfo {author} {\bibfnamefont {F.-W.}\ \bibnamefont {{Zhang}}},\
  and\ \bibinfo {author} {\bibfnamefont {S.-Y.}\ \bibnamefont {{Zhu}}},\ }\href
  {https://doi.org/10.1088/1674-4527/21/10/254} {\bibfield  {journal} {\bibinfo
   {journal} {Research in Astronomy and Astrophysics}\ }\textbf {\bibinfo
  {volume} {21}},\ \bibinfo {eid} {254} (\bibinfo {year} {2021})}\BibitemShut
  {NoStop}%
\bibitem [{\citenamefont {Janka}\ \emph {et~al.}(2006)\citenamefont {Janka},
  \citenamefont {Mazzali}, \citenamefont {Aloy},\ and\ \citenamefont
  {Pian}}]{Janka:2005yh}%
  \BibitemOpen
  \bibfield  {author} {\bibinfo {author} {\bibfnamefont {H.-T.}\ \bibnamefont
  {Janka}}, \bibinfo {author} {\bibfnamefont {P.~A.}\ \bibnamefont {Mazzali}},
  \bibinfo {author} {\bibfnamefont {M.~A.}\ \bibnamefont {Aloy}},\ and\
  \bibinfo {author} {\bibfnamefont {E.}~\bibnamefont {Pian}},\ }\href
  {https://doi.org/10.1086/504580} {\bibfield  {journal} {\bibinfo  {journal}
  {Astrophys. J.}\ }\textbf {\bibinfo {volume} {645}},\ \bibinfo {pages} {1305}
  (\bibinfo {year} {2006})},\ \Eprint {https://arxiv.org/abs/astro-ph/0509722}
  {arXiv:astro-ph/0509722} \BibitemShut {NoStop}%
\bibitem [{\citenamefont {{Troja}}\ \emph {et~al.}(2010)\citenamefont
  {{Troja}}, \citenamefont {{Rosswog}},\ and\ \citenamefont
  {{Gehrels}}}]{2010ApJ...723.1711T}%
  \BibitemOpen
  \bibfield  {author} {\bibinfo {author} {\bibfnamefont {E.}~\bibnamefont
  {{Troja}}}, \bibinfo {author} {\bibfnamefont {S.}~\bibnamefont {{Rosswog}}},\
  and\ \bibinfo {author} {\bibfnamefont {N.}~\bibnamefont {{Gehrels}}},\ }\href
  {https://doi.org/10.1088/0004-637X/723/2/1711} {\bibfield  {journal}
  {\bibinfo  {journal} {\apj}\ }\textbf {\bibinfo {volume} {723}},\ \bibinfo
  {pages} {1711} (\bibinfo {year} {2010})},\ \Eprint
  {https://arxiv.org/abs/1009.1385} {arXiv:1009.1385 [astro-ph.HE]}
  \BibitemShut {NoStop}%
\bibitem [{\citenamefont {Xiao}\ \emph {et~al.}(2024)\citenamefont {Xiao} \emph
  {et~al.}}]{Xiao:2022quv}%
  \BibitemOpen
  \bibfield  {author} {\bibinfo {author} {\bibfnamefont {S.}~\bibnamefont
  {Xiao}} \emph {et~al.},\ }\href {https://doi.org/10.3847/1538-4357/ad4ee1}
  {\bibfield  {journal} {\bibinfo  {journal} {Astrophys. J.}\ }\textbf
  {\bibinfo {volume} {970}},\ \bibinfo {pages} {6} (\bibinfo {year} {2024})},\
  \Eprint {https://arxiv.org/abs/2205.02186} {arXiv:2205.02186 [astro-ph.HE]}
  \BibitemShut {NoStop}%
\bibitem [{\citenamefont {Kiuchi}\ \emph {et~al.}(2009)\citenamefont {Kiuchi},
  \citenamefont {Sekiguchi}, \citenamefont {Shibata},\ and\ \citenamefont
  {Taniguchi}}]{Kiuchi:2009jt}%
  \BibitemOpen
  \bibfield  {author} {\bibinfo {author} {\bibfnamefont {K.}~\bibnamefont
  {Kiuchi}}, \bibinfo {author} {\bibfnamefont {Y.}~\bibnamefont {Sekiguchi}},
  \bibinfo {author} {\bibfnamefont {M.}~\bibnamefont {Shibata}},\ and\ \bibinfo
  {author} {\bibfnamefont {K.}~\bibnamefont {Taniguchi}},\ }\href
  {https://doi.org/10.1103/PhysRevD.80.064037} {\bibfield  {journal} {\bibinfo
  {journal} {Phys. Rev.}\ }\textbf {\bibinfo {volume} {D80}},\ \bibinfo {pages}
  {064037} (\bibinfo {year} {2009})},\ \Eprint
  {https://arxiv.org/abs/0904.4551} {arXiv:0904.4551 [gr-qc]} \BibitemShut
  {NoStop}%
\bibitem [{\citenamefont {Dietrich}\ \emph {et~al.}(2017)\citenamefont
  {Dietrich}, \citenamefont {Ujevic}, \citenamefont {Tichy}, \citenamefont
  {Bernuzzi},\ and\ \citenamefont {Bruegmann}}]{Dietrich:2016hky}%
  \BibitemOpen
  \bibfield  {author} {\bibinfo {author} {\bibfnamefont {T.}~\bibnamefont
  {Dietrich}}, \bibinfo {author} {\bibfnamefont {M.}~\bibnamefont {Ujevic}},
  \bibinfo {author} {\bibfnamefont {W.}~\bibnamefont {Tichy}}, \bibinfo
  {author} {\bibfnamefont {S.}~\bibnamefont {Bernuzzi}},\ and\ \bibinfo
  {author} {\bibfnamefont {B.}~\bibnamefont {Bruegmann}},\ }\href
  {https://doi.org/10.1103/PhysRevD.95.024029} {\bibfield  {journal} {\bibinfo
  {journal} {Phys. Rev.}\ }\textbf {\bibinfo {volume} {D95}},\ \bibinfo {pages}
  {024029} (\bibinfo {year} {2017})},\ \Eprint
  {https://arxiv.org/abs/1607.06636} {arXiv:1607.06636 [gr-qc]} \BibitemShut
  {NoStop}%
\bibitem [{\citenamefont {Schianchi}\ \emph {et~al.}(2024)\citenamefont
  {Schianchi}, \citenamefont {Ujevic}, \citenamefont {Neuweiler}, \citenamefont
  {Gieg}, \citenamefont {Markin},\ and\ \citenamefont
  {Dietrich}}]{Schianchi:2024vvi}%
  \BibitemOpen
  \bibfield  {author} {\bibinfo {author} {\bibfnamefont {F.}~\bibnamefont
  {Schianchi}}, \bibinfo {author} {\bibfnamefont {M.}~\bibnamefont {Ujevic}},
  \bibinfo {author} {\bibfnamefont {A.}~\bibnamefont {Neuweiler}}, \bibinfo
  {author} {\bibfnamefont {H.}~\bibnamefont {Gieg}}, \bibinfo {author}
  {\bibfnamefont {I.}~\bibnamefont {Markin}},\ and\ \bibinfo {author}
  {\bibfnamefont {T.}~\bibnamefont {Dietrich}},\ }\href
  {https://doi.org/10.1103/PhysRevD.109.123011} {\bibfield  {journal} {\bibinfo
   {journal} {Phys. Rev. D}\ }\textbf {\bibinfo {volume} {109}},\ \bibinfo
  {pages} {123011} (\bibinfo {year} {2024})},\ \Eprint
  {https://arxiv.org/abs/2402.16626} {arXiv:2402.16626 [astro-ph.HE]}
  \BibitemShut {NoStop}%
\bibitem [{\citenamefont {Karakas}\ \emph {et~al.}(2024)\citenamefont
  {Karakas}, \citenamefont {Matur},\ and\ \citenamefont
  {Ruffert}}]{Karakas:2024avr}%
  \BibitemOpen
  \bibfield  {author} {\bibinfo {author} {\bibfnamefont {B.}~\bibnamefont
  {Karakas}}, \bibinfo {author} {\bibfnamefont {R.}~\bibnamefont {Matur}},\
  and\ \bibinfo {author} {\bibfnamefont {M.}~\bibnamefont {Ruffert}},\
  }\href@noop {} {\bibfield  {journal} {\bibinfo  {journal} {arXiv}\ }
  (\bibinfo {year} {2024})},\ \Eprint {https://arxiv.org/abs/2405.13687}
  {arXiv:2405.13687 [astro-ph.HE]} \BibitemShut {NoStop}%
\bibitem [{\citenamefont {Hotokezaka}\ \emph {et~al.}(2013)\citenamefont
  {Hotokezaka}, \citenamefont {Kiuchi}, \citenamefont {Kyutoku}, \citenamefont
  {Muranushi}, \citenamefont {Sekiguchi}, \citenamefont {Shibata},\ and\
  \citenamefont {Taniguchi}}]{Hotokezaka:2013iia}%
  \BibitemOpen
  \bibfield  {author} {\bibinfo {author} {\bibfnamefont {K.}~\bibnamefont
  {Hotokezaka}}, \bibinfo {author} {\bibfnamefont {K.}~\bibnamefont {Kiuchi}},
  \bibinfo {author} {\bibfnamefont {K.}~\bibnamefont {Kyutoku}}, \bibinfo
  {author} {\bibfnamefont {T.}~\bibnamefont {Muranushi}}, \bibinfo {author}
  {\bibfnamefont {Y.-i.}\ \bibnamefont {Sekiguchi}}, \bibinfo {author}
  {\bibfnamefont {M.}~\bibnamefont {Shibata}},\ and\ \bibinfo {author}
  {\bibfnamefont {K.}~\bibnamefont {Taniguchi}},\ }\href
  {https://doi.org/10.1103/PhysRevD.88.044026} {\bibfield  {journal} {\bibinfo
  {journal} {Phys. Rev. D}\ }\textbf {\bibinfo {volume} {88}},\ \bibinfo
  {pages} {044026} (\bibinfo {year} {2013})},\ \Eprint
  {https://arxiv.org/abs/1307.5888} {arXiv:1307.5888 [astro-ph.HE]}
  \BibitemShut {NoStop}%
\bibitem [{\citenamefont {Kiuchi}\ \emph {et~al.}(2019)\citenamefont {Kiuchi},
  \citenamefont {Kyutoku}, \citenamefont {Shibata},\ and\ \citenamefont
  {Taniguchi}}]{Kiuchi:2019lls}%
  \BibitemOpen
  \bibfield  {author} {\bibinfo {author} {\bibfnamefont {K.}~\bibnamefont
  {Kiuchi}}, \bibinfo {author} {\bibfnamefont {K.}~\bibnamefont {Kyutoku}},
  \bibinfo {author} {\bibfnamefont {M.}~\bibnamefont {Shibata}},\ and\ \bibinfo
  {author} {\bibfnamefont {K.}~\bibnamefont {Taniguchi}},\ }\href
  {https://doi.org/10.3847/2041-8213/ab1e45} {\bibfield  {journal} {\bibinfo
  {journal} {Astrophys. J.}\ }\textbf {\bibinfo {volume} {876}},\ \bibinfo
  {pages} {L31} (\bibinfo {year} {2019})},\ \Eprint
  {https://arxiv.org/abs/1903.01466} {arXiv:1903.01466 [astro-ph.HE]}
  \BibitemShut {NoStop}%
\bibitem [{\citenamefont {Gottlieb}\ \emph
  {et~al.}(2022{\natexlab{b}})\citenamefont {Gottlieb}, \citenamefont
  {Moseley}, \citenamefont {Ramirez-Aguilar}, \citenamefont {Murguia-Berthier},
  \citenamefont {Liska},\ and\ \citenamefont
  {Tchekhovskoy}}]{Gottlieb:2022sis}%
  \BibitemOpen
  \bibfield  {author} {\bibinfo {author} {\bibfnamefont {O.}~\bibnamefont
  {Gottlieb}}, \bibinfo {author} {\bibfnamefont {S.}~\bibnamefont {Moseley}},
  \bibinfo {author} {\bibfnamefont {T.}~\bibnamefont {Ramirez-Aguilar}},
  \bibinfo {author} {\bibfnamefont {A.}~\bibnamefont {Murguia-Berthier}},
  \bibinfo {author} {\bibfnamefont {M.}~\bibnamefont {Liska}},\ and\ \bibinfo
  {author} {\bibfnamefont {A.}~\bibnamefont {Tchekhovskoy}},\ }\href
  {https://doi.org/10.3847/2041-8213/ac7728} {\bibfield  {journal} {\bibinfo
  {journal} {Astrophys. J. Lett.}\ }\textbf {\bibinfo {volume} {933}},\
  \bibinfo {pages} {L2} (\bibinfo {year} {2022}{\natexlab{b}})},\ \Eprint
  {https://arxiv.org/abs/2205.01691} {arXiv:2205.01691 [astro-ph.HE]}
  \BibitemShut {NoStop}%
\bibitem [{\citenamefont {Foucart}(2018)}]{Foucart:2017mbt}%
  \BibitemOpen
  \bibfield  {author} {\bibinfo {author} {\bibfnamefont {F.}~\bibnamefont
  {Foucart}},\ }\href {https://doi.org/10.1093/mnras/sty108} {\bibfield
  {journal} {\bibinfo  {journal} {Mon. Not. Roy. Astron. Soc.}\ }\textbf
  {\bibinfo {volume} {475}},\ \bibinfo {pages} {4186} (\bibinfo {year}
  {2018})},\ \Eprint {https://arxiv.org/abs/1708.08452} {arXiv:1708.08452
  [astro-ph.HE]} \BibitemShut {NoStop}%
\bibitem [{\citenamefont {Kyutoku}\ \emph {et~al.}(2015)\citenamefont
  {Kyutoku}, \citenamefont {Ioka}, \citenamefont {Okawa}, \citenamefont
  {Shibata},\ and\ \citenamefont {Taniguchi}}]{Kyutoku:2015gda}%
  \BibitemOpen
  \bibfield  {author} {\bibinfo {author} {\bibfnamefont {K.}~\bibnamefont
  {Kyutoku}}, \bibinfo {author} {\bibfnamefont {K.}~\bibnamefont {Ioka}},
  \bibinfo {author} {\bibfnamefont {H.}~\bibnamefont {Okawa}}, \bibinfo
  {author} {\bibfnamefont {M.}~\bibnamefont {Shibata}},\ and\ \bibinfo {author}
  {\bibfnamefont {K.}~\bibnamefont {Taniguchi}},\ }\href
  {https://doi.org/10.1103/PhysRevD.92.044028} {\bibfield  {journal} {\bibinfo
  {journal} {Phys. Rev.}\ }\textbf {\bibinfo {volume} {D92}},\ \bibinfo {pages}
  {044028} (\bibinfo {year} {2015})},\ \Eprint
  {https://arxiv.org/abs/1502.05402} {arXiv:1502.05402 [astro-ph.HE]}
  \BibitemShut {NoStop}%
\bibitem [{\citenamefont {Kyutoku}\ \emph {et~al.}(2011)\citenamefont
  {Kyutoku}, \citenamefont {Okawa}, \citenamefont {Shibata},\ and\
  \citenamefont {Taniguchi}}]{Kyutoku:2011vz}%
  \BibitemOpen
  \bibfield  {author} {\bibinfo {author} {\bibfnamefont {K.}~\bibnamefont
  {Kyutoku}}, \bibinfo {author} {\bibfnamefont {H.}~\bibnamefont {Okawa}},
  \bibinfo {author} {\bibfnamefont {M.}~\bibnamefont {Shibata}},\ and\ \bibinfo
  {author} {\bibfnamefont {K.}~\bibnamefont {Taniguchi}},\ }\href
  {https://doi.org/10.1103/PhysRevD.84.064018} {\bibfield  {journal} {\bibinfo
  {journal} {Phys. Rev. D}\ }\textbf {\bibinfo {volume} {84}},\ \bibinfo
  {pages} {064018} (\bibinfo {year} {2011})},\ \Eprint
  {https://arxiv.org/abs/1108.1189} {arXiv:1108.1189 [astro-ph.HE]}
  \BibitemShut {NoStop}%
\bibitem [{\citenamefont {Rosswog}(2005)}]{Rosswog:2005su}%
  \BibitemOpen
  \bibfield  {author} {\bibinfo {author} {\bibfnamefont {S.}~\bibnamefont
  {Rosswog}},\ }\href {https://doi.org/10.1086/497062} {\bibfield  {journal}
  {\bibinfo  {journal} {Astrophys. J.}\ }\textbf {\bibinfo {volume} {634}},\
  \bibinfo {pages} {1202} (\bibinfo {year} {2005})},\ \Eprint
  {https://arxiv.org/abs/astro-ph/0508138} {arXiv:astro-ph/0508138 [astro-ph]}
  \BibitemShut {NoStop}%
\bibitem [{\citenamefont {Shibata}\ and\ \citenamefont
  {Taniguchi}(2008)}]{Shibata:2007zm}%
  \BibitemOpen
  \bibfield  {author} {\bibinfo {author} {\bibfnamefont {M.}~\bibnamefont
  {Shibata}}\ and\ \bibinfo {author} {\bibfnamefont {K.}~\bibnamefont
  {Taniguchi}},\ }\href {https://doi.org/10.1103/PhysRevD.77.084015} {\bibfield
   {journal} {\bibinfo  {journal} {Phys. Rev.}\ }\textbf {\bibinfo {volume}
  {D77}},\ \bibinfo {pages} {084015} (\bibinfo {year} {2008})},\ \Eprint
  {https://arxiv.org/abs/0711.1410} {arXiv:0711.1410 [gr-qc]} \BibitemShut
  {NoStop}%
\bibitem [{\citenamefont {Etienne}\ \emph {et~al.}(2009)\citenamefont
  {Etienne}, \citenamefont {Liu}, \citenamefont {Shapiro},\ and\ \citenamefont
  {Baumgarte}}]{Etienne:2008re}%
  \BibitemOpen
  \bibfield  {author} {\bibinfo {author} {\bibfnamefont {Z.~B.}\ \bibnamefont
  {Etienne}}, \bibinfo {author} {\bibfnamefont {Y.~T.}\ \bibnamefont {Liu}},
  \bibinfo {author} {\bibfnamefont {S.~L.}\ \bibnamefont {Shapiro}},\ and\
  \bibinfo {author} {\bibfnamefont {T.~W.}\ \bibnamefont {Baumgarte}},\ }\href
  {https://doi.org/10.1103/PhysRevD.79.044024} {\bibfield  {journal} {\bibinfo
  {journal} {Phys. Rev.}\ }\textbf {\bibinfo {volume} {D79}},\ \bibinfo {pages}
  {044024} (\bibinfo {year} {2009})},\ \Eprint
  {https://arxiv.org/abs/0812.2245} {arXiv:0812.2245 [astro-ph]} \BibitemShut
  {NoStop}%
\bibitem [{\citenamefont {Lovelace}\ \emph {et~al.}(2013)\citenamefont
  {Lovelace}, \citenamefont {Duez}, \citenamefont {Foucart}, \citenamefont
  {Kidder}, \citenamefont {Pfeiffer}, \citenamefont {Scheel},\ and\
  \citenamefont {Szilegyi}}]{Lovelace:2013vma}%
  \BibitemOpen
  \bibfield  {author} {\bibinfo {author} {\bibfnamefont {G.}~\bibnamefont
  {Lovelace}}, \bibinfo {author} {\bibfnamefont {M.~D.}\ \bibnamefont {Duez}},
  \bibinfo {author} {\bibfnamefont {F.}~\bibnamefont {Foucart}}, \bibinfo
  {author} {\bibfnamefont {L.~E.}\ \bibnamefont {Kidder}}, \bibinfo {author}
  {\bibfnamefont {H.~P.}\ \bibnamefont {Pfeiffer}}, \bibinfo {author}
  {\bibfnamefont {M.~A.}\ \bibnamefont {Scheel}},\ and\ \bibinfo {author}
  {\bibfnamefont {B.}~\bibnamefont {Szilegyi}},\ }\href
  {https://doi.org/10.1088/0264-9381/30/13/135004} {\bibfield  {journal}
  {\bibinfo  {journal} {Class. Quant. Grav.}\ }\textbf {\bibinfo {volume}
  {30}},\ \bibinfo {pages} {135004} (\bibinfo {year} {2013})},\ \Eprint
  {https://arxiv.org/abs/1302.6297} {arXiv:1302.6297 [gr-qc]} \BibitemShut
  {NoStop}%
\bibitem [{\citenamefont {Foucart}\ \emph {et~al.}(2018)\citenamefont
  {Foucart}, \citenamefont {Hinderer},\ and\ \citenamefont
  {Nissanke}}]{Foucart:2018rjc}%
  \BibitemOpen
  \bibfield  {author} {\bibinfo {author} {\bibfnamefont {F.}~\bibnamefont
  {Foucart}}, \bibinfo {author} {\bibfnamefont {T.}~\bibnamefont {Hinderer}},\
  and\ \bibinfo {author} {\bibfnamefont {S.}~\bibnamefont {Nissanke}},\ }\href
  {https://doi.org/10.1103/PhysRevD.98.081501} {\bibfield  {journal} {\bibinfo
  {journal} {Phys. Rev.}\ }\textbf {\bibinfo {volume} {D98}},\ \bibinfo {pages}
  {081501} (\bibinfo {year} {2018})},\ \Eprint
  {https://arxiv.org/abs/1807.00011} {arXiv:1807.00011 [astro-ph.HE]}
  \BibitemShut {NoStop}%
\bibitem [{\citenamefont {Fernández}\ \emph {et~al.}(2019)\citenamefont
  {Fernández}, \citenamefont {Tchekhovskoy}, \citenamefont {Quataert},
  \citenamefont {Foucart},\ and\ \citenamefont {Kasen}}]{Fernandez:2018kax}%
  \BibitemOpen
  \bibfield  {author} {\bibinfo {author} {\bibfnamefont {R.}~\bibnamefont
  {Fernández}}, \bibinfo {author} {\bibfnamefont {A.}~\bibnamefont
  {Tchekhovskoy}}, \bibinfo {author} {\bibfnamefont {E.}~\bibnamefont
  {Quataert}}, \bibinfo {author} {\bibfnamefont {F.}~\bibnamefont {Foucart}},\
  and\ \bibinfo {author} {\bibfnamefont {D.}~\bibnamefont {Kasen}},\ }\href
  {https://doi.org/10.1093/mnras/sty2932} {\bibfield  {journal} {\bibinfo
  {journal} {Mon. Not. Roy. Astron. Soc.}\ }\textbf {\bibinfo {volume} {482}},\
  \bibinfo {pages} {3373} (\bibinfo {year} {2019})},\ \Eprint
  {https://arxiv.org/abs/1808.00461} {arXiv:1808.00461 [astro-ph.HE]}
  \BibitemShut {NoStop}%
\bibitem [{\citenamefont {Christie}\ \emph {et~al.}(2019)\citenamefont
  {Christie}, \citenamefont {Lalakos}, \citenamefont {Tchekhovskoy},
  \citenamefont {Fern\'andez}, \citenamefont {Foucart}, \citenamefont
  {Quataert},\ and\ \citenamefont {Kasen}}]{Christie:2019lim}%
  \BibitemOpen
  \bibfield  {author} {\bibinfo {author} {\bibfnamefont {I.~M.}\ \bibnamefont
  {Christie}}, \bibinfo {author} {\bibfnamefont {A.}~\bibnamefont {Lalakos}},
  \bibinfo {author} {\bibfnamefont {A.}~\bibnamefont {Tchekhovskoy}}, \bibinfo
  {author} {\bibfnamefont {R.}~\bibnamefont {Fern\'andez}}, \bibinfo {author}
  {\bibfnamefont {F.}~\bibnamefont {Foucart}}, \bibinfo {author} {\bibfnamefont
  {E.}~\bibnamefont {Quataert}},\ and\ \bibinfo {author} {\bibfnamefont
  {D.}~\bibnamefont {Kasen}},\ }\href {https://doi.org/10.1093/mnras/stz2552}
  {\bibfield  {journal} {\bibinfo  {journal} {Mon. Not. Roy. Astron. Soc.}\
  }\textbf {\bibinfo {volume} {490}},\ \bibinfo {pages} {4811} (\bibinfo {year}
  {2019})},\ \Eprint {https://arxiv.org/abs/1907.02079} {arXiv:1907.02079
  [astro-ph.HE]} \BibitemShut {NoStop}%
\bibitem [{\citenamefont {Fujibayashi}\ \emph
  {et~al.}(2020{\natexlab{b}})\citenamefont {Fujibayashi}, \citenamefont
  {Shibata}, \citenamefont {Wanajo}, \citenamefont {Kiuchi}, \citenamefont
  {Kyutoku},\ and\ \citenamefont {Sekiguchi}}]{Fujibayashi:2020jfr}%
  \BibitemOpen
  \bibfield  {author} {\bibinfo {author} {\bibfnamefont {S.}~\bibnamefont
  {Fujibayashi}}, \bibinfo {author} {\bibfnamefont {M.}~\bibnamefont
  {Shibata}}, \bibinfo {author} {\bibfnamefont {S.}~\bibnamefont {Wanajo}},
  \bibinfo {author} {\bibfnamefont {K.}~\bibnamefont {Kiuchi}}, \bibinfo
  {author} {\bibfnamefont {K.}~\bibnamefont {Kyutoku}},\ and\ \bibinfo {author}
  {\bibfnamefont {Y.}~\bibnamefont {Sekiguchi}},\ }\href
  {https://doi.org/10.1103/PhysRevD.102.123014} {\bibfield  {journal} {\bibinfo
   {journal} {Phys. Rev. D}\ }\textbf {\bibinfo {volume} {102}},\ \bibinfo
  {pages} {123014} (\bibinfo {year} {2020}{\natexlab{b}})},\ \Eprint
  {https://arxiv.org/abs/2009.03895} {arXiv:2009.03895 [astro-ph.HE]}
  \BibitemShut {NoStop}%
\end{thebibliography}

\appendix

\section{Grid resolution and density floor effects}\label{app:err}

The finite grid resolution and presence of floor density in the simulations induce the errors in the results. To evaluate the size of these errors, we perform simulations varying the grid resolution (MT01s08LR, MT01s08HR) and floor density setup (MT01s08DF01).

We find that the difference in the total emitted neutrino luminosity and energy among the runs with different grid resolutions as well as the different floor density setup are minor compared to the MC shot noise, which is typically less than 10\% percent in our simulations (see Table~\ref{tb:jetene}). The total pair annihilation deposition rate and energy are more significantly affected by the error due to the finite grid resolutions. 
Nevertheless, the difference among the different resolution runs are within $\sim 10\%$. We also find that the floor density setup has only a minor effect on the total pair annihilation deposition rate and energy.

\begin{figure*}
 	 \includegraphics[width=0.49\linewidth]{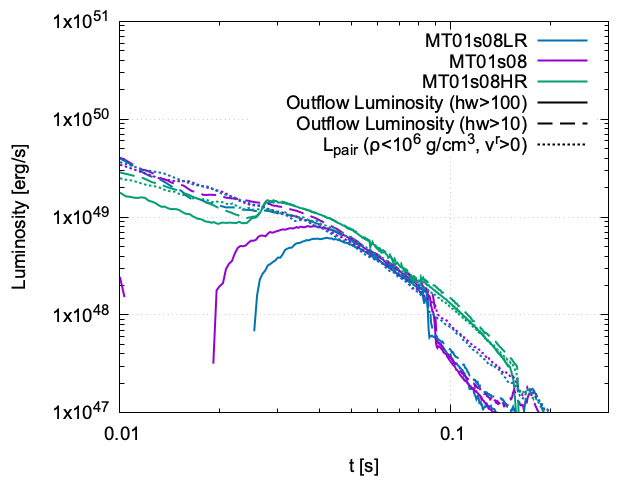}
 	 \includegraphics[width=0.49\linewidth]{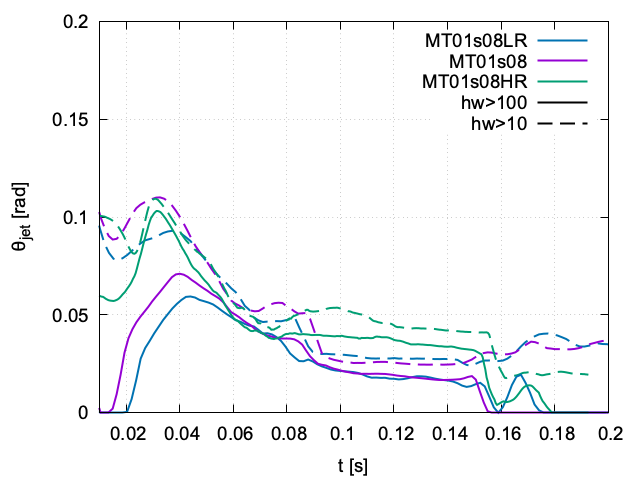}
 	 \caption{The same as Fig.~\ref{fig:relout_avis} but for the fiducial models with different grid resolution runs.}
	 \label{fig:relout_conv}
\end{figure*}

On the other hand, the relativistic outflow luminosity and energy are significantly affected by the grid resolution. Figure~\ref{fig:relout_conv} compares the results for the fiducial model among different grid resolution runs. The delayed increase and earlier sudden shut down found in the relativistic outflow luminosity in $t\lesssim 20\,{\rm ms}$ and at $\approx 90\,{\rm ms}$, respectively, for the fiducial and low resolution runs seem to be the artifact of the insufficient grid resolution. During the developing phase of the relativistic fireball ($t\lesssim 20\,{\rm ms}$) and after the pair annihilation deposition luminosity drops ($t\gtrsim 90\,{\rm ms}$), the opening angle of the relativistic outflow is small. With a low grid resolution run, the numerical diffusion of the rest-mass density in such a highly collimated outflow becomes strong and leads to the enhancement of the artificial baryon loading effect. As a consequence, the region of the outflow with the reduced terminal Lorentz factor causes the sudden drop in the relativistic outflow luminosity. For the same reason, the opening angle of the relativistic outflow becomes small as the resolution becomes low. In fact, for the high resolution run, we can see that the location of the sudden decrease in the relativistic outflow luminosity shifts in a later epoch ($\approx 150\,{\rm ms}$). Also the opening angle of the relativistic outflow shows the largest for the highest resolution run.

The total relativistic outflow energy increases approximately by a factor of 2 as the grid resolution increases. This reflects the more delayed increase and earlier sudden shut down in the relativistic outflow luminosity seen in Fig.~\ref{fig:relout_conv} for the lower grid resolution runs. Hence, we should note that more than a factor of 2 error is present in the results of the relativistic outflow luminosity and energy due to the finite grid resolution.

On the other hand, the total isotropic energy of the relativistic outflow as well as the launching duration is less affected by these errors and uncertainty. The total isotropic relativistic outflow energy shows only the variation within $\sim 20\%$ among the different resolution and different floor density runs (see Table~\ref{tb:jetene}). We also do not find a significant difference among the runs in the the time scale over which 90\% of the isotropic relativistic outflow energy is emitted. These reflect that the isotropic relativistic outflow energy is likely to be determined primarily by the isotropic equivalent deposition energy of pair annihilation (see also~\cite{Just:2015dba}). 

\begin{figure*}
    \includegraphics[width=0.49\linewidth]{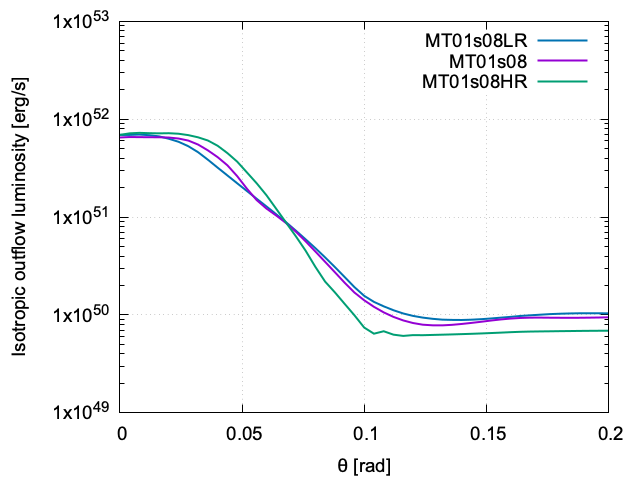}
    \includegraphics[width=0.49\linewidth]{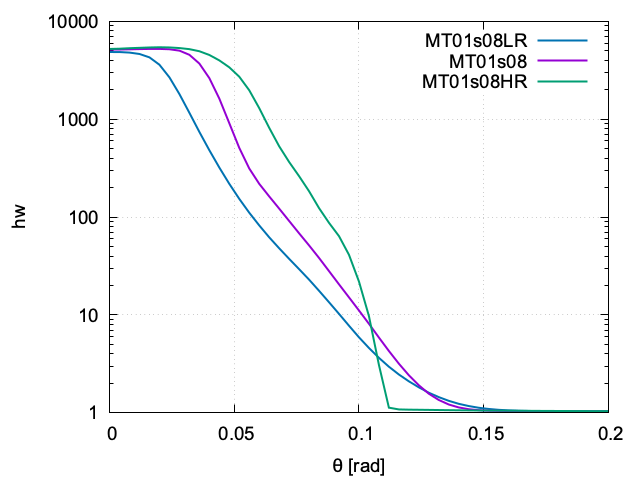}
 	 \caption{(Left panel) Latitudinal profile of the isotropic outflow luminosity averaged over $r=1500$--$3000\,{\rm km}$ obtained for the snapshot at $t=50\,{\rm ms}$ for the fiducial with different grid resolution runs. (Right panel) the same as the left panel but for the terminal Lorentz factor.}
	 \label{fig:jet_prof_chk}
\end{figure*}

The left panel of Fig.~\ref{fig:jet_prof_chk} compares the latitudinal profile of the isotropic outflow luminosity for the fiducial model with different grid resolutions. The core region of the relativistic outflow, where the isotropic luminosity is approximately constant within $<0.05\,{\rm rad}$, is extended as the grid resolution is improved, while the slope of the skirt of the mountain-like profile becomes steep. The same or even more pronounced trend is found in the latitudinal profile of the terminal Lorentz factor profile shown in the right panel of Fig.~\ref{fig:jet_prof_chk}. 

As is mentioned above, the more developed relativistic outflow component in the higher grid resolutions can be understood by the suppression of the artificial baryon loading from the high-density inner edge of the torus due to the numerical diffusion, which basically suppresses the relativistic fireball to develop. However, we note that while this trend holds within the range of resolutions explored here, different model setups and even higher grid resolutions could introduce additional effects. For example, a much higher resolution domain may resolve the KH instability around the contact surface between the relativistic outflow and non-relativistic ambient matters, which drives more significant physical baryon loading~\cite{Aloy:2004nh}. The limited computational domain of our simulations can also be the factor to underestimate the effect of the matter entrainment by the relativistic outflow: we solve the evolution of the relativistic outflow only in the domain within $\leq 4\times 10^{8}\,{\rm cm}$ in contrast to the work of~\cite{Aloy:2004nh}, which covers an extended region of $\geq 3\times 10^{9}\,{\rm cm}$. Furthermore, the terminal Lorentz factor should be significantly altered in the presence of the preceding ejecta launching during the merger (e.g., dynamical ejecta).

\begin{figure*}
    \includegraphics[width=0.49\linewidth]{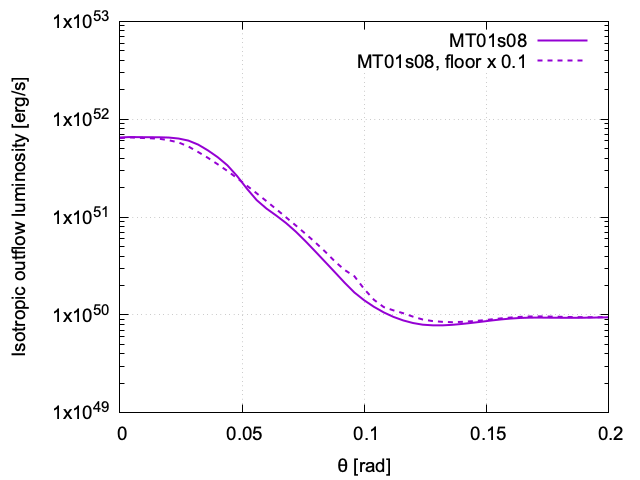} 	 \includegraphics[width=0.49\linewidth]{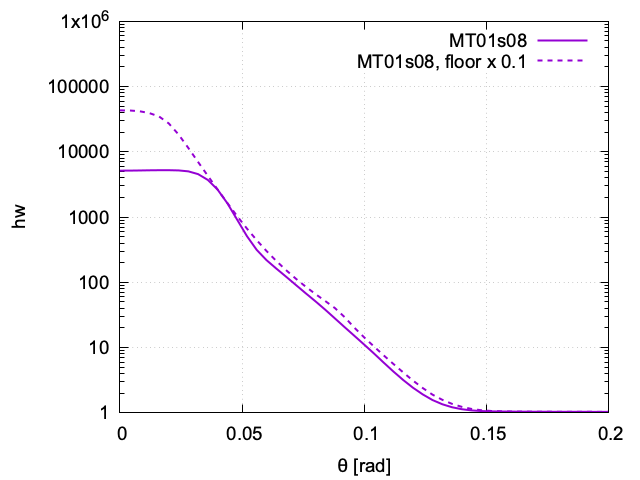}
 	 \caption{The same as Fig.~\ref{fig:jet_prof_chk} but for the fiducial model with the default and 10 times lower floor density runs.}
	 \label{fig:jet_prof_chk2}
\end{figure*}

Figure~\ref{fig:jet_prof_chk2} shows the same as Fig.~\ref{fig:jet_prof_chk} but for the fiducial model with the default and 10 times lower floor density runs. The left panel of Fig.~\ref{fig:jet_prof_chk2} indicates that the floor density setup has relatively a minor effect on the latitudinal profile of the isotropic outflow luminosity. Indeed, the total isotropic relativistic outflow energy agrees within $\approx 20\%$ for two runs. On the other hand, the right panel of Fig.~\ref{fig:jet_prof_chk2} suggests that the value of the terminal Lorentz factor is strongly affected by the setup of the floor density. The region with $hw > 100$ increases by about an order of magnitude as a result of the reduced floor density. This clearly shows that the terminal Lorentz factor is affected by the floor density. Hence, we should only take the results of the terminal Lorentz factor shown in the right panel of Fig.~\ref{fig:jet_prof} as results for a given floor density. The investigation of longer and further evolution of the relativistic outflow with sufficiently high grid resolution and low floor density is needed to precisely predict the terminal Lorentz factor of the relativistic outflow.
\end{document}